%% file: PlanckXXXIII_arXiv_July2015.tex
\documentclass[twocolumn,longauth,traditabstract]{aa}

\usepackage{natbib}
\usepackage{url}
\usepackage{graphicx}
\usepackage{subfigure}
\usepackage{amsmath}
\usepackage{txfonts}
\usepackage{color}
\usepackage{listings} 
\usepackage{times}
\usepackage{multirow}
\usepackage{float}
\usepackage{epstopdf}
\usepackage{longtable}
\usepackage{soul}
\usepackage[colorlinks=true,linkcolor=red,citecolor=blue, breaklinks=true]{hyperref}

\input{Planck.tex}

\newcommand{\hi}{\ion{H}{i}}

\newcommand{\healpix}{{\tt  HEALPix}}
\newcommand{\planck}{\textit{Planck}}  
\newcommand{\herschel}{{\it Herschel}}
\newcommand{\NH}{N_{\rm H}} % NH
\def\NHUNIT{\ifmmode {\rm \,cm^{-2}} \else $\rm \,cm^{-2}$ \fi} % NH units
\def\nhh{\ifmmode N_{\rm H_{2}}\else $N_{\rm H_{2}}$\fi} 
\def\nh{\ifmmode N_{\rm H}\else $N_{\rm H}$\fi}
\newcommand{\I}{{$I$}}
\newcommand{\U}{{$U$}}
\newcommand{\Q}{{$Q$}}
\def\arcm{\ifmmode {^{\scriptstyle\prime}}
          \else $^{\scriptstyle\prime}$\fi}
\newdimen\sa  \newdimen\sb
\def\parcs{\sa=.07em \sb=.03em
     \ifmmode \hbox{\rlap{.}}^{\scriptstyle\prime\kern -\sb\prime}\hbox{\kern -\sa}
     \else \rlap{.}$^{\scriptstyle\prime\kern -\sb\prime}$\kern -\sa\fi}
\def\parcm{\sa=.08em \sb=.03em
     \ifmmode \hbox{\rlap{.}\kern\sa}^{\scriptstyle\prime}\hbox{\kern-\sb}
     \else \rlap{.}\kern\sa$^{\scriptstyle\prime}$\kern-\sb\fi}

\def\n{}
\def\rev{}
\begin{document} 

\input{PIP_108_Arzoumanian_authors_and_institutes.tex}
\title{{\planck} intermediate results. XXXIII. Signature of the magnetic field geometry  of   interstellar filaments in dust polarization maps }
  
  \titlerunning{Signature of the magnetic field geometry  of   interstellar filaments in  dust polarization maps  }
\authorrunning{$Planck$  collaboration}

%\date{\today}

\abstract{
\planck\  observations  at 353\,GHz provide the first fully-sampled maps of the
polarized dust emission towards interstellar filaments and their backgrounds (i.e., 
the emission observed in the surroundings of the filaments). The data  allow us to  determine the intrinsic polarization properties
of the filaments and therefore to provide insight on the structure of their magnetic field ($\vec{B}$).
We present the polarization maps of three nearby (several parsecs long) star
forming filaments of moderate column density ($\NH$ about $10^{22}$\,cm$^{-2}$): Musca,  B211, and L1506. These three filaments 
are detected above the background in dust total and polarized emission.   
We use the spatial information to separate Stokes \I, \Q, and \U\  of the
filaments from those of their  backgrounds, an essential step to measure  the intrinsic
polarization fraction ($p$) and angle ($\psi$) of each emission component. 
We find that the polarization angles in the three filaments ($\psi_{\rm fil}$) are coherent along their lengths and  
 not the same  as in  their backgrounds ($\psi_{\rm bg}$). 
The differences between $\psi_{\rm fil}$ and $\psi_{\rm bg}$ are $12^\circ$ and $54^\circ$ for Musca and L1506, respectively, and only $6^\circ$ in the case of B211.
  These differences for Musca and L1506 are larger than 
the dispersions of  $\psi$, both along the filaments and in their backgrounds.
The observed changes of $\psi$ are direct evidence for variations of the orientation of the plane of the sky (POS) projection of the magnetic field. 
As in previous studies, we find a decrease of several percent of $p$
with $\NH$ from the bacgrounds to the
crest of the filaments. We show that the bulk of the drop in $p$  cannot be explained by random fluctuations %
of the orientation of the magnetic field within the filaments because they are too small ($\sigma_{\psi}<10^\circ$).
We recognize the degeneracy between the dust alignment efficiency and the 
structure of the \vec{B}-field  in causing variations in $p$, but 
we argue that the decrease of $p$  from the backgrounds to the filaments results in part
from depolarization associated with the  3D structure of the \vec{B}-field: both its orientation  in the POS and with respect to the POS. 
 We do not resolve the inner structure of the
filaments, but  at the smallest scales accessible with \planck\ ($\sim$0.2\,pc), the observed changes of $\psi$ and $p$ hold information on the
magnetic field structure within filaments.  They show that both the mean  field and its fluctuations in the filaments are different 
from those of their backgrounds, confirming that the magnetic field has an active role
in the formation of filaments.}


\maketitle
%________________________________________________________________
%________________________________________________________________
\clearpage

\section{Introduction}\label{intro}

The interstellar medium (ISM)
has been observed to be filamentary for more than three decades. 
Filamentary structures have been observed in extinction  \citep[e.g.,][]{Schneider1979,Myers2009}, in dust emission \citep[e.g.,][]{Abergel1994}, in \hi\   \citep[e.g.,][]{Joncas1992,McClure-Griffiths2006}, 
in CO emission from  diffuse molecular gas \citep{Falgarone2001,HilyBlant2009} and  dense star forming regions \citep[e.g.,][]{Bally1987,Cambresy1999}.  
Filaments are striking features  of Galactic images from the far-infrared/submm \herschel\ space observatory \citep{Andre2010,Motte2010,Molinari2010}. 
They are ubiquitous both in the diffuse ISM and  in star-forming molecular clouds, and  the densest ones  are observed to be associated with prestellar cores \citep[e.g.,][]{Arzoumanian2011,Palmeirim2013}. %Konyves et al. 2015
Their formation and dynamical evolution  has become a central research topic in the field of star formation
\citep[see the review by][and references therein]{Andre2014}. In particular, the role played by the magnetic field is a main outstanding question.

The importance of the Galactic magnetic field for the dynamics of molecular clouds is supported by  
Zeeman measurements, which show that there is rough equipartition between their magnetic, gravitational, and kinetic energies \citep[e.g.,][]{Myers1988, Crutcher2004}.
Dust polarization observations provide an additional means to study the structure of the magnetic field. For a uniform magnetic field, 
the observed polarization angle is perpendicular to the component of the magnetic field  on the plane of the sky (POS) in emission and parallel in extinction.
The polarization fraction depends on
the dust polarization properties and  the grain alignment efficiency, but also on the structure of the magnetic field \citep{Hildebrand1983}.  
A number of  studies have investigated the relative orientation between  the magnetic field  and  filaments in molecular clouds using starlight polarization   \citep[e.g.][]{Goodman1990,Pereyra2004,Chapman2011,Sugitani2011}.   
Dust polarized emission from the densest regions of  molecular clouds, i.e., mainly dense cores and the brightest structures in nearby molecular clouds, has been observed from the ground 
at  sub-mm wavelengths  \citep[e.g.,][]{Ward-Thompson2000,Crutcher2004,Attard2009,Matthews2001,Matthews2009} and more recently using balloon-borne experiments \citep[e.g.,][]{Pascale2012,Matthews2014}.
However, due to the limited  sensitivity and  range of angular scales probed by  these observations, 
detection of polarization from filaments and  their lower column density surroundings  has not been achieved.        
This is a limitation because, as polarization is a pseudo-vector, the polarized emission 
observed towards a filament depend on those of the background. This effect is all the more important that the contrast between the filament and its
background is low.

\planck\footnote{\Planck\ (\url{http://www.esa.int/Planck}) is a project of the  European Space Agency  (ESA) with instruments provided by two scientific 
consortia funded by ESA member states and led by Principal Investigators from France and Italy, telescope reflectors provided through a collaboration 
between ESA and a scientific consortium led and funded by Denmark, and additional contributions from NASA (USA).} 
has produced the first all-sky map of the polarized emission from dust at sub-mm wavelengths \citep{planck2014-a01}.  
The \Planck\ maps of  polarization angle, $\psi$, and fraction, $p$,  encode information on the  magnetic field structure \citep{planck2014-XIX}. 
The observations have been compared to synthetic polarized emission maps computed from
simulations of anisotropic magnetohydrodynamical turbulence assuming simply a uniform intrinsic 
polarization fraction of dust grains \citep{planck2014-XX}.  In these simulations, 
the turbulent structure of the magnetic field is able to reproduce the main statistical 
properties of  $p$ and $\psi$ that are observed directly in a variety of
nearby clouds, dense cores excluded \citep[see also][]{Falceta-Goncalves2008,Falceta-Goncalves2009}. 
\citet{planck2014-XX} concludes that  the large scatter of $p$ at $\NH$ smaller than 
$10^{22}\NHUNIT$  in the observations  is due mainly to fluctuations in the magnetic field orientation along
the line of sight (LOS), rather than to changes in grain shape and/or the efficiency of grain alignment.   
They also show that the large-scale field orientation with respect 
to the LOS plays a major role in the quantitative analysis of these statistical properties.

 The  \planck\ maps of total intensity, as well as polarized intensity, display the filamentary structure of the  ISM \citep[][]{planck2013-p06b,planck2014-XIX,
 planck2014-a12}. 
 \citet{planck2014-XXXII} identify interstellar filaments over  the intermediate latitude sky,  and present a statistical analysis of their orientation with respect to the component of the magnetic field on the POS ($\vec{B}_\mathrm{POS}$). In the diffuse ISM, filaments are preferentially aligned with $\vec{B}_\mathrm{POS}$.
Towards nearby molecular clouds the relative orientation changes progressively from preferentially parallel 
to preferentially perpendicular from the lowest to the highest column  densities \citep{planck2015-XXXV}.  

In this paper, we make use of the \planck\ polarization data to study the structure of the magnetic field within three fields comprising 
the archetype examples of star forming filaments of moderate column density: B211, L1506, and Musca. 
Characterizing the magnetic field structure in such filaments, and its connection with that of the surrounding cloud, is a step towards understanding
the role of the magnetic field in their formation and evolution.  Stellar polarization data in 
these fields have been reported by several authors \citep{Goodman1990,Pereyra2004,Chapman2011}.  With \planck\, we have now access to 
fully-sampled maps of the dust polarized emission of both the filaments and their surrounding environment. 
The spatial information allow us to derive the polarization properties intrinsic to the filaments after subtracting the contribution of the surrounding 
background to the Stokes \I, \Q, and \U\ maps. We relate the results of our data analysis to the structure of the magnetic field. 

The paper is organized as follows. 
The \planck\ data  at 353\,GHz and the relations used to derive the polarization parameters from the Stokes  \I, \Q, and \U\  maps are presented in Sect.\,\ref{data}. 
In Sect.\,\ref{obs}, we present the   \I, \Q, and \U\ maps of the three filaments and their profiles perpendicular to their  axes. 
In Sect.\,\ref{sec:pol_prop} we quantify the variations of the polarization angle and fraction from the background to the filaments, and within the filaments.
Section\,\ref{Interp} discusses possible interpretations of the observed decrease of $p$ from the background to the filaments. 
Section\,\ref{conclusion} summarizes  the results of our data analysis and presents perspectives to further studies. 
In Appendix\,\ref{App2Layers},  we 
present a two-component model that applies as a first approximation to the polarized emission from interstellar filaments and their backgrounds.

\begin{table*}
\begingroup
\newdimen\tblskip \tblskip=5pt
    \caption{Observed properties of the filaments. 
    Column 3 gives the  length along  the filament crest. 
    Column 4 gives the filament  full width at half maximum (FWHM)    derived from a Gaussian fit to the  total intensity 
    radial profile. The  values given are for the observations, i.e., without beam deconvolution.   The outer radius (given in Column 5) is defined as the radial distance from the  filament axis
 at which its radial profile amplitude is equal to that of the   background (without beam deconvolution). 
 {\rev Columns 6 and 7 give the observed column density values at the central part of the filament ($r=0$) and at $r=R_{\rm out}$, respectively. The latter corresponds to the mean value of the  background. Column 8 gives  the   mass per unit length  estimated from the radial column density profiles of the filaments.} 
Column 9 gives the position angle (PA)  of the segment of the filament that is used to derive the mean  profile.  The   PA (measured positively from North to East)  is the angle between the Galactic North (GN) and the tangential direction to the filament crest derived  from the \I\ map. }   \label{table_param}\nointerlineskip
\vskip -3mm
\setbox\tablebox=\vbox{
   \newdimen\digitwidth 
   \setbox0=\hbox{\rm 0} 
   \digitwidth=\wd0 
   \catcode`*=\active 
   \def*{\kern\digitwidth}
   \newdimen\signwidth 
   \setbox0=\hbox{+} 
   \signwidth=\wd0 
   \catcode`!=\active 
   \def!{\kern\signwidth}
   \newdimen\pointwidth
   \setbox0=\hbox{{.}}
   \pointwidth=\wd0
   \catcode`?=\active
   \def?{\kern\pointwidth}
\halign{
\hbox to 1.0 in{#\leaderfil}\tabskip=1.8em&
\hfil #\hfil&
\hfil #\hfil&
\hfil #\hfil&
\hfil #\hfil&
\hfil #\hfil&
\hfil #\hfil&
\hfil #\hfil&
\hfil #\hfil\tabskip=0pt\cr
\noalign{\doubleline \vskip 2pt}
\omit Filament &$d$& Length & Width &$R_{\rm out}$ &$N_{\rm H}^{\rm fil}$&$N_{\rm H}^{\rm bg}$ &$M_{\rm line}$  & PA \cr
\omit &[pc]&[pc]&[pc]&[pc ]& [10$^{21}$ cm$^{-2}$] &[10$^{21}$ cm$^{-2}$] &[M$_{\odot}$ pc$^{-1}$]&[deg]\cr
\noalign{\vskip 4pt\hrule\vskip 6pt}
   Musca&200&10&0.75&1&8&2.2&30&30\cr
   B211&140&2.6&0.33&0.5&17&3&47&90\cr 
  L1506&140&3&0.35&0.5&7&2.3&26&55\cr    
\noalign{\vskip 5pt\hrule\vskip 3pt}}}
\endPlancktablewide
\endgroup
\end{table*}

%____________________________________________________________

\section{\Planck\  observations}\label{data}

\planck\ observed the sky in nine frequency bands from $30$ to $857$\,GHz in   total intensity, and in polarization up to  $353$\,GHz \citep{planck2013-p01}. 
Here, we only use  the intensity and polarization data at 353 GHz, which is the highest frequency \planck\ channel with polarization capabilities and the one with best signal-to-noise ratio (S/N) for dust polarization  {\n \citep{planck2014-XXII}}. 
The in-flight performance of 
the   High Frequency Instrument (HFI) 
is described in  \cite{planck2011-1.5} and \citet{planck2013-p02}. 
We use a \planck\   internal release  data set  (DR3, delta-DX9)  at 353\,GHz, presented {\n and analysed  in  \citet[][]{planck2014-XIX}, \citet{planck2014-XX},  \citet{planck2014-XXI}, and \citet{planck2014-XXXII}}. 
 We ignore the polarization of the CMB, which at 353\,GHz is a negligible contribution to the sky polarization towards molecular clouds {\n \citep{planck2014-XXX}}.
The  \I, \Q, and \U\   maps at the resolution of 4\parcm8 analysed here have been constructed using the gnomonic projection of the \healpix\footnote{\url{http://healpix.sourceforge.net}}  \citep{Gorski2005}  all-sky maps.  The regions that we study in this paper are within the regions of high S/N, which are not masked in  \citet[][]{planck2014-XIX}.

   \begin{figure*}
   \centering
 \resizebox{0.9\hsize}{!}{\includegraphics{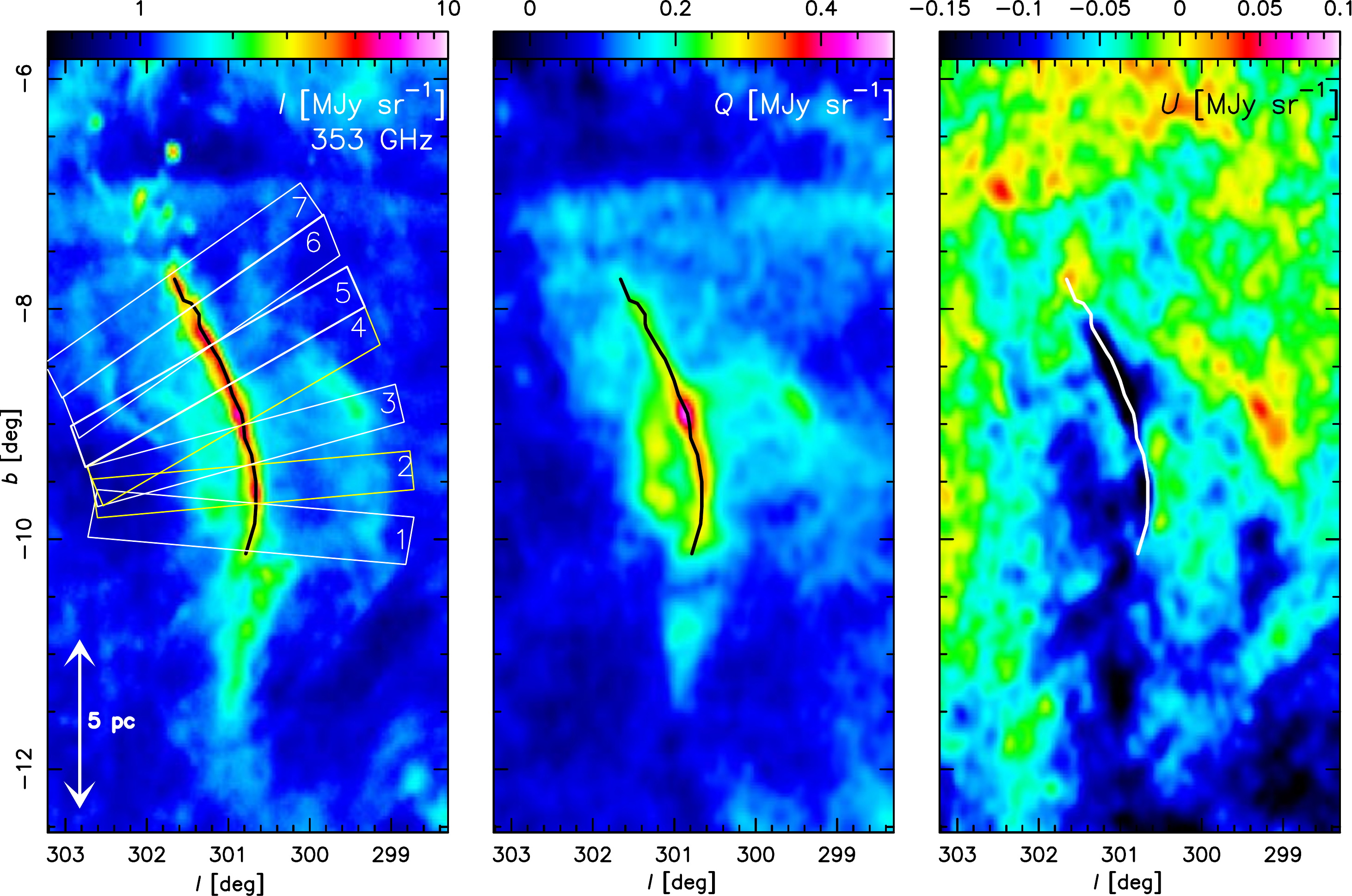}}
   \caption{\planck\ 353\,GHz Stokes parameter maps of  Musca (in MJy sr$^{-1}$). The total intensity  map is at the resolution of 4\parcm8, while the \Q\ and \U\ maps are smoothed to a resolution of 9\parcm6 for better visualization. 
 The crest of the filament  traced on the \I\ map is drawn in  black (on the \I\ and \Q\ maps) and white (on the \U\ map). 
     The boxes drawn on the \I\ map, numbered from 1\,to\,7,  show the regions from which the mean profiles are derived  (see Fig.\,\ref{Musca_profiles}).}
              \label{Musca_maps}
    \end{figure*}
 
       \begin{figure*}
   \centerline{
   \hspace*{-0.40cm}
 \resizebox{0.33\hsize}{!}{\includegraphics{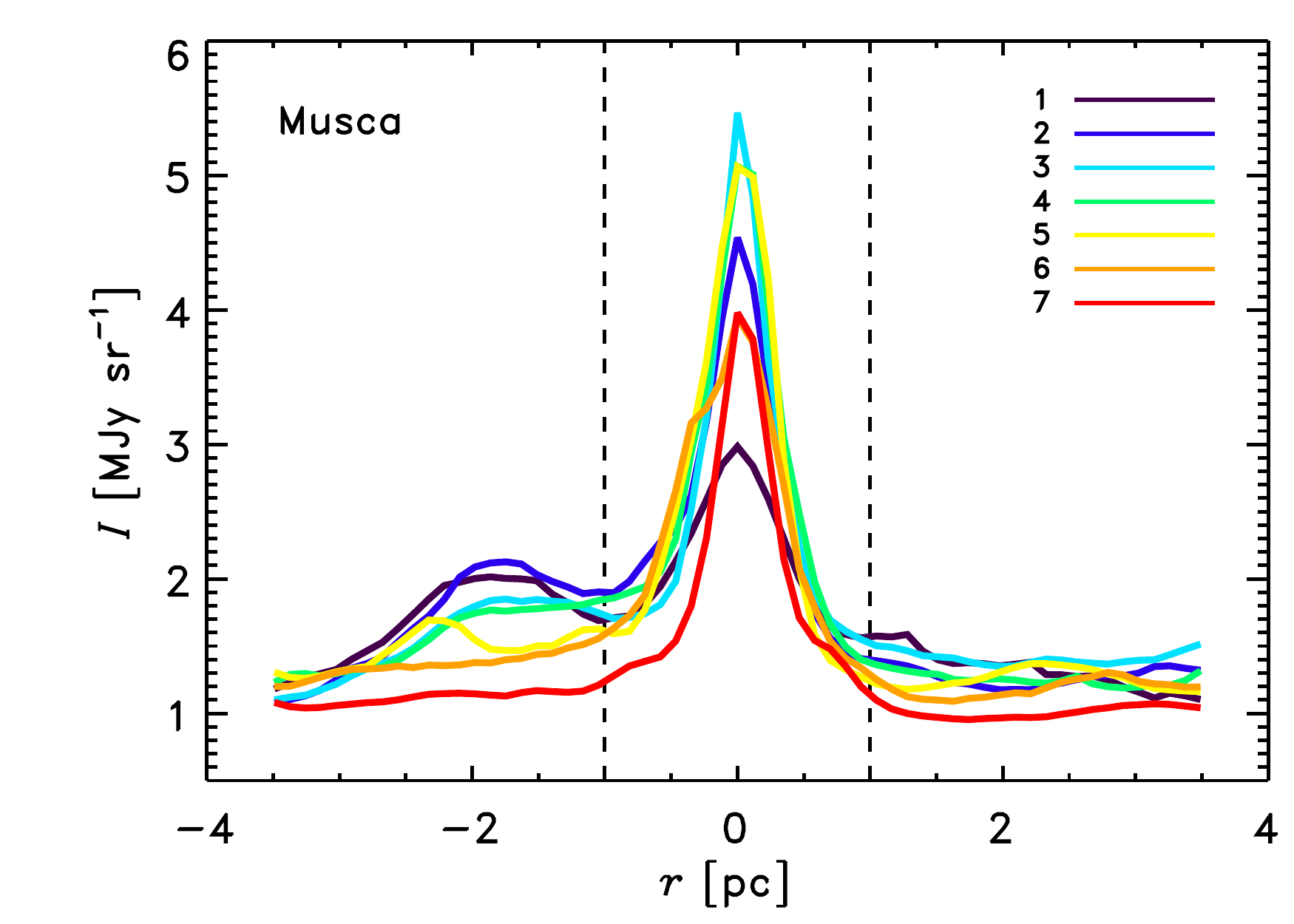}}
   \hspace*{-0.20cm}
  \resizebox{0.33\hsize}{!}{\includegraphics{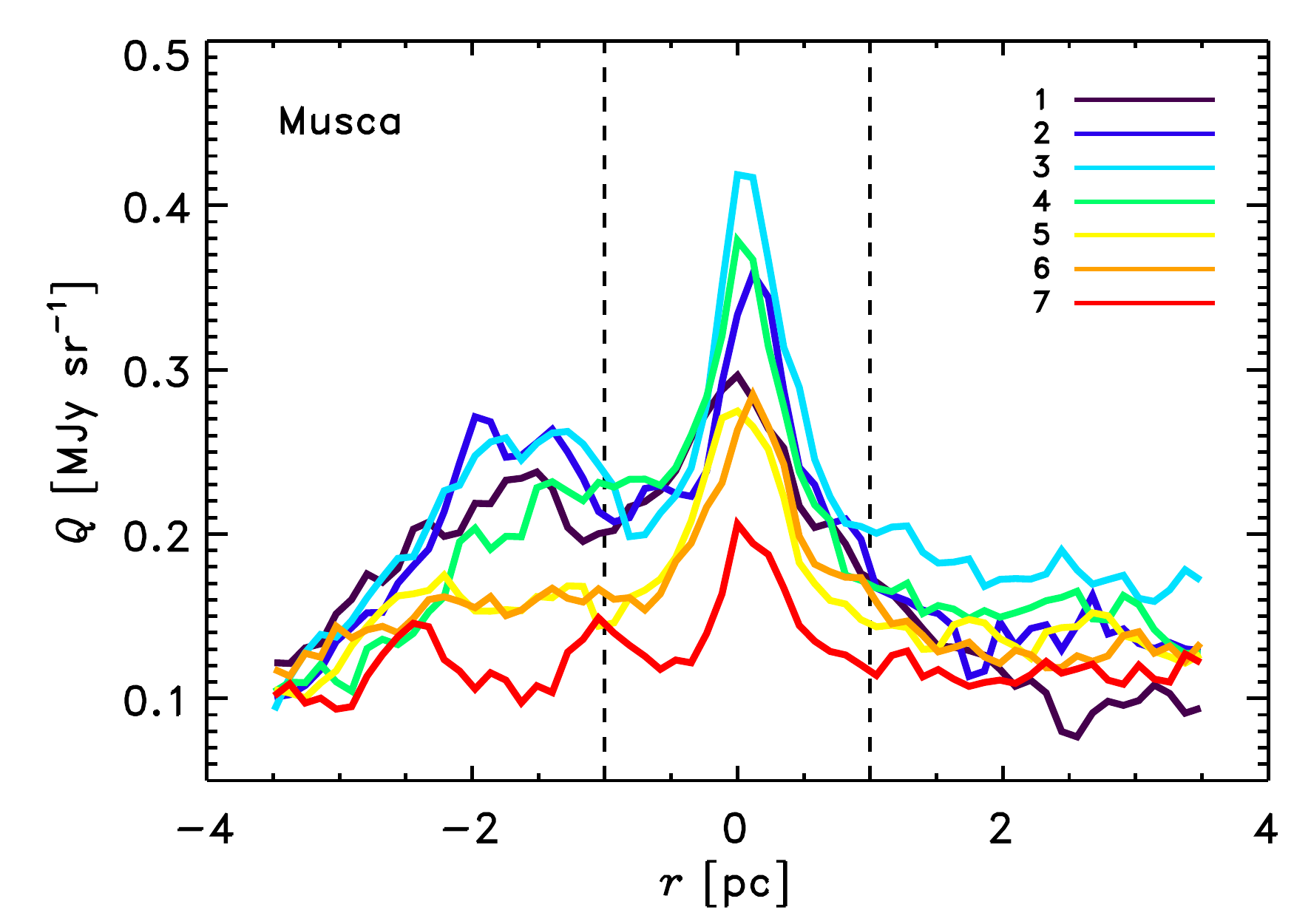}}
    \resizebox{0.33\hsize}{!}{\includegraphics{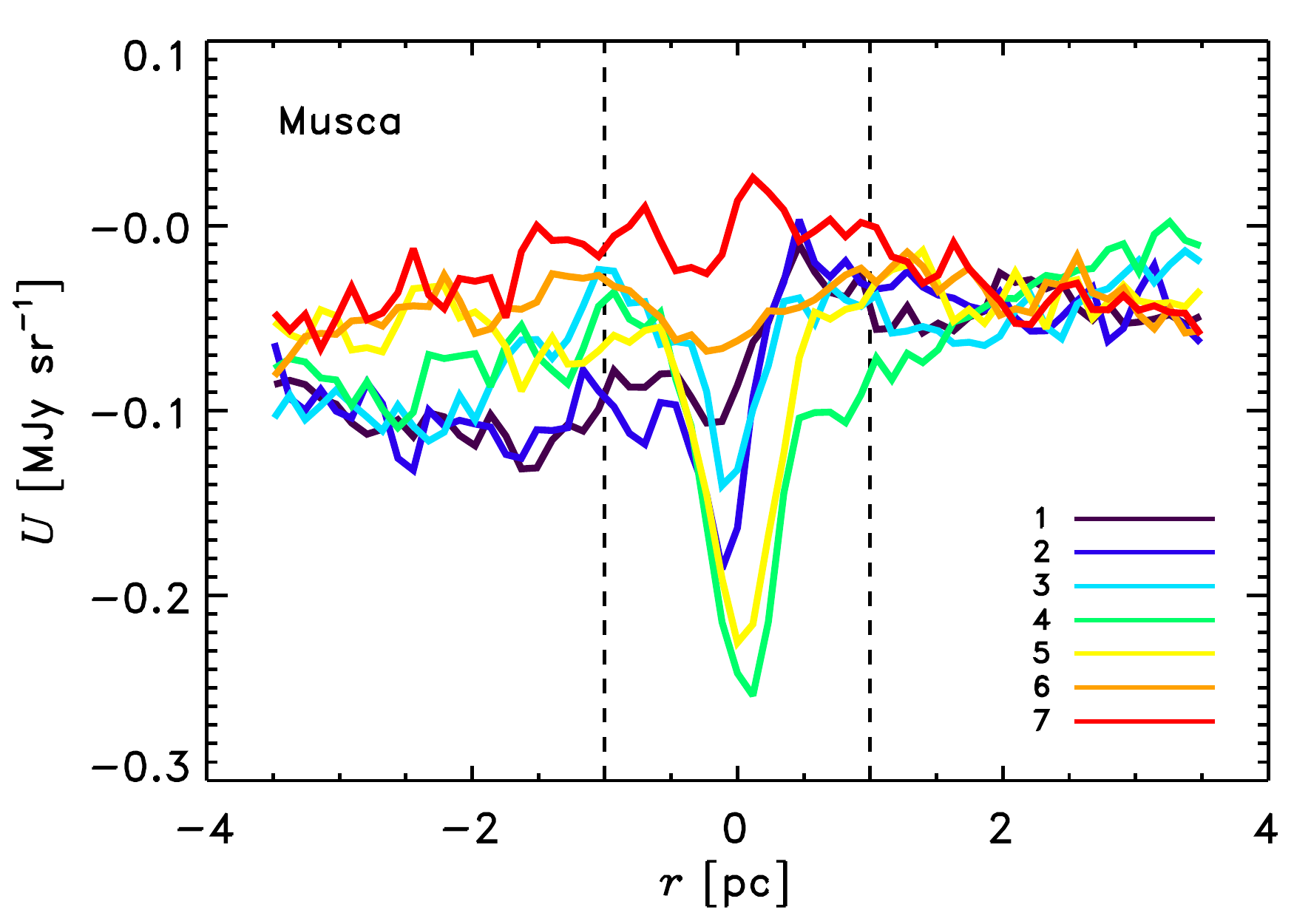}}}
   \caption{ Observed  radial profiles  perpendicular to the   axis of the Musca filament. The \I, \Q, and \U\  radial profiles are shown in the left, middle, and right panels, respectively. Values $r<0$  correspond to the eastern side of the filament axis. 
   The vertical dashed lines indicate  the position of the  outer radius $R_{\rm out}$.
   The  numbers of the profiles correspond to the cuts shown on the left  panel of Fig.\,\ref{Musca_maps}. 
     }
              \label{Musca_profiles}
    \end{figure*}    
           \begin{figure*}
   \centerline{
 \resizebox{0.33\hsize}{!}{\includegraphics{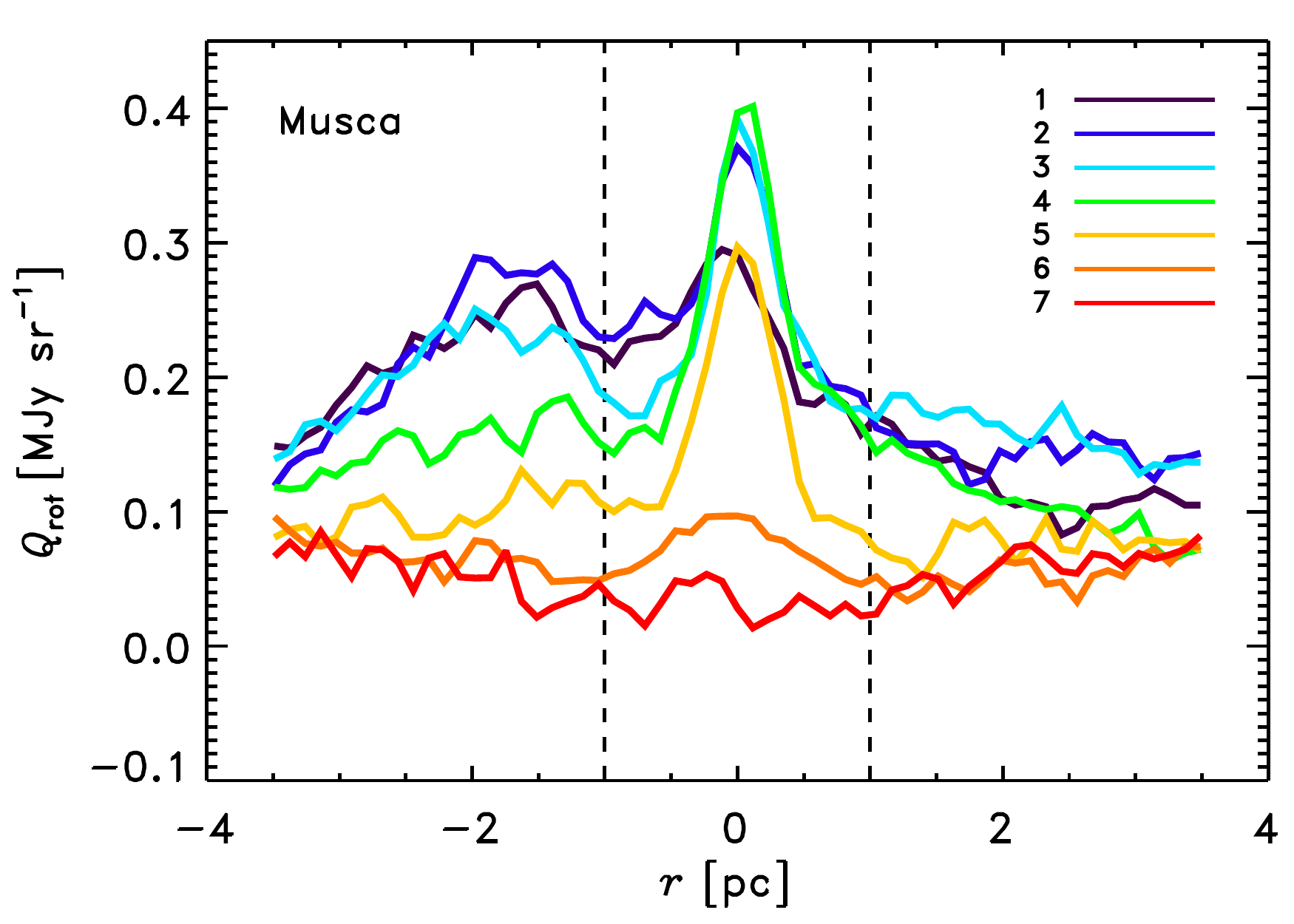}}
 \resizebox{0.33\hsize}{!}{\includegraphics{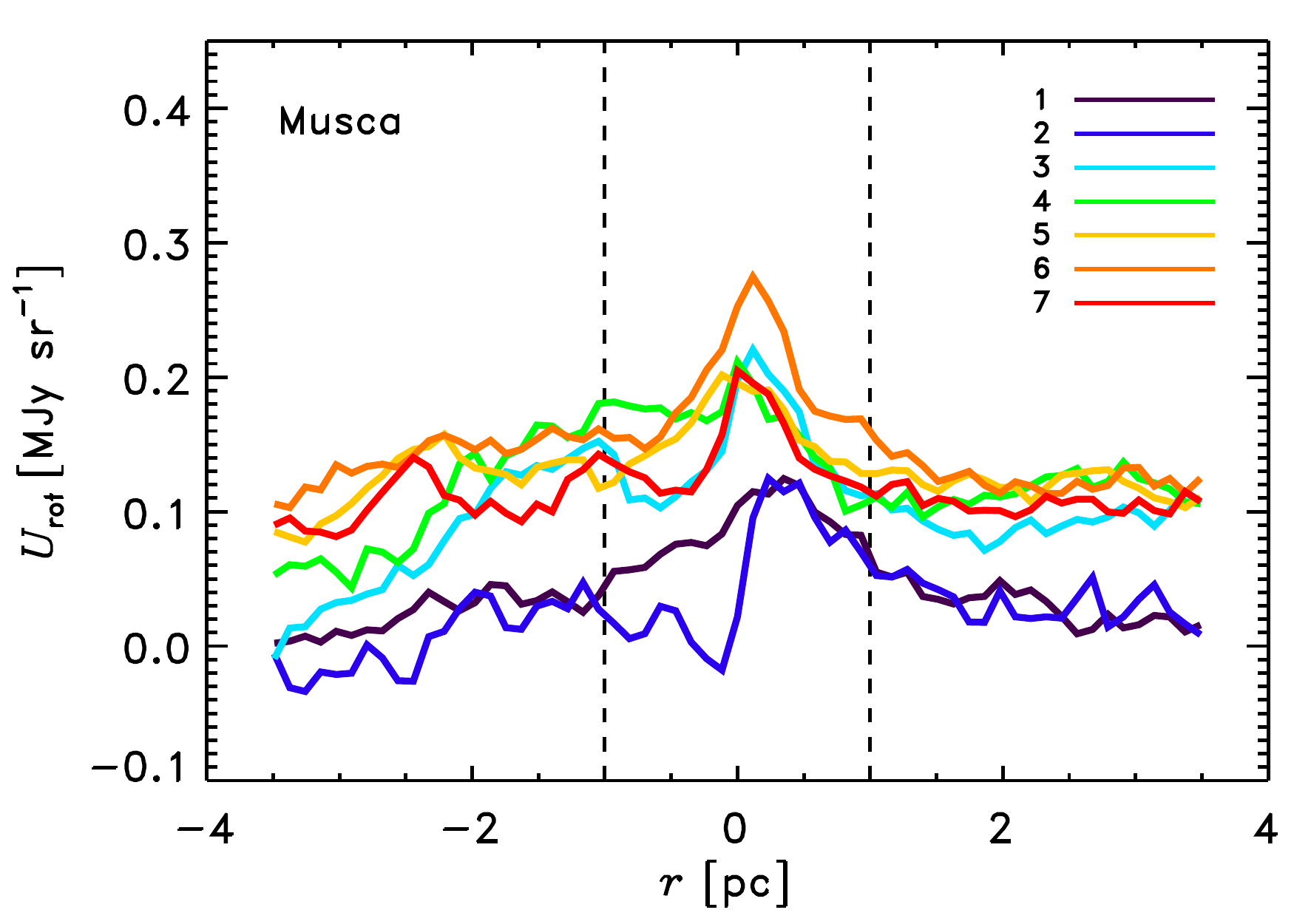}}}
  \vspace*{-0.25 cm}
   \caption{ {\rev Observed  radial $Q_{\rm rot}$ ({\it left}) and $U_{\rm rot}$  ({\it right}) profiles} of the Musca filament (same as Fig.\,\ref{Musca_profiles}) computed so that the reference direction is aligned with the filament axis. 
     }
              \label{Musca_RotQU}
    \end{figure*}
 Stokes  \I, \Q, and \U\ parameters are derived from \planck\ observations.
 Stokes \I\ is  the total dust intensity. The Stokes  \Q\ and \U\  parameters are the two components of the linearly polarized dust emission resulting from  LOS integration   and are related as 
\begin{eqnarray}
Q&=&I p \cos(2\psi), \label{eq1}\\
U&=&I p \sin(2\psi), \,\,\label{eq2}\\
P&=&\sqrt{Q^2+U^2},\label{P} \\
p&=&P/I, \label{PI}\\
\psi &=& 0.5\,\arctan(U,Q).\label{psi} 
\end{eqnarray}
\noindent
where $P$ is the total polarized intensity, $p$ is the  polarization fraction (see Eq.\,\ref{p_eq}), and $\psi$  is the polarization  angle {\rev given in the IAU convention \citep[see][]{planck2014-XIX}.}
 The $\arctan(U,Q)$ is used to compute $\arctan(U/Q)$ avoiding the $\pi$ ambiguity. 
 The POS  magnetic field orientation ($\chi$) is obtained by adding 90$^\circ$  to the polarization angle   ($\chi = \psi + 90^\circ$).
 {\rev In the paper we show the Stokes parameter maps as provided in the  \healpix\ convention, where the \planck\ \U\ Stokes map 
 is given by  $U=-I p \sin(2\psi)$. 
 Due to the noise present in the observed Stokes parameter maps,  $P$ and $p$ are biased positively and the errors  on $\psi$ are not Gaussian \citep{planck2014-XIX,Montier2014}. 
 We debias $P$ and $p$ according to the method proposed by \citet{Plaszczynski2014}, by taking into account
 the full noise covariance matrix of the \Planck\ data.

  \begin{figure*}
   \centering
 \resizebox{1.\hsize}{!}{\includegraphics{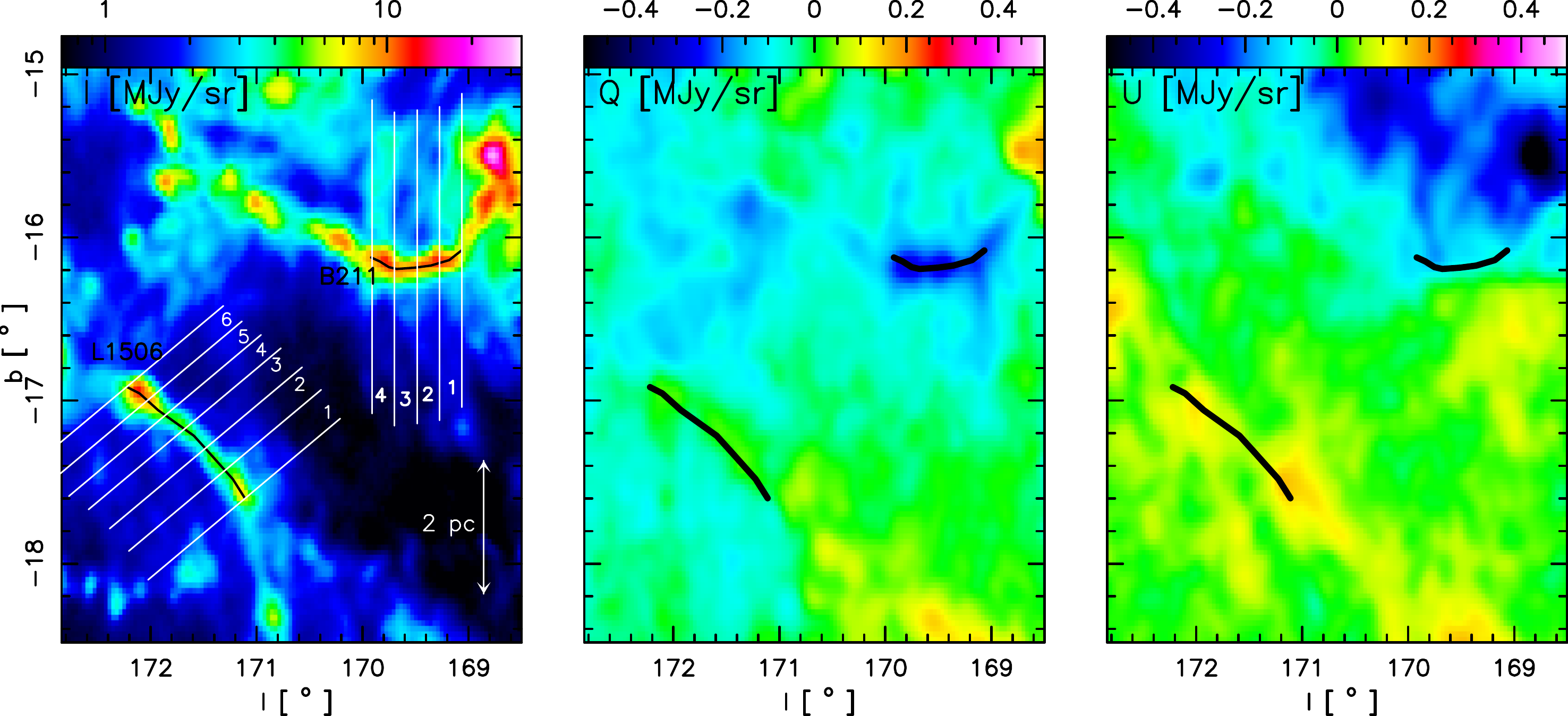}}
   \caption{ Same as Fig.\,\ref{Musca_maps}
   for part of the Taurus molecular cloud around the B211 (Northwest) and the L1506 (Southeast) filaments.
   The numbers and white lines on the total intensity map correspond to  the positions of the different cuts used to derive the radial profiles shown in Figs.\,\ref{Tau_profiles} and \ref{L1506_profiles}.
}
              \label{Tau_map2}
    \end{figure*}

   \begin{figure*}
   \centerline{
  \resizebox{0.33\hsize}{!}{\includegraphics{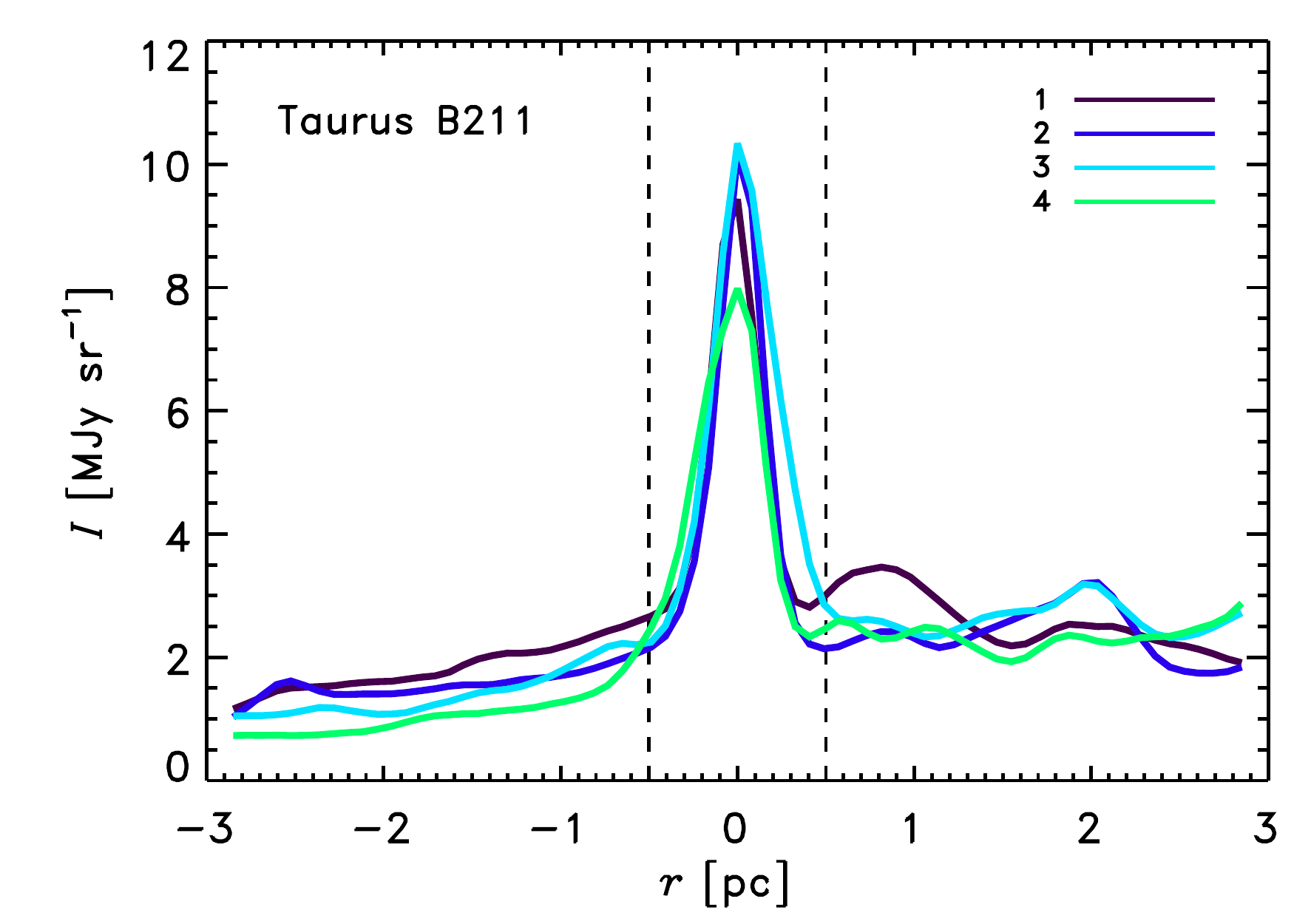}}
  \resizebox{0.33\hsize}{!}{\includegraphics{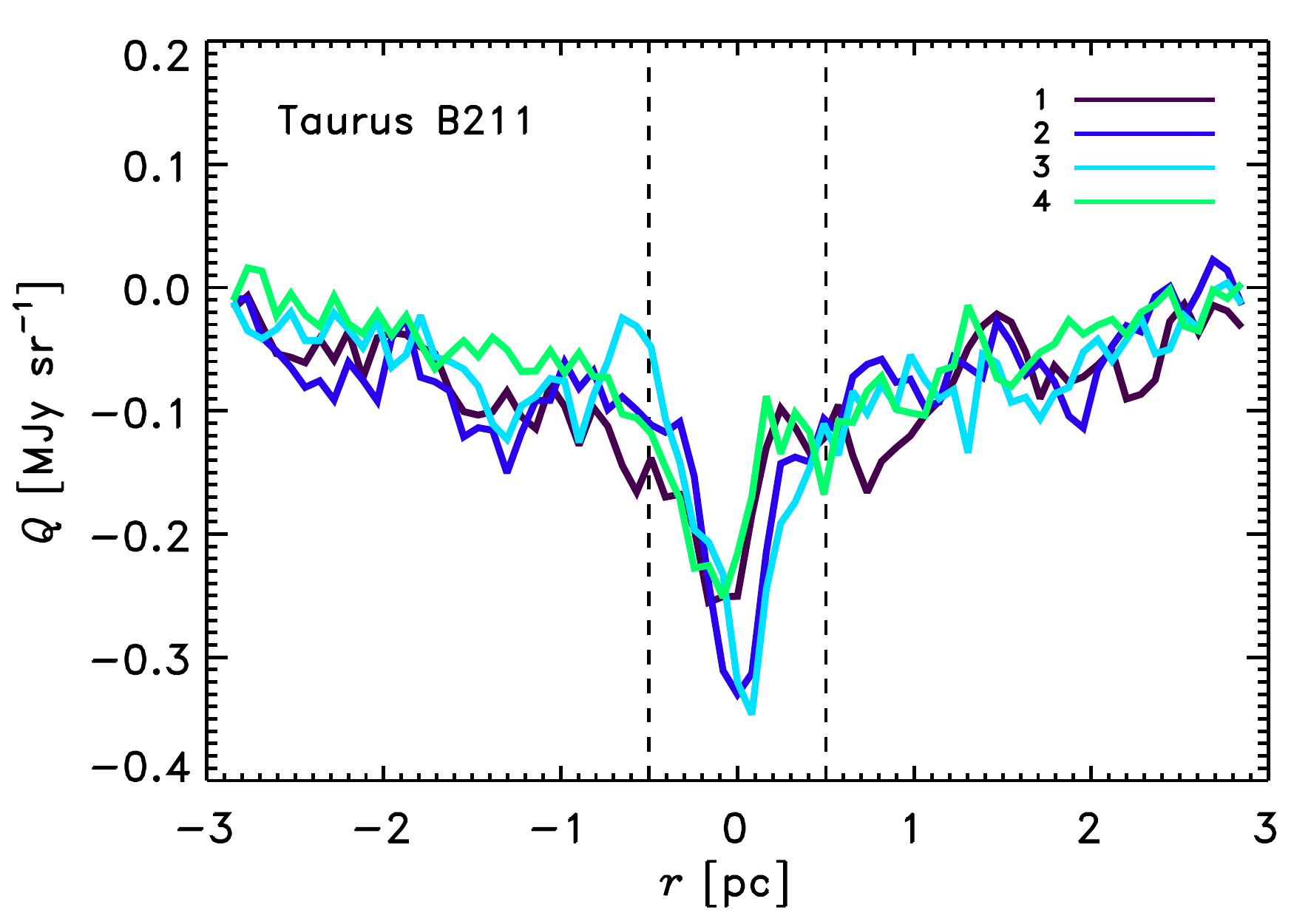}}
 \resizebox{0.33\hsize}{!}{\includegraphics{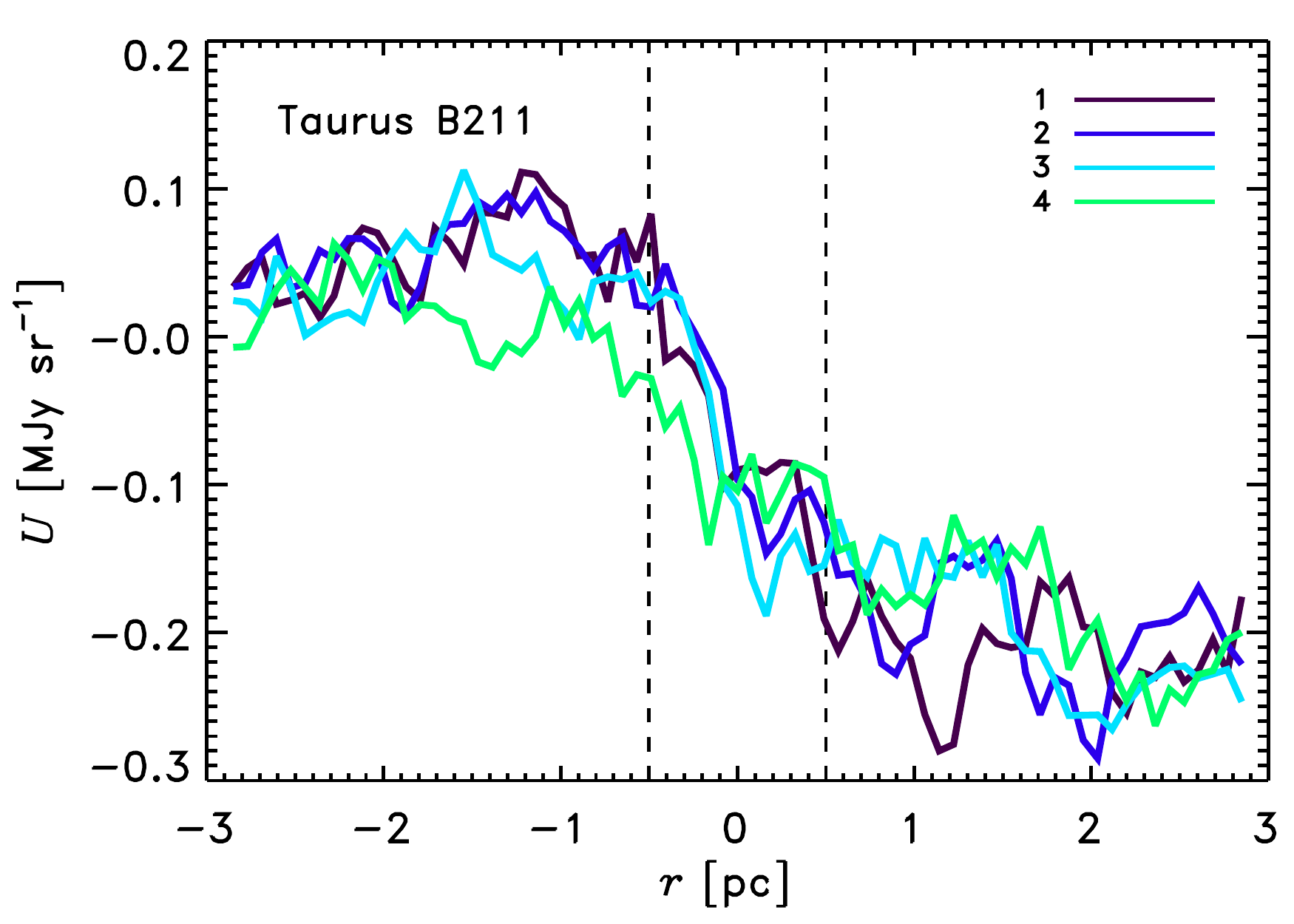}}}
  %\vspace*{-0.25 cm}
   \caption{Same as Fig.\,\ref{Musca_profiles} for the Taurus B211 filament. 
 Here $r<0$  corresponds to the southern  side of the filament axis.  
   The filament is clearly seen in \Q\ and is located in the area of a steep variation of the \U\ emission. The  dispersion of the emission is small along the %2.6\,pc 
   length of the filament.  }
              \label{Tau_profiles}
    \end{figure*}

   \begin{figure*}
   \centerline{
  \resizebox{0.33\hsize}{!}{\includegraphics{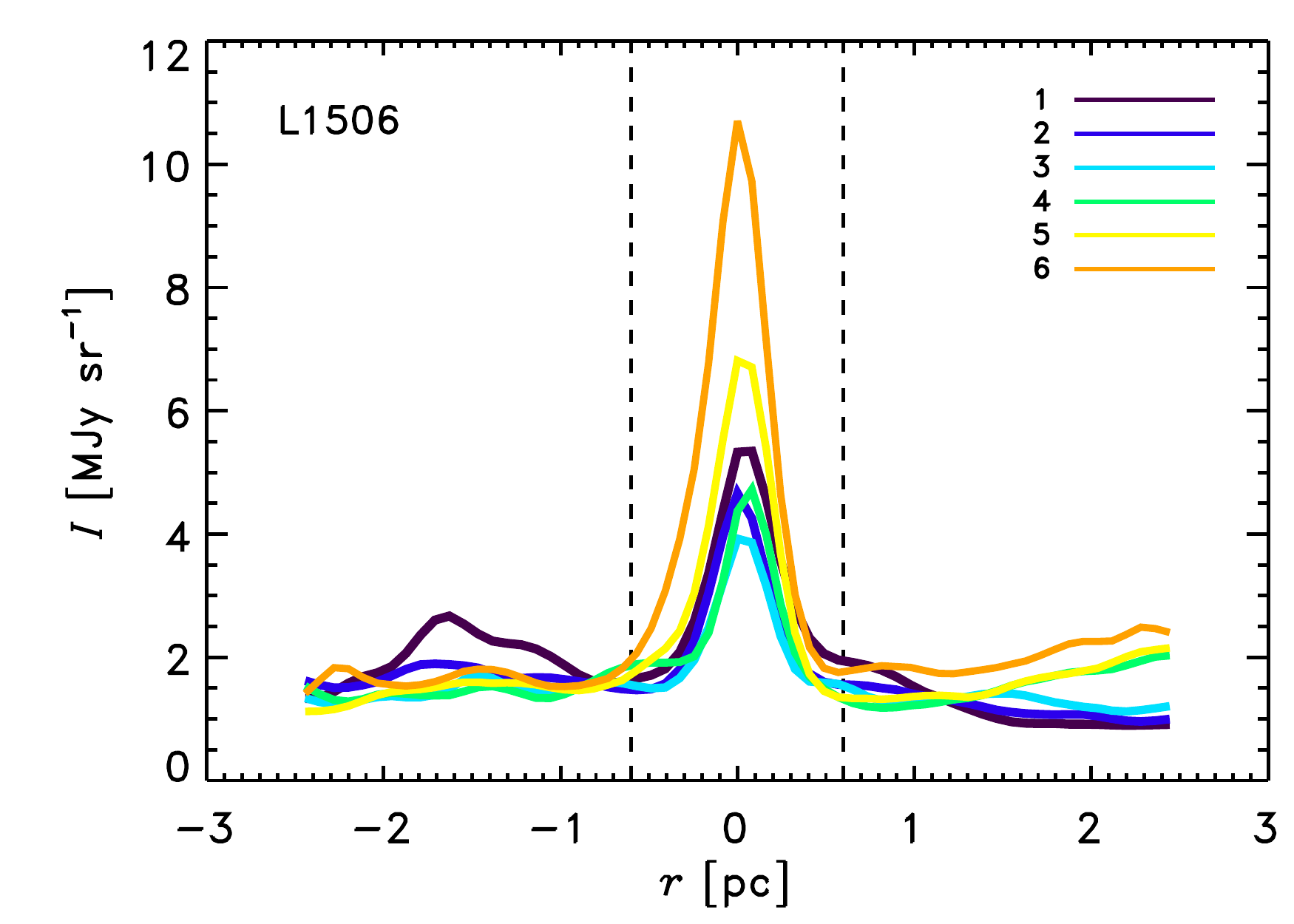}}
  \resizebox{0.33\hsize}{!}{\includegraphics{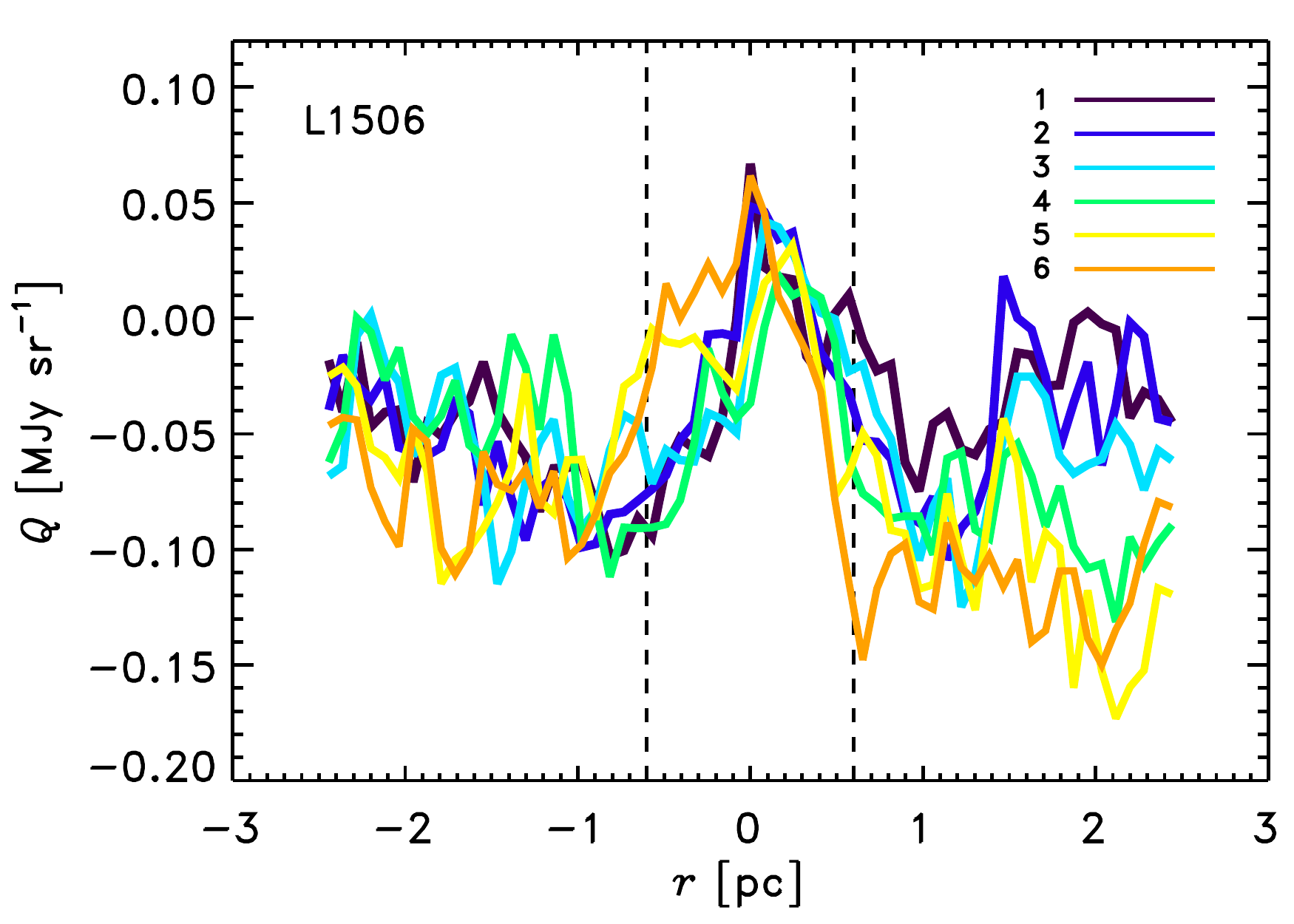}}
   \resizebox{0.33\hsize}{!}{\includegraphics{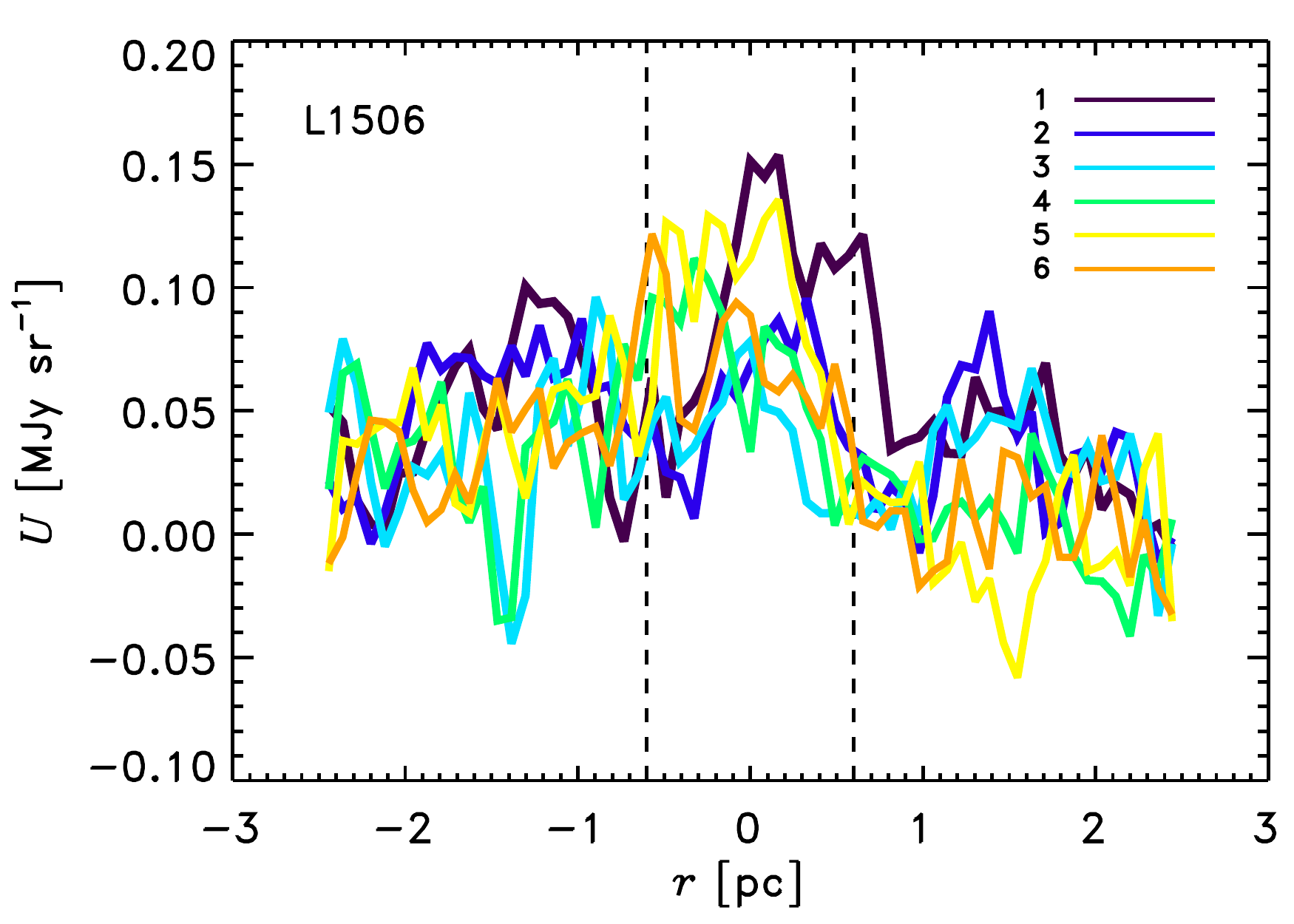}}}
  %\vspace*{-0.25 cm}
   \caption{Same as Fig.\,\ref{Musca_profiles} for L1506. Here    $r<0$  corresponds to the southeastern  side of the filament axis.}
              \label{L1506_profiles}
    \end{figure*}
 
\section{The  filaments as seen by \planck}\label{obs}

In the following, we present the \planck\   \I, \Q, and \U\ maps of three filaments  in two nearby molecular clouds: the Musca filament, and  the Taurus B211 and   L1506 filaments.  
The angular resolution of \planck\  (4\parcm8) translates into a linear resolution of 0.2\,pc and 0.3\,pc at  the distances  of  the Taurus and the Musca clouds, 
140 and 200\,pc respectively \citep{Franco1991,Schlafly2014}.  Table\,\ref{table_param} summarizes the main characteristics of the three   filaments. 
We describe each of them in the following sections. 

\subsection{The Musca filament}

Musca   is  a 10\,pc long filament located at a distance of 200\,pc  from the sun, in the North of the  Chamaeleon region \citep{Gregorio-hetem1988,Franco1991}.   
  The mean column density along  the crest of the filament is   $8\times10^{21}$\,cm$^{-2}$ as derived from the \planck\  data
   \citep{planck2014-XXIX}.  
 The magnetic field in the neighbourhood of the Musca  cloud has been traced using optical polarization measurements of background stars by \citet{Pereyra2004}.  
  $Herschel$ SPIRE images show  $hair$-$like$ striations  of matter perpendicular to the main Musca filament and aligned with the magnetic field lines.  %(Cox et al. in prep). 

Figure \ref{Musca_maps} shows the \planck\ 353\,GHz Stokes parameter maps of the Musca cloud. The filament 
is well detected in  total intensity and  polarization. 
To quantify the polarized intensity  observed towards the filament  we derive  radial profiles perpendicular to its axis.
The  crest of the filament  is traced using  the {\tt DisPerSE} algorithm  \citep{Sousbie2011,Arzoumanian2011}. Cuts perpendicular to the  axis of the filament  are then constructed at each pixel position along the filament crest.   The profiles centred on neighbouring pixels along  the filament crest,  corresponding to six times the beam size (6\,$\times$\,4\parcm8), are  averaged to increase the S/N.  The position of each of the profiles is shown on the intensity map of Fig.\,\ref{Musca_maps}. 
  The mean profiles are numbered from 1 to 7, running from the South to the North of the filament. 
 Figure \ref{Musca_profiles} shows  the radial profiles  of the Stokes parameters where hereafter $r$ corresponds to the radial distance from the  axis of the filament. 
  The profiles in Fig.\,\ref{Musca_profiles} illustrate the variability of the emission along the different cuts.
  The presence of neighbouring structures  
  next (in projection) to the main Musca filament, 
  can be seen in the profiles (e.g., the bumps at $r\simeq-$2\,pc correspond to the elongated structure to the East of  the  axis of the Musca filament).

 We observe a variation of the Stokes \Q\ and \U\ profiles of the filament associated with the change of its orientation on the POS. 
 Thus the observed variations in the Stokes \Q\ and \U\ profiles are not necessarily due to variations of the \vec{B}-field orientation with respect to the filament axis.
{\rev In Fig.\,\ref{Musca_RotQU} the \Q\ and \U\ parameters, $Q_{\rm rot}$ and $U_{\rm rot}$, are computed  using the filament axis as the reference direction; 
the position angle (PA) of each segment  (1 to 7) is estimated from the mean tangential direction to the filament crest. 
The parameter $Q_{\rm rot}$ is positive and $U_{\rm rot}=0$ if the magnetic field is perpendicular to the filament axis. 
The data  (Fig.\,\ref{Musca_RotQU}) show that most of the polarized emission of  segments 1 to 5 is associated with $Q_{\rm rot}>0$ and $Q_{\rm rot}>|U_{\rm rot}|$.
 This is consistent with a magnetic  field close to perpendicular to the filament axis. 
 For the segments 6 and 7, the relative orientation between the $\vec{B}_\mathrm{POS}$ angle and the filament is different 
  as indicated by the $Q_{\rm rot}$ and $U_{\rm rot}$ profiles.}

\subsection{The Taurus {\rm B211}  filament}\label{B211obs}

The B211 filament is located in the Taurus molecular cloud (TMC). It is one of the closest star forming regions in our Galaxy located at a distance of only 140\,pc  from the Sun {\n \citep{Elias1978,Kenyon1994,Schlafly2014}.} This region has been the target  of numerous observations,  has long been considered as a prototypical molecular cloud of isolated low-mass star formation,  and has inspired magnetically-regulated models of star formation \citep[e.g.,][]{Shu1987,Nakamura2008}.
The B211 filament  is one of the well-studied nearby star-forming filaments that shows a number of young stars and prestellar cores along its ridge \citep[][]{Schmalzl2010,Li2012,Palmeirim2013}. Recently, \citet{Li2012} studied the Taurus B211 filament and found that the measured densities and column densities indicate a filament width along the LOS that is equal to the width observed on the POS ($\sim$0.1pc).  Studies previous to \planck, using polarization observations of background stars, found that the structure of the Taurus cloud and that of the magnetic field are related  \citep[e.g.,][]{Heyer2008,Chapman2011}. Using the Chandrasekhar-Fermi method, \citet{Chapman2011} estimated a magnetic field strength of about 25\,$\mu$G in the cloud surrounding the B211 filament, concluding that the former is magnetically supported. 

Figure \ref{Tau_map2} shows the \planck\ 353\,GHz Stokes parameter maps of the TMC around the B211 and L1506  filaments.  The  B211 filament is well detected in the \I\ and \Q\ maps, 
 as a structure distinct  from its surrounding. On the \U\ map, on the other hand, the filament is not seen as an elongated structure, 
 but it is perpendicular to a  large \U\ gradient  that separates two regions of almost uniform \U. 
The \U\ emission is negative on the northern side of the filament while it is positive on the southern side. This indicates that the \vec{B}-field orientation varies in the cloud surrounding the B211 filament. This variation can be seen very clearly  in the radial profiles perpendicular to the filament axis, shown in Fig.\,\ref{Tau_profiles}. 
These profiles  are derived as explained in the previous section (for the Musca filament).
The four  \Q\ and \U\ profiles  (shown in the middle and right-hand  panels of Fig.\,\ref{Tau_profiles}) 
are derived by averaging the cuts within a distance along the filament crest of 3 times the \planck\ beam (3\,$\times$\,4\parcm8). 
The cloud intensity (\I)   increases  from the southern to the northern  side of the filament, while the \Q\ emission is similar on  both sides of the filament axis. The (negative) \Q\ emission of B211 is  very clearly seen in the radial profiles.  The different  profiles show the  approximate 
invariance  of the emission along the filament crest. The total and polarized emission components are remarkably constant along the  length of  B211.

\subsection{The Taurus {\rm L1506} filament}\label{L1506obs}

The L1506 filament is located  on the Southeast side of B211  (see Fig.~\ref{Tau_map2}). 
Stellar polarization data are presented by \citet{Goodman1990}.
The density structure and the dust emission properties have been studied by \citet{Stepnik2003}, and more recently by \citet{Ysard2013} using \herschel\ data. This filament 
has mean column densities comparable to those of the Musca filament. 
Star formation at both ends of L1506 has been observed with the detection of a few candidate prestellar cores \citep{Stepnik2003,Pagani2010}. 
Figure \ref{L1506_profiles} shows the  radial profiles perpendicular to its axis. 
The  colour profiles numbered from 1 to 5 (and derived as explained in Sect.\,\ref{B211obs}), correspond to mean profiles at different positions along its crest. 
Profiles 1, 5, and 6 trace the emission corresponding to the star forming cores at the two ends of the filament. The other profiles (2 to 4) trace the emission associated with the filament, not affected by  emission of star-forming cores. 
The fluctuations seen in the \Q\ and \U\ profiles are of the same order as the fluctuations  of the emission of  B211   located  a few parsecs Northwest of L1506, but the polarized emission associated with the filament is much smaller (the scale of the plots in Figs.\,\ref{Tau_profiles} and \ref{L1506_profiles} is not the same). The polarized intensity observed towards L1506 is smaller than that associated with the Musca and B211 filaments,  while the total intensity is of the same order of magnitude.

 \subsection{Polarized intensity  and polarization fraction}\label{compfil}

Figure\,\ref{Pmaps}  presents the  polarized  intensity ($P$) maps of the two studied molecular clouds, derived from the \Q\ and \U\ maps using Eq.\,(\ref{P}) and debiased according to the method proposed by \citet{Plaszczynski2014}, as mentioned in Sect.\,\ref{data}.
The POS  angle of the magnetic field ($\chi$)   is overplotted on the maps;  the length of the pseudo-vectors is   proportional to the observed {\n(debiased)} polarization fraction.  
These maps show that the Musca and B211 filaments are detected in polarized emission,  while the L1506 filament is not seen as an enhanced structure in polarized intensity 
unlike in total intensity.

The polarization fraction ($p$) is plotted as a function of  the column density ($N_{\rm H}$) for the filaments and their backgrounds  in Fig.\,\ref{pVsnh_3fil}. 
This plot shows a large scatter of  $p$ for the lowest column density values corresponding to the background in L1506 and B211. 
Such a scatter is  present in  the statistical analysis of \planck\ polarization data towards  Galactic molecular clouds (see in particular Fig.\,2 in \color{blue}{Planck Collaboration Int. XX 2014}\color{black}). For the filaments, $p$ decreases as a function of column density with different slopes (see Fig.\,\ref{pVsnh_3fil}). 
We have drawn three lines, one for each filament. They represent the linear fit log$p$ versus log\nh\ for \nh\ larger than $2.2\times10^{21}$, $6\times10^{21}$, and $3\times10^{21}$\,cm$^{-2}$. The best fit slopes are 
  $-0.35, -0.36,$ and $-1.15$, for Musca, B211 and L1506, respectively.
A decrease of $p$ with $N_{\rm H}$ has been reported in previous studies and ascribed to 
the loss of grain alignment efficiency  \citep[e.g.][]{Lazarian1997,Whittet2008} and/or the random component of the magnetic field \citep[e.g.][]{Jones1992,Ostriker2001,planck2014-XIX,planck2014-XX}.  In the next sections, we take advantage of the imaging capability and sensitivity of \planck\ 
to further characterize the origins of the polarization properties of the
filaments.

\begin{figure*}
   \centerline{
 \resizebox{0.46\hsize}{!}{\includegraphics{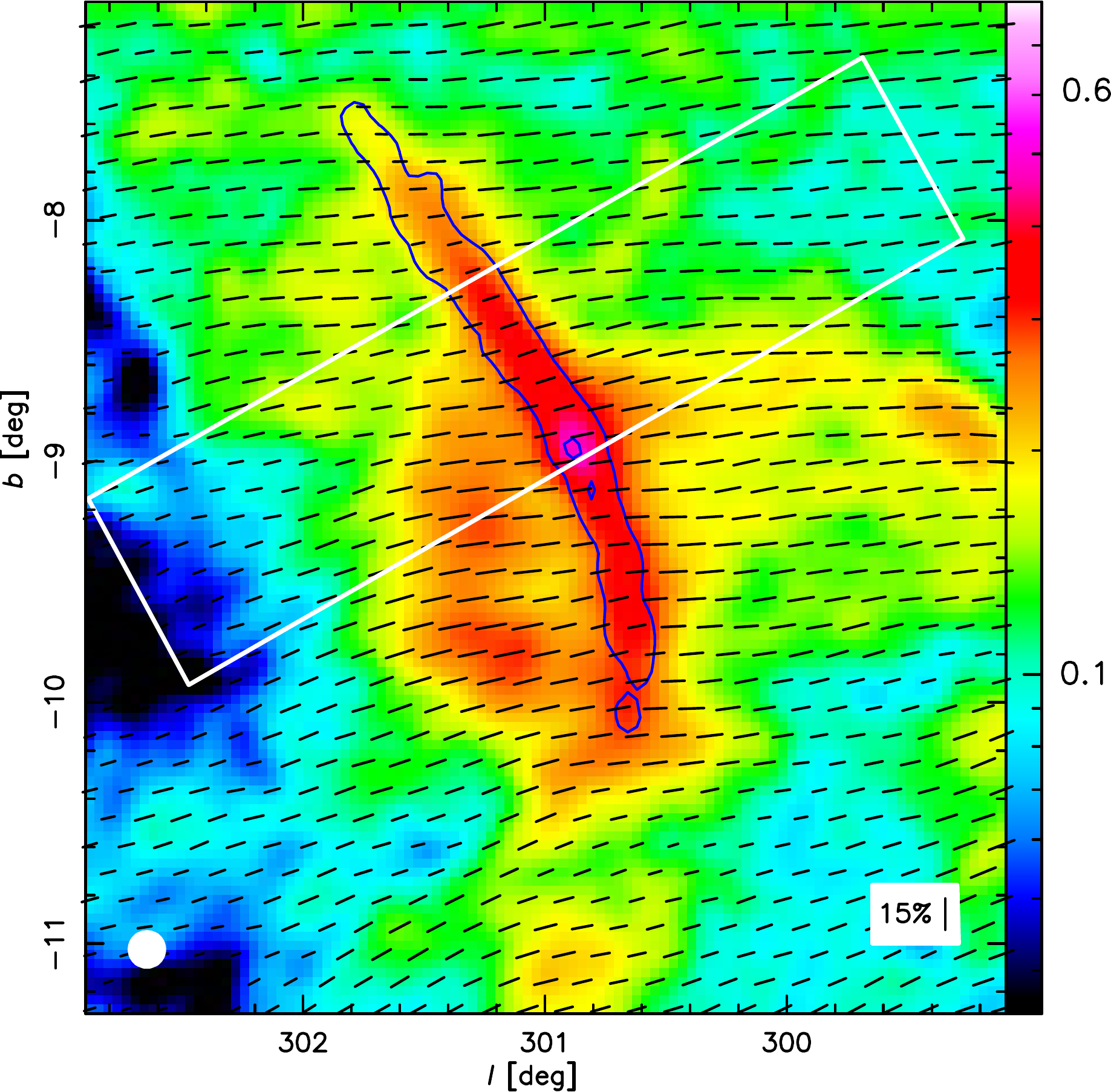}}
 \hspace{0.2cm}
  \resizebox{0.54\hsize}{!}{\includegraphics{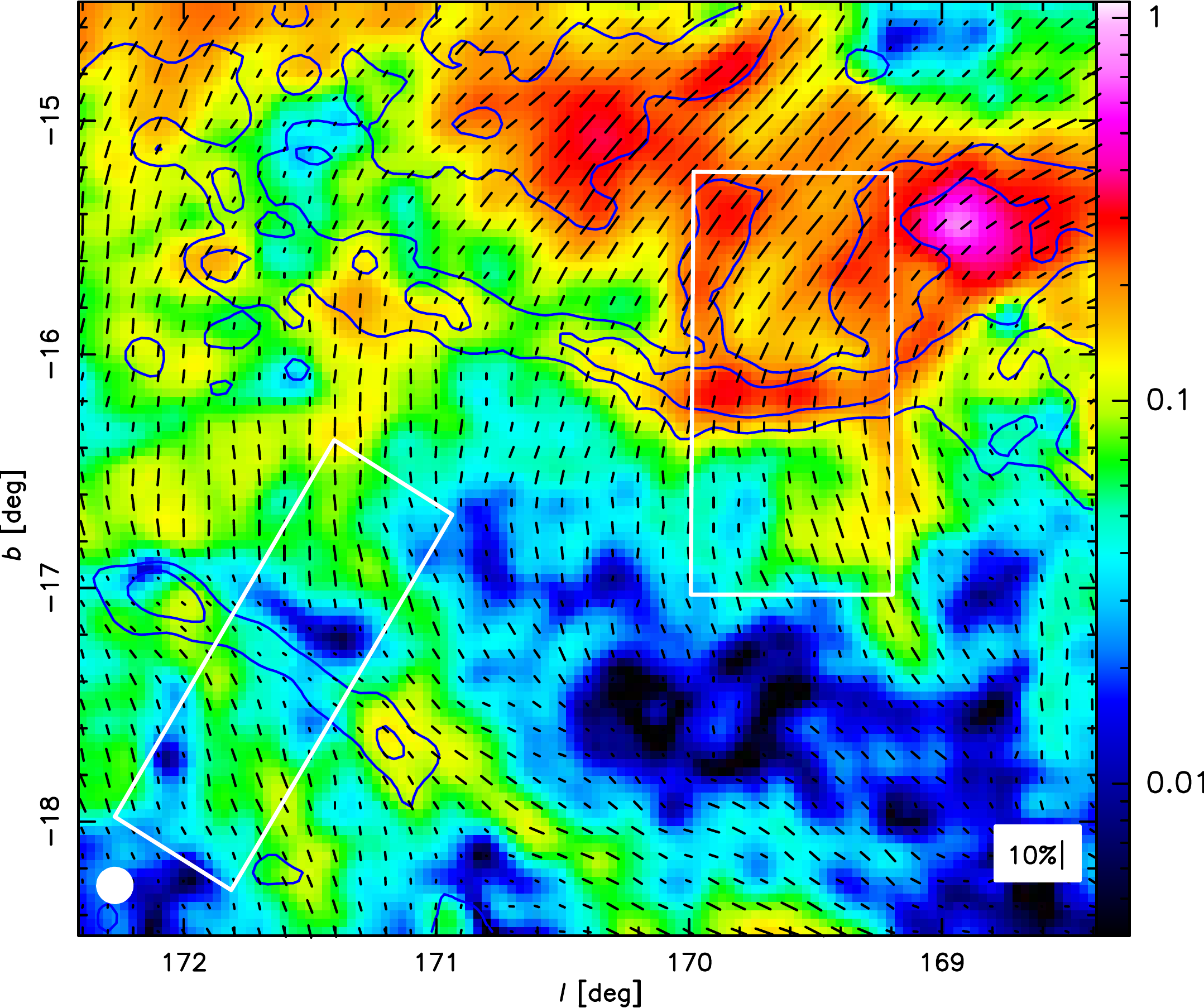}}}
   \caption{  Observed  polarized emission at 353\,GHz (in MJy\,sr$^{-1}$) of the Musca (left) and Taurus (right) clouds. 
   The maps are at the resolution of  
   9\parcm6 (indicated by the white filled circles)  for increased S/N. 
   The black segments show the $\vec{B_{\rm POS}}$-field orientation ($\psi$+90$^{\circ}$). The length of the pseudo-vectors is proportional to the polarization fraction. The blue contours show the total dust intensity at levels of 3 and 6 MJy\,sr$^{-1}$, at the resolution of 4\parcm8. The white boxes correspond to the area of the filaments and their backgrounds that is analysed in the rest of the paper. 
       }
              \label{Pmaps}
    \end{figure*}

      \begin{figure}
   \centerline{
 \resizebox{1.\hsize}{!}{\includegraphics{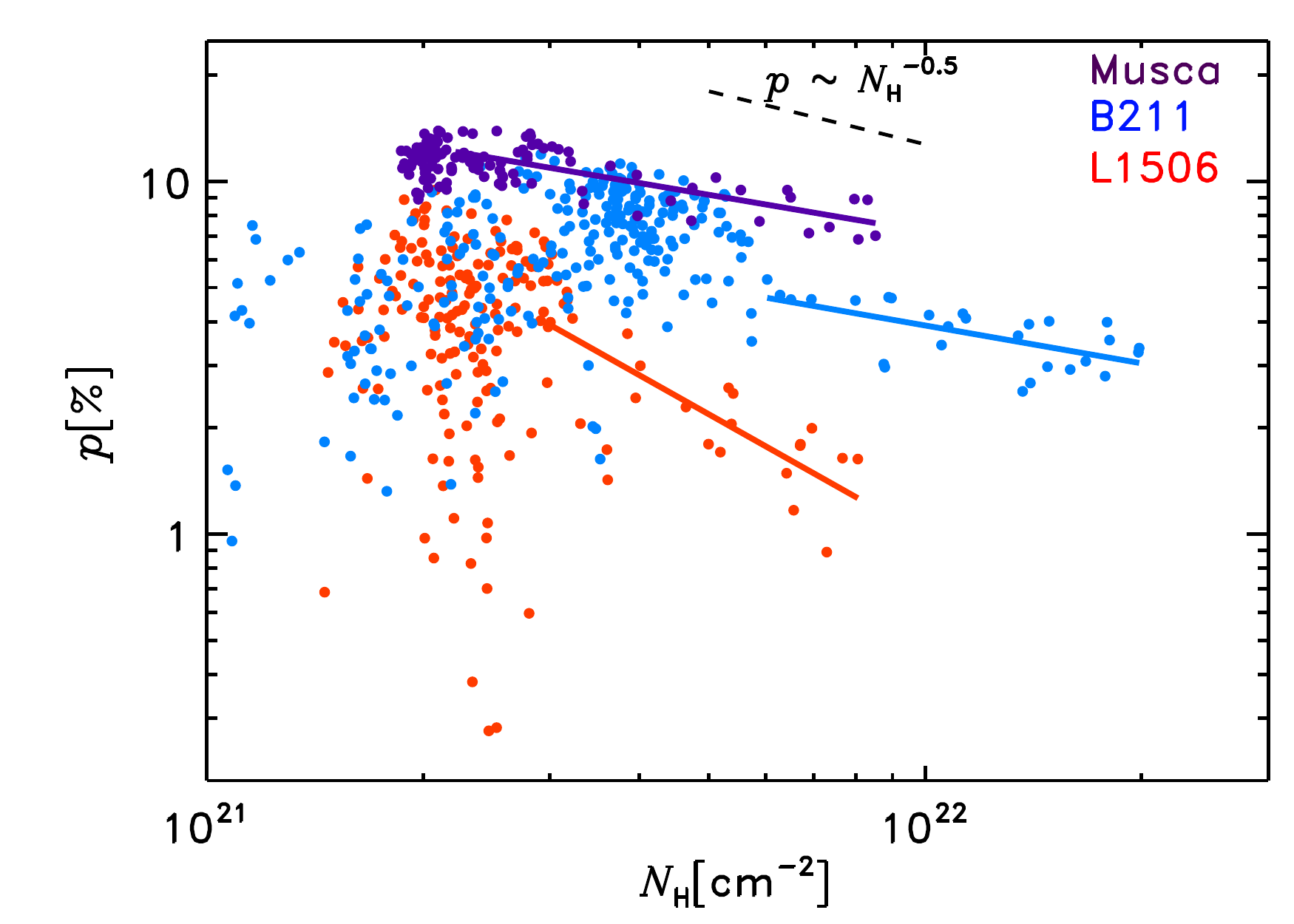}}}
   \caption{  Observed polarization fraction ($p$) as a function of  column density ($N_{\rm H}$), for  
   the cuts across the axis of the  filaments derived from the boxes shown in Fig.\,\ref{Pmaps}. 
    The three lines show  linear fits of log$p$ versus log$N_{\rm H}$  described in Sect.\,\ref{compfil}. 
    The data uncertainties depend on the intensity of the polarized emission. They are the largest 
    for low $p$ and $N_{\rm H}$ values. The 
    mean error-bar on $p$ is $1.2\,\%$ for the data points used for the linear fits, and $2.5\,\%$  for the points where $p<3\,\%$ and $N_{\rm H}<3\times10^{21}$\,cm$^{-2}$.
    }   
              \label{pVsnh_3fil}
    \end{figure}

\section{Polarization properties}
\label{sec:pol_prop}

 In the following, we introduce a two-component model that uses the spatial information in \planck\ images to 
 separate the emission of the filaments from the  surrounding emission (Sect.\,\ref{TwoLayers}).  This allows us to characterize and 
 compare the polarization properties of each emission component (Sects.\,\ref{pol_maps} and \ref{subsec:comparison}).
  
 \begin{figure*}
   \centerline{
    \vspace*{-0.25 cm}
 \resizebox{0.33\hsize}{!}{\includegraphics{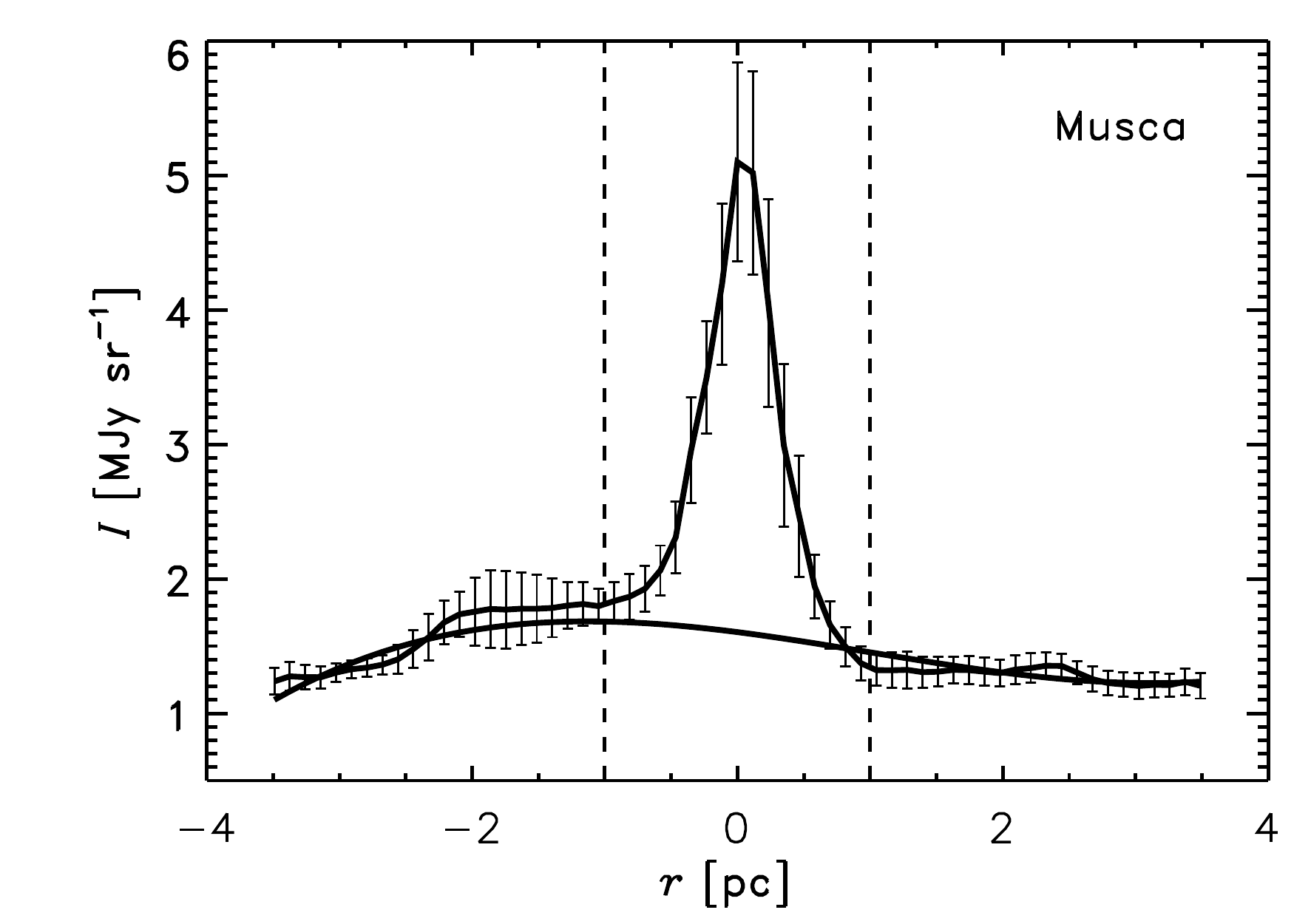}}
\resizebox{0.33\hsize}{!}{\includegraphics{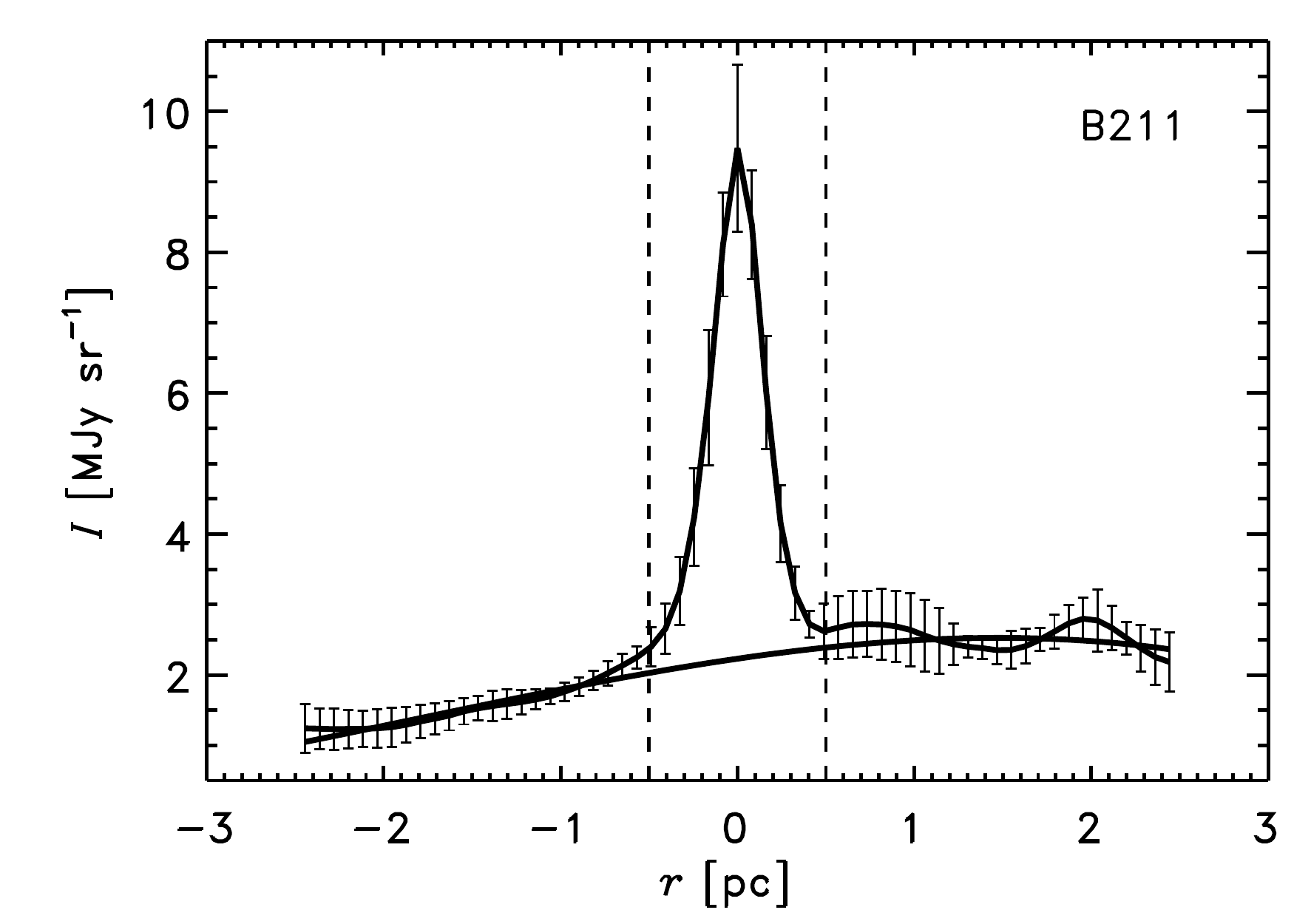}}
 \resizebox{0.33\hsize}{!}{\includegraphics{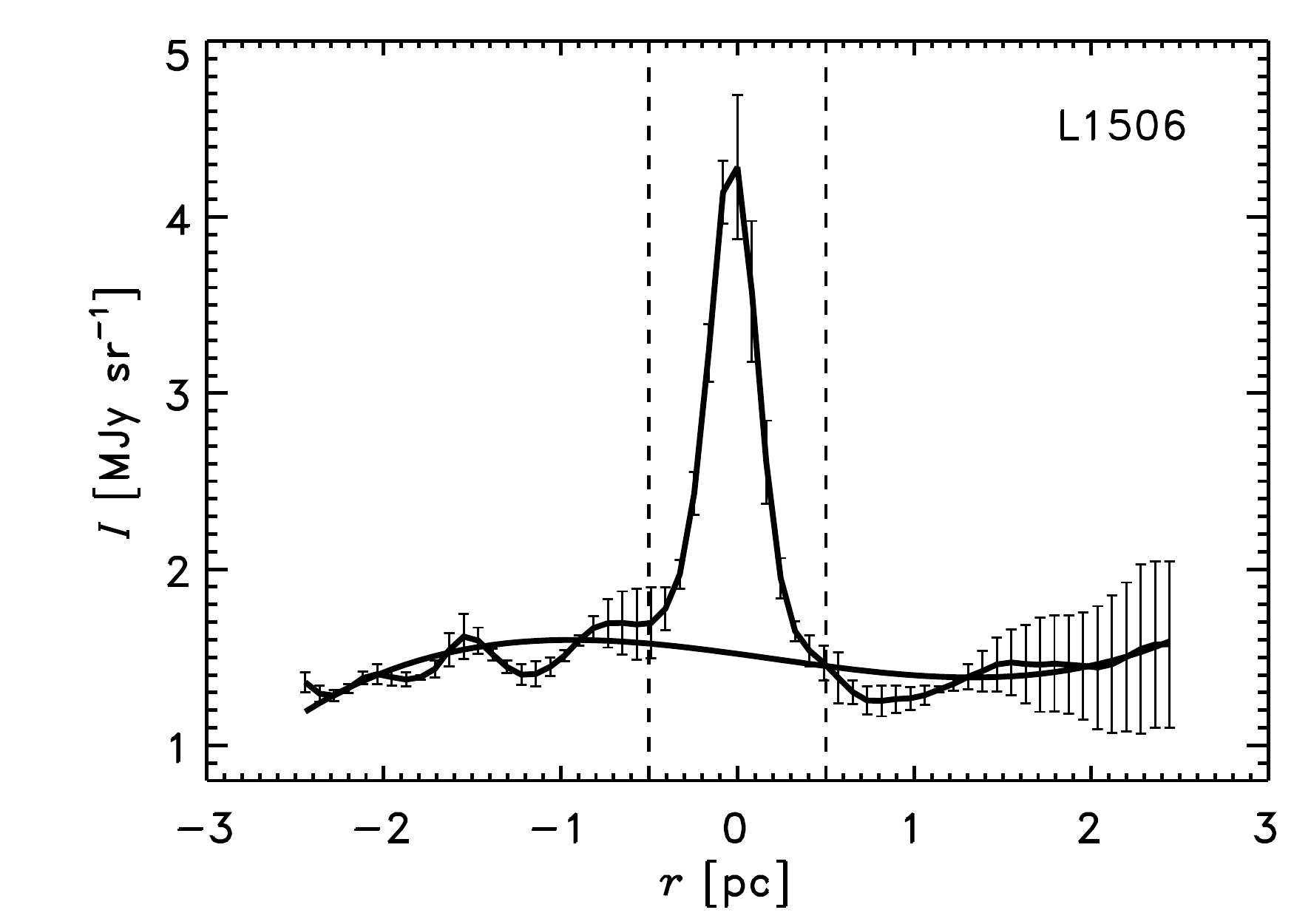}}}
  \centerline{    
  \resizebox{0.33\hsize}{!}{\includegraphics{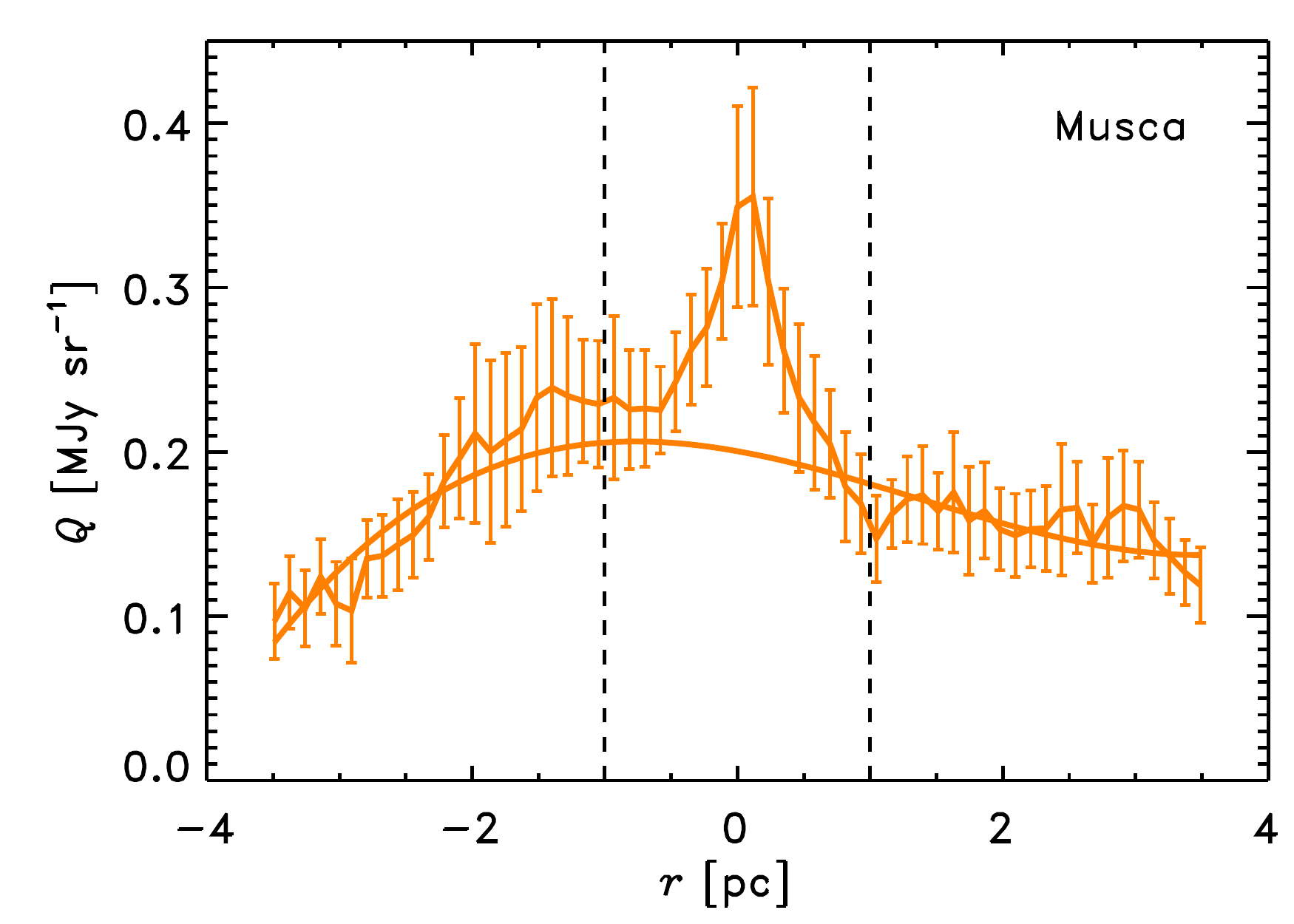}}
      \resizebox{0.33\hsize}{!}{\includegraphics{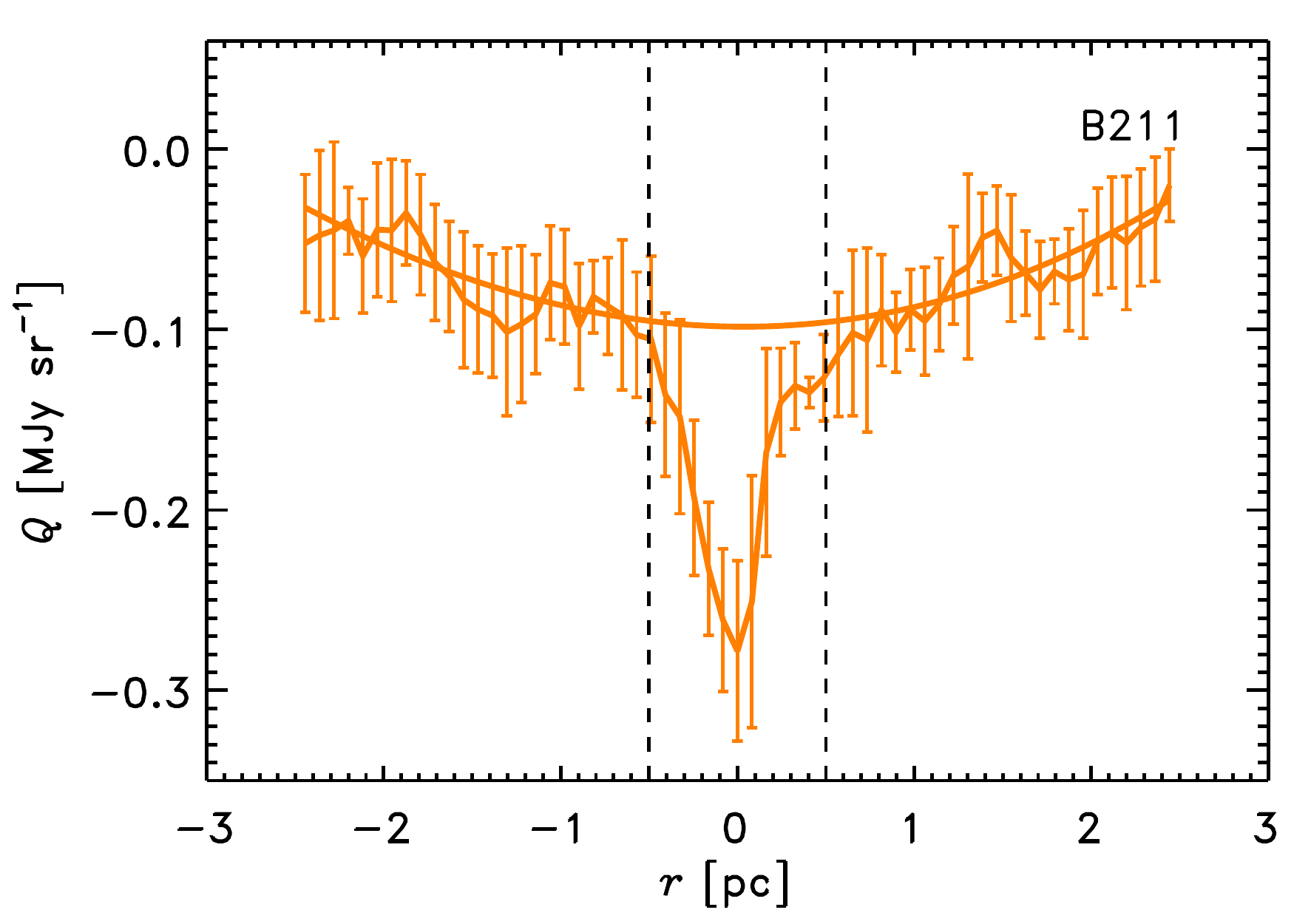}}
  \resizebox{0.33\hsize}{!}{\includegraphics{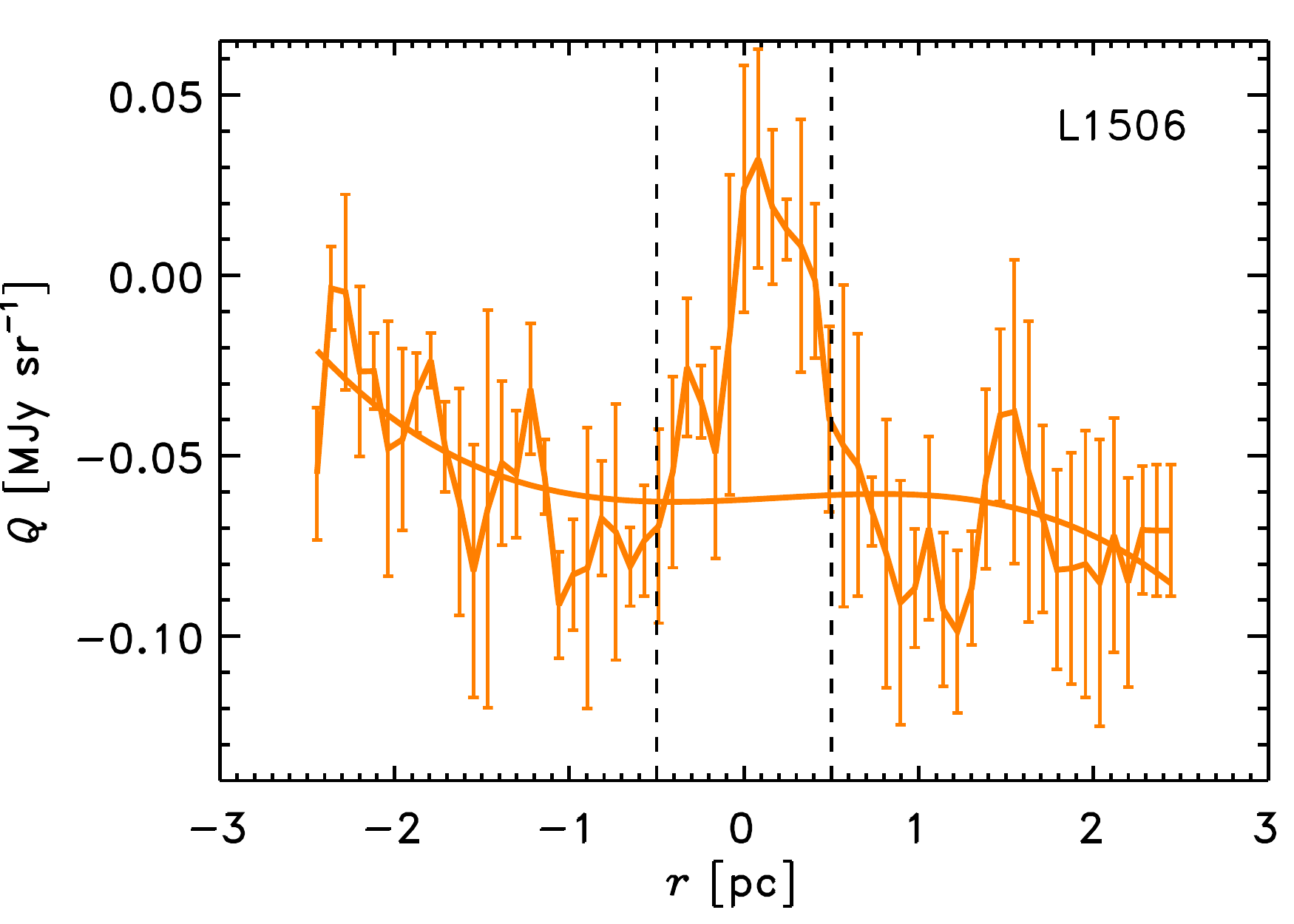}}}
   \centerline{
 \resizebox{0.33\hsize}{!}{\includegraphics{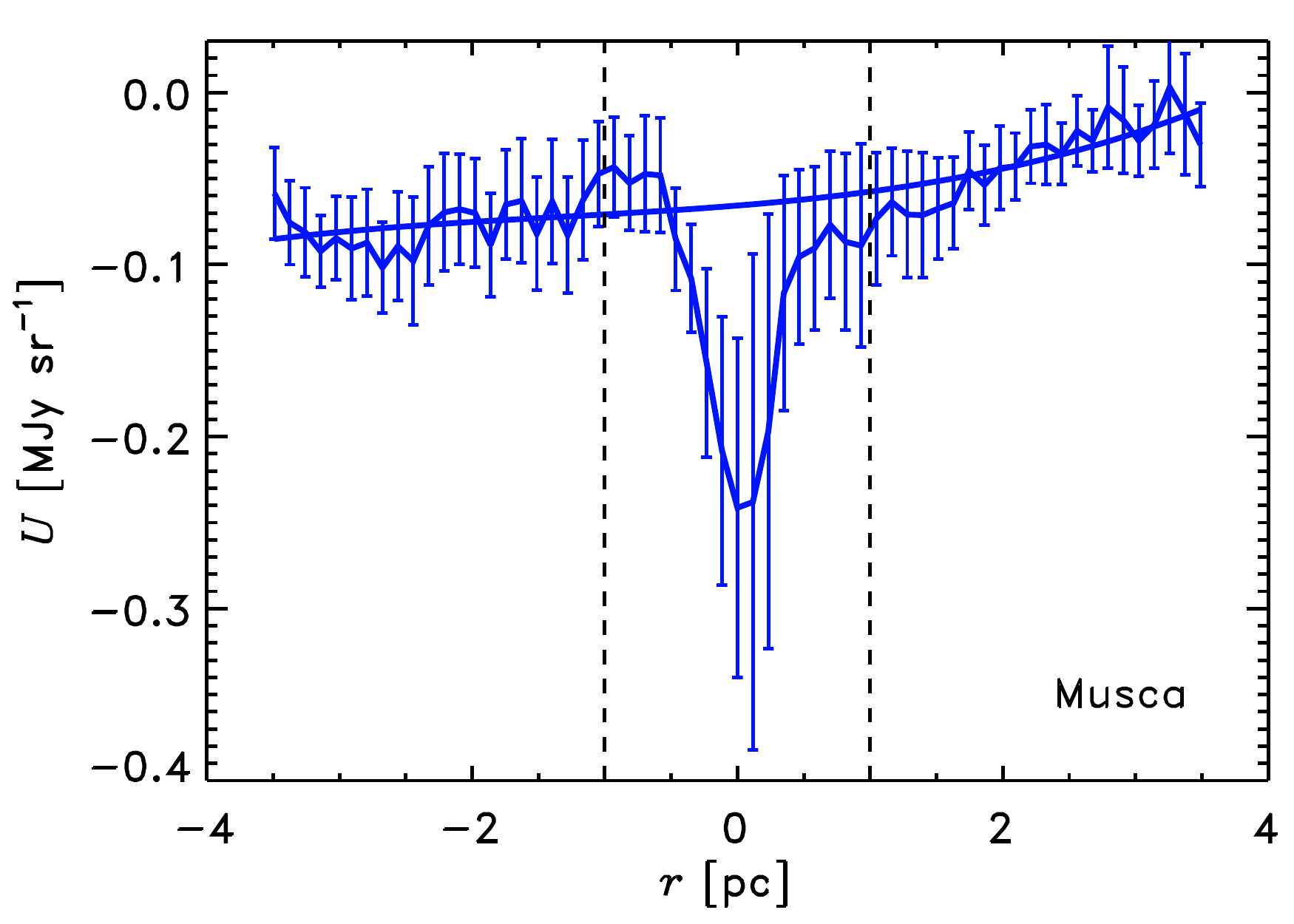}}
   \resizebox{0.33\hsize}{!}{\includegraphics{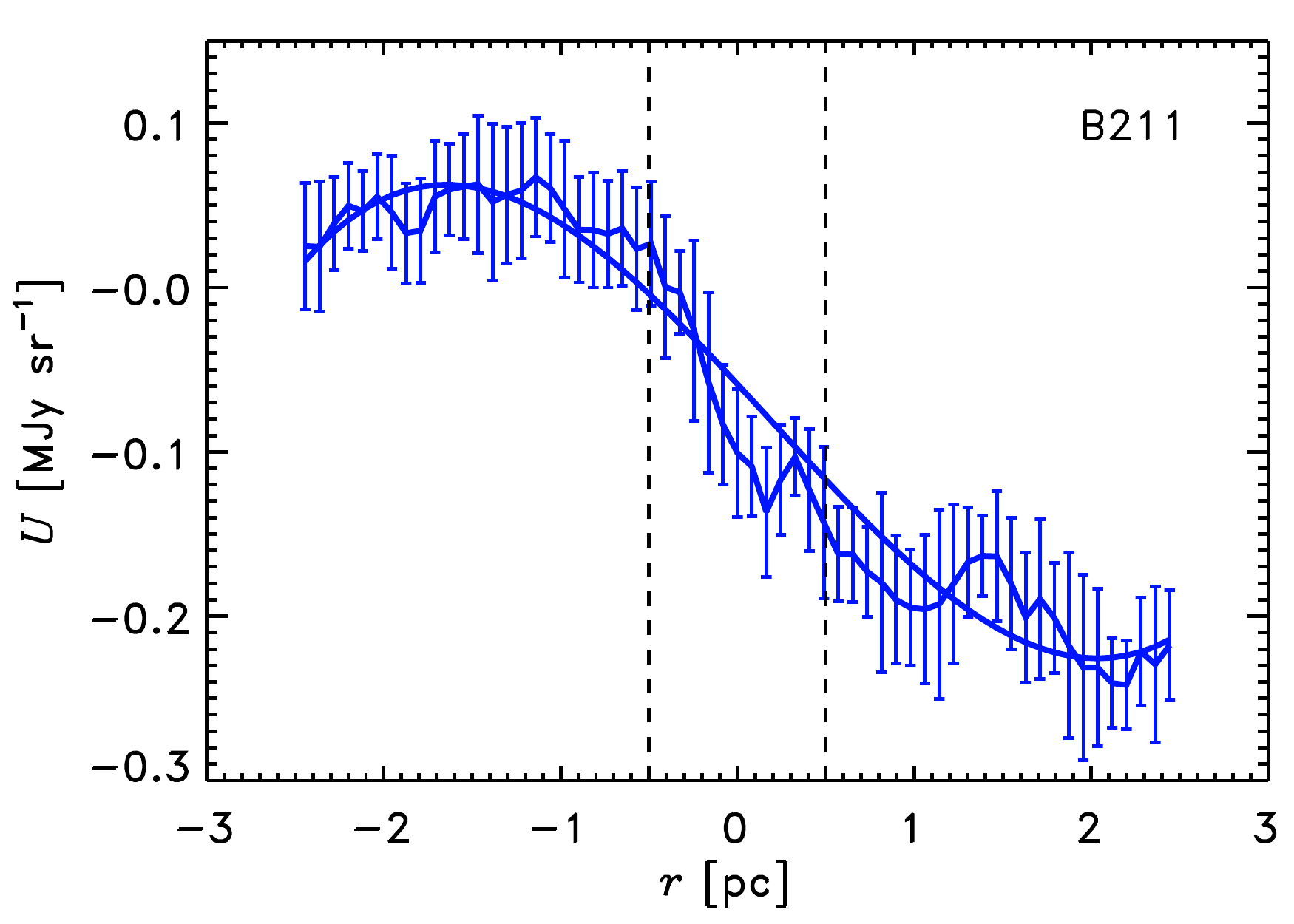}}
    \resizebox{0.33\hsize}{!}{\includegraphics{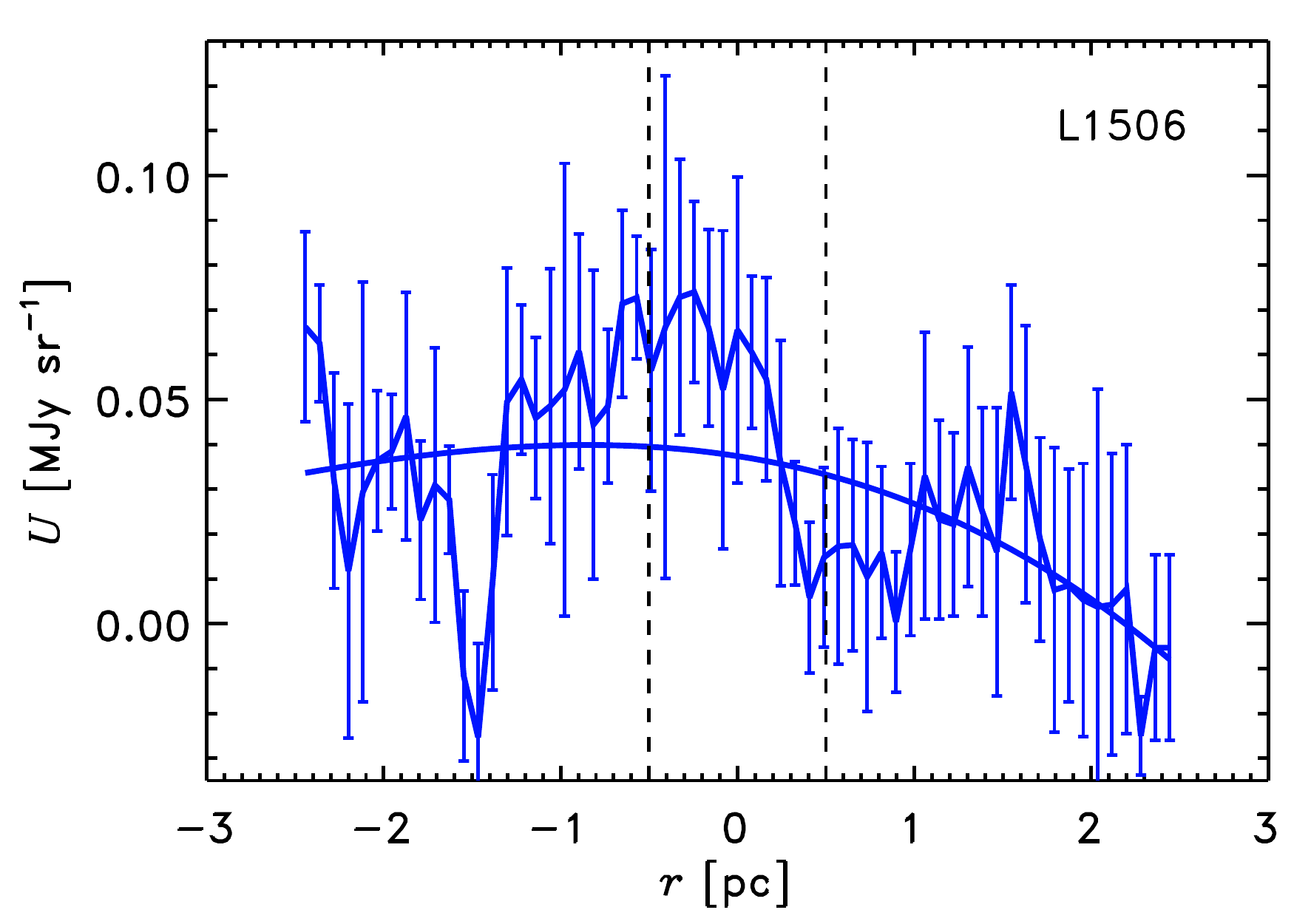}} }
 % \vspace*{-0.25 cm}
   \caption{Stokes parameters observed towards  Musca (left), B211(middle), and L1506 (right).  The profiles  
   correspond to the observed  \I\ (top), \Q\ (middle) and \U\ (bottom) emission averaged along the filament crest as explained in Sect.\,\ref{TwoLayers}.
    The error-bars represent the dispersion of the pixel values that have been averaged  at a given $r$.   
   The polynomial fits to the background are also shown. The vertical dashed lines indicate the position of the  outer radius $R_{\rm out}$ for each filament.
      }
              \label{IUQprof_bg}
    \end{figure*}

 \begin{figure*}
      \centerline{
 \resizebox{0.33\hsize}{!}{\includegraphics{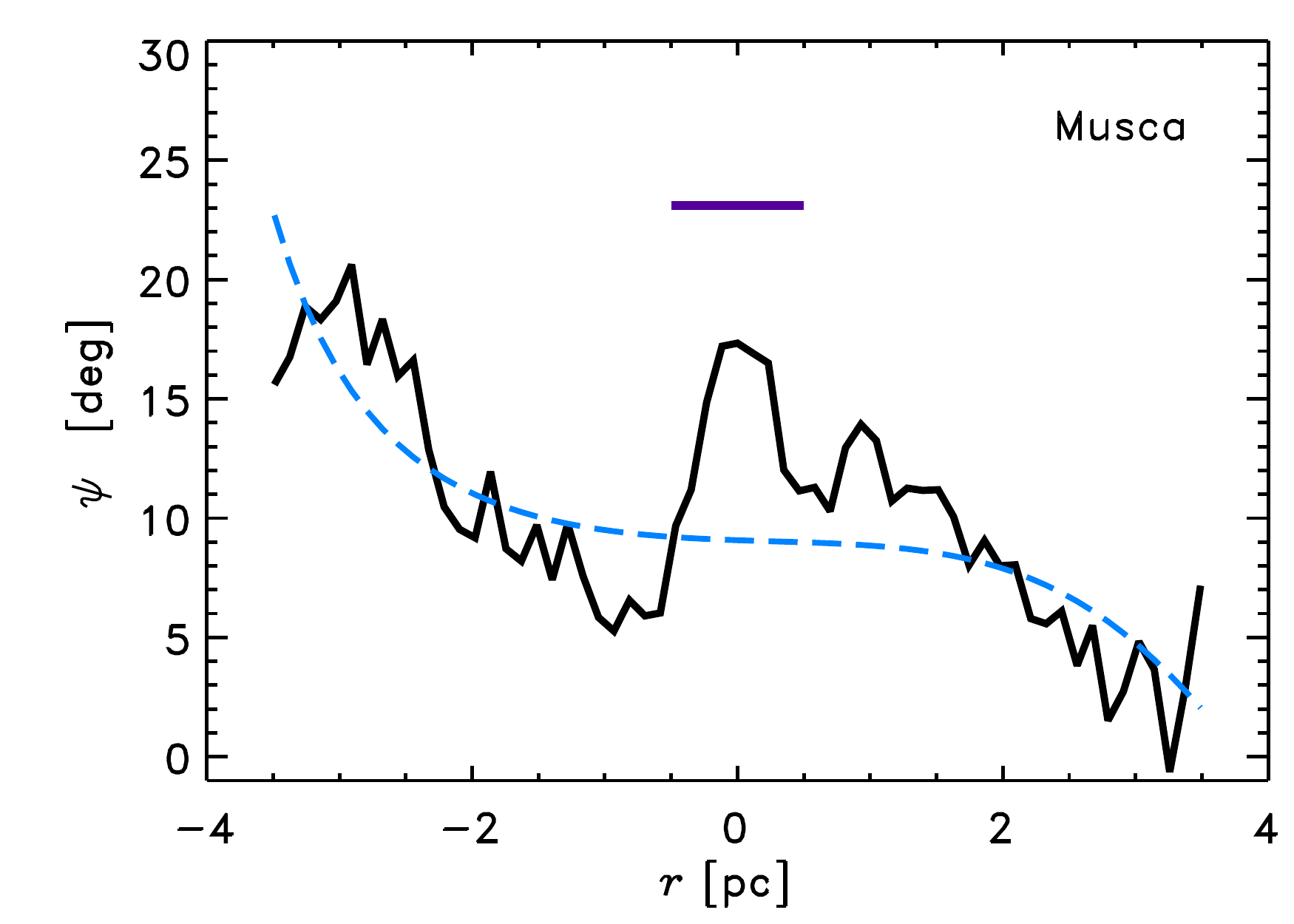}} 
  \resizebox{0.33\hsize}{!}{\includegraphics{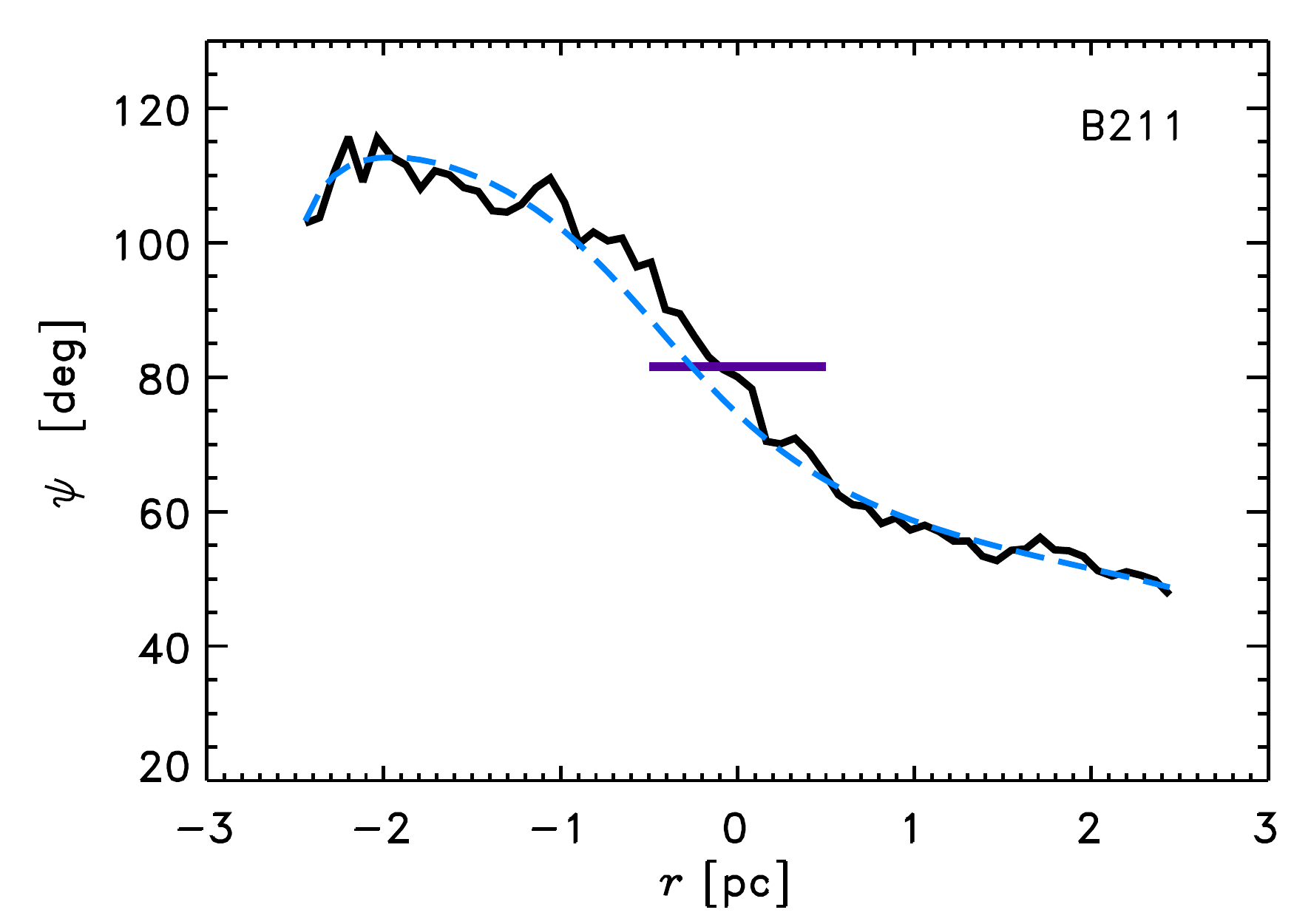}} 
   \resizebox{0.33\hsize}{!}{\includegraphics{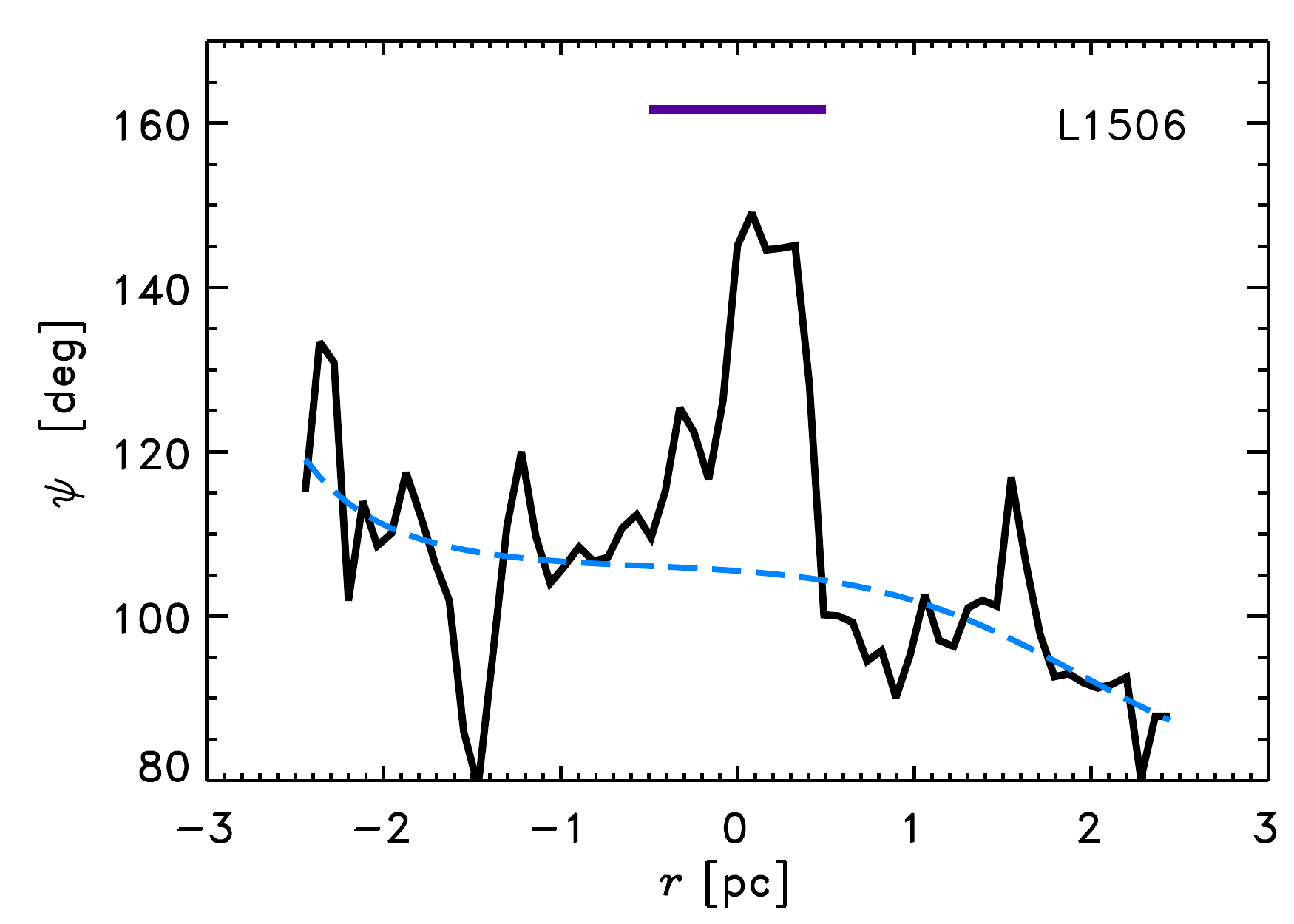}}}
   \centerline{
 \resizebox{0.33\hsize}{!}{\includegraphics{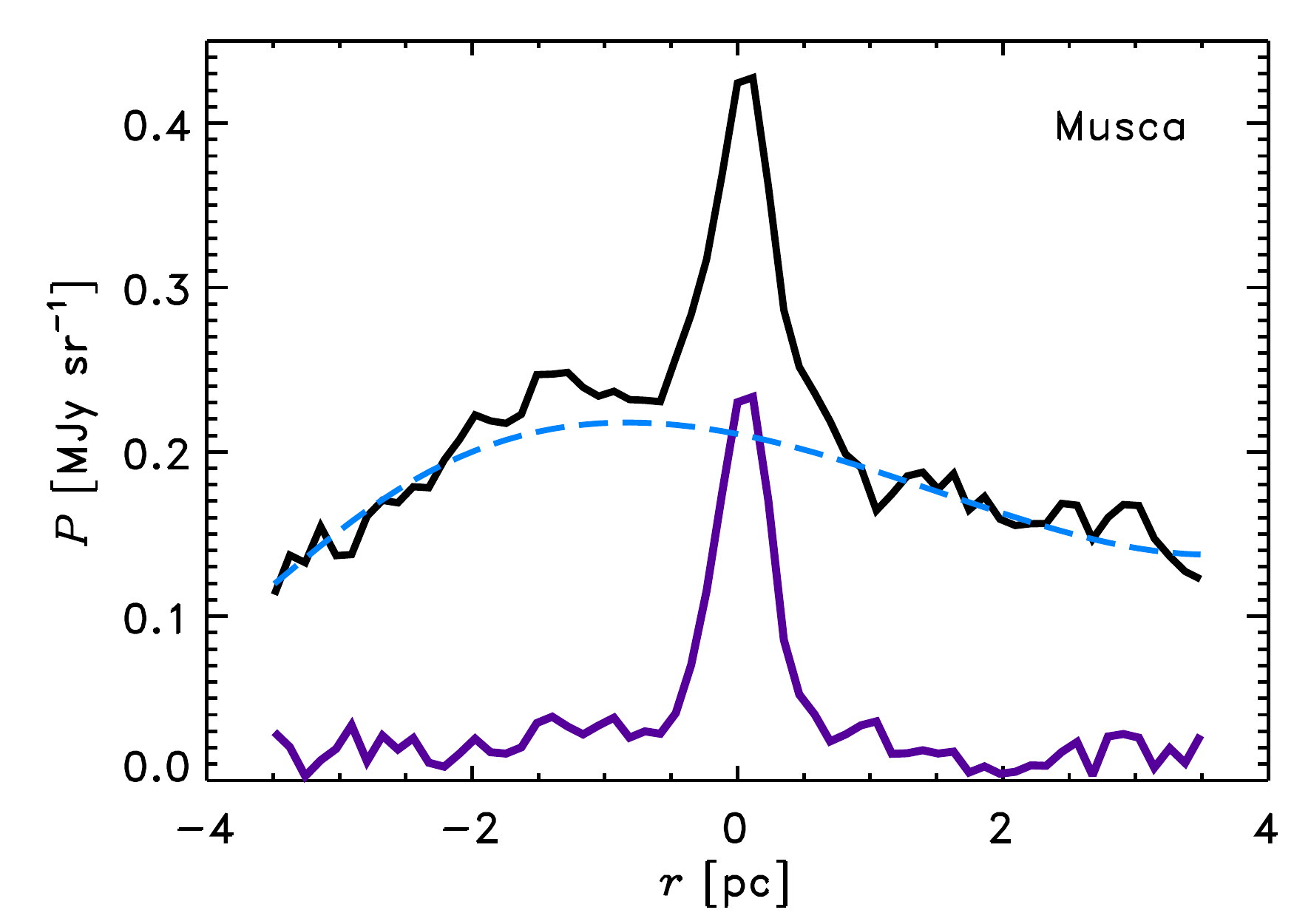}}
  \resizebox{0.33\hsize}{!}{\includegraphics{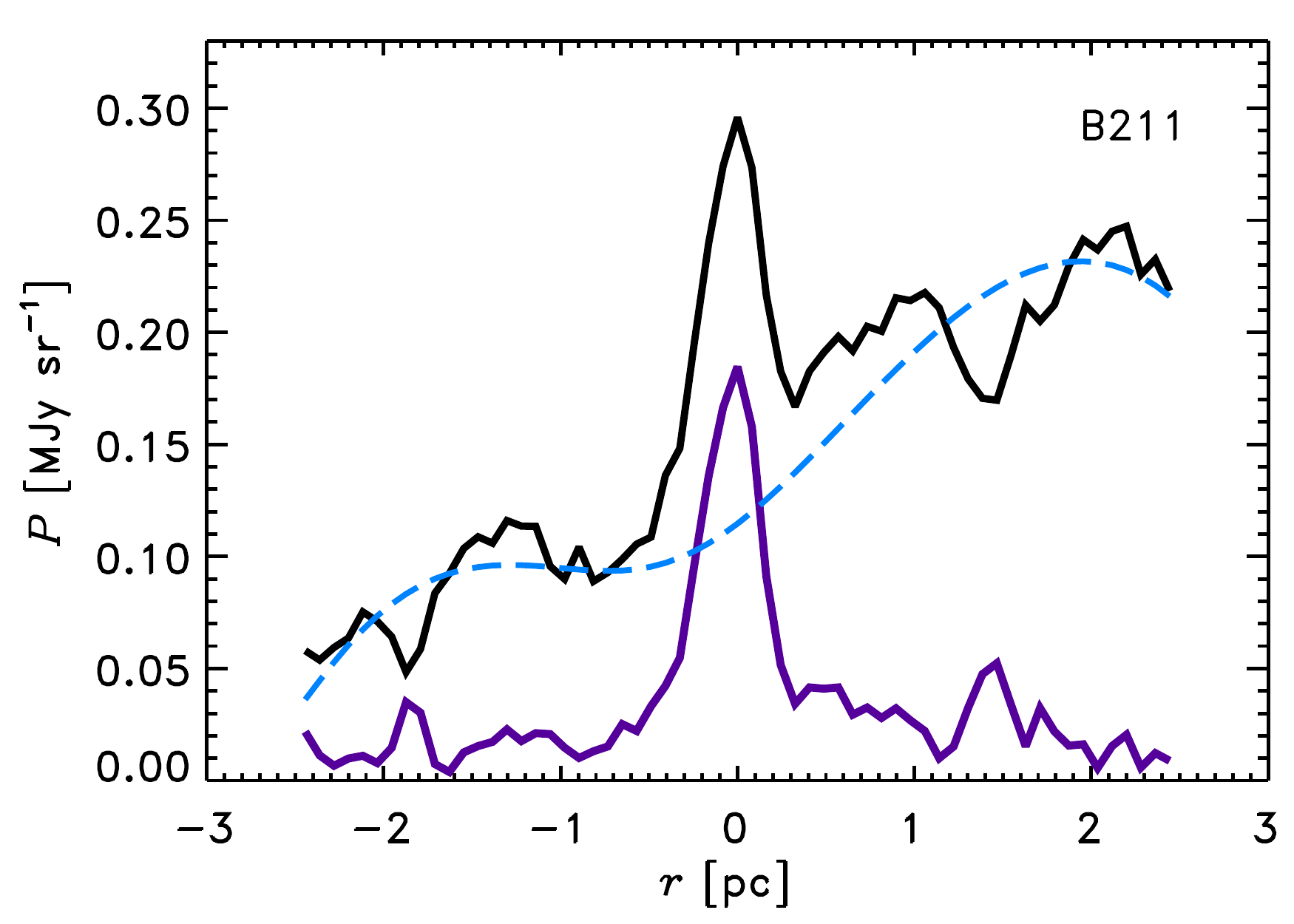}}
    \resizebox{0.33\hsize}{!}{\includegraphics{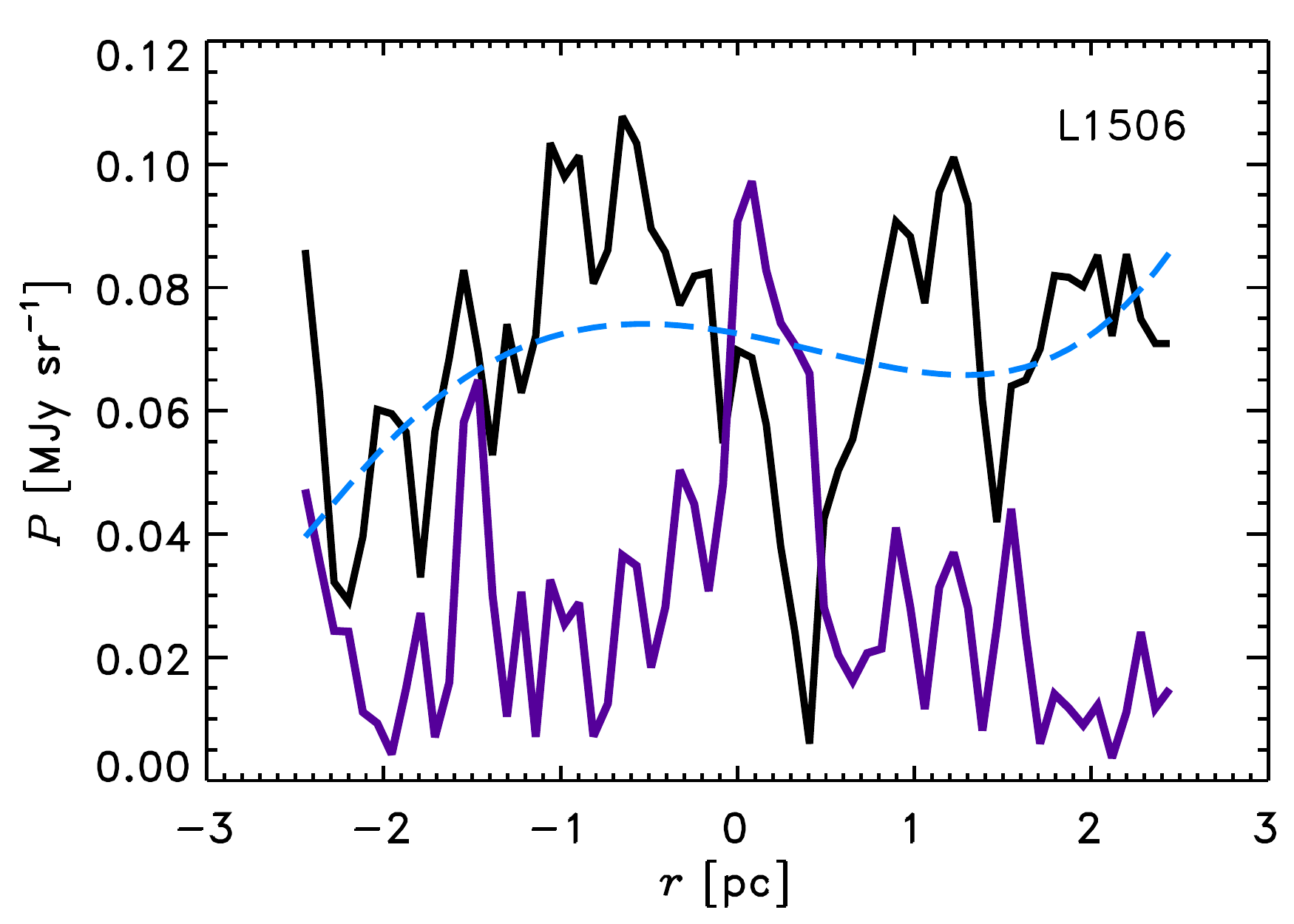}} }
        \centerline{
 \resizebox{0.33\hsize}{!}{\includegraphics{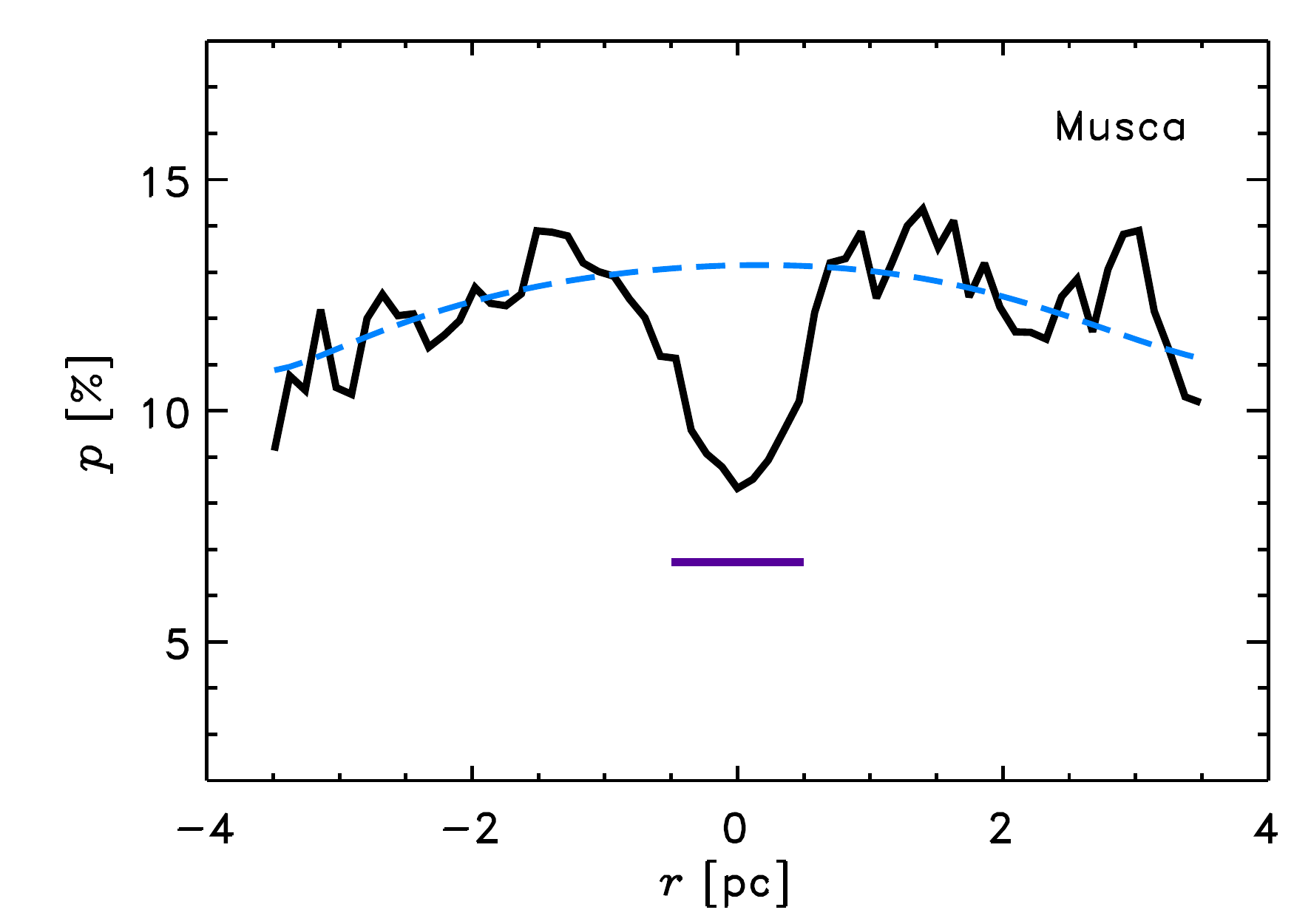}}
 \resizebox{0.33\hsize}{!}{\includegraphics{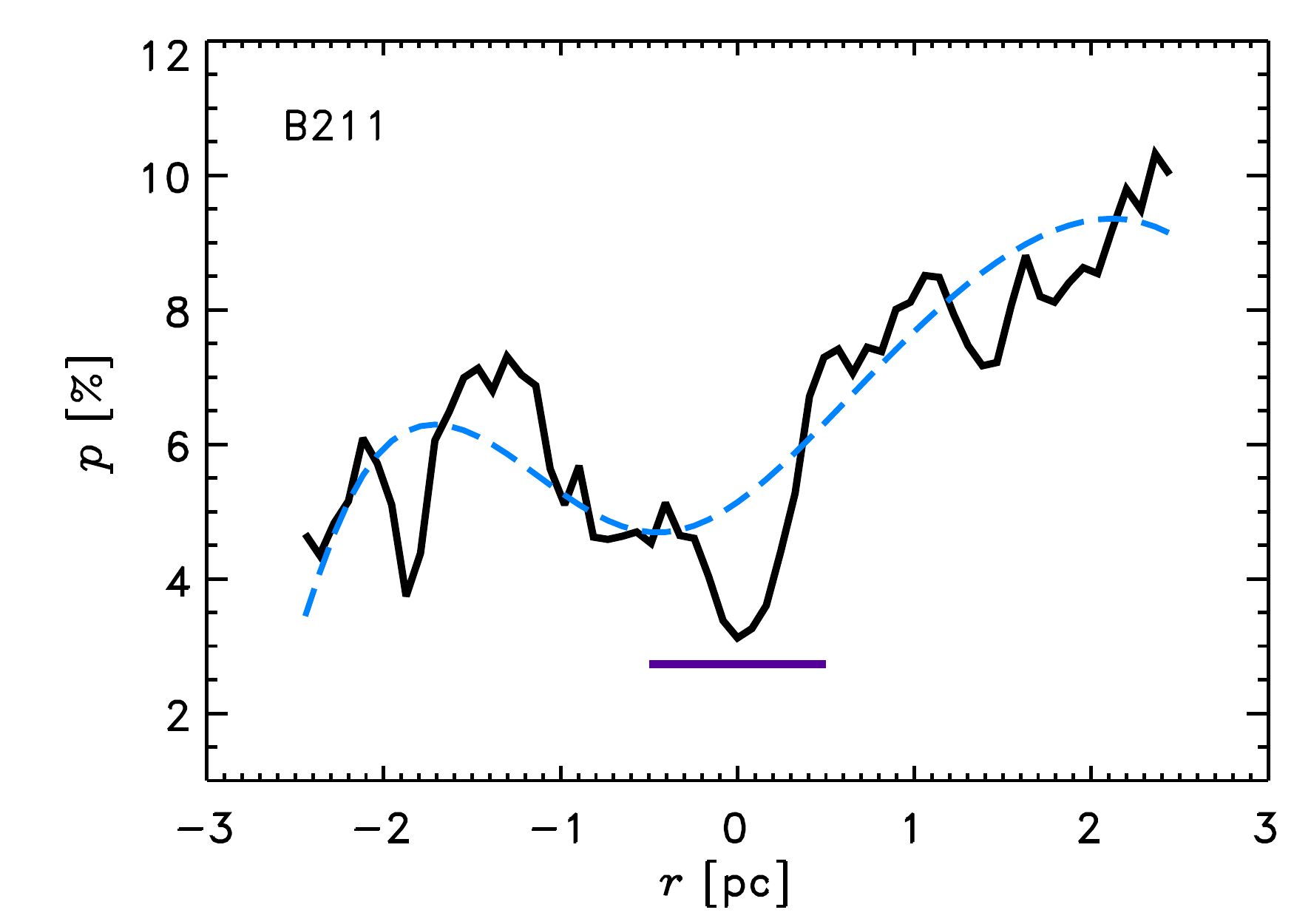}}
  \resizebox{0.33\hsize}{!}{\includegraphics{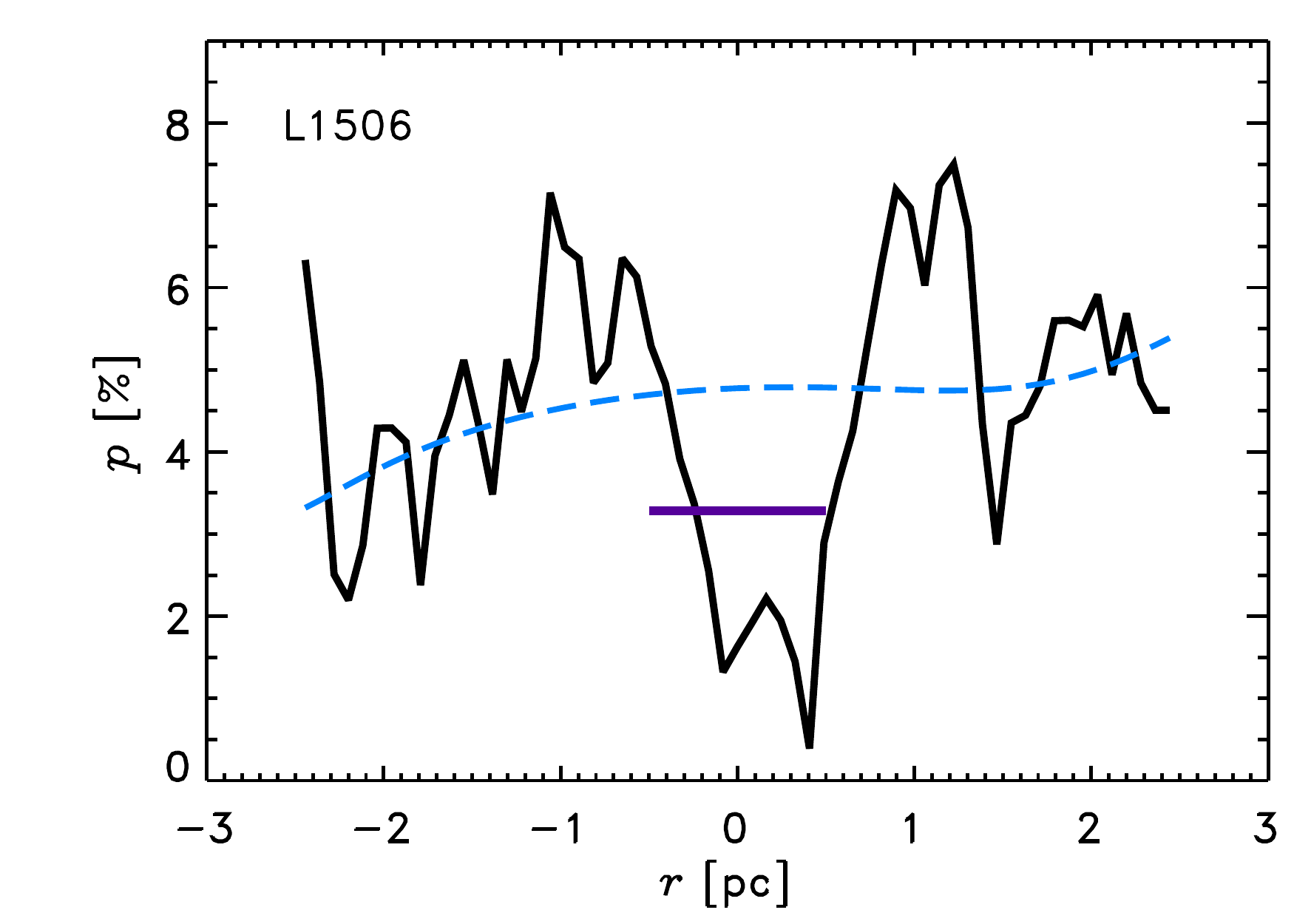}}
 }
 % \vspace*{-0.25 cm}
   \caption{Observed  profiles (in black) of the polarization angle, the  polarized intensity,  and the polarization fraction, from top to bottom.  
        The dashed blue curves show the variation of the background values of $\psi_{\rm bg}$, $P_{\rm bg}$, and $p_{\rm bg}$ derived from \I$_{\rm bg}$, \Q$_{\rm bg}$, and \U$_{\rm bg}$.
     The $P_{\rm fil}$ profiles in purple are derived from the polynomial fits   \I$_{\rm fil}$, \Q$_{\rm fil}$, and  \U$_{\rm fil}$.
      The horizontal purple lines indicate the  mean  polarization angle and fraction of the filaments  for $|r|\le R_{\rm out}$.   
    }
              \label{polarprof}
    \end{figure*}

\subsection{A two-component model: filament and background}\label{TwoLayers}

The observed polarized emission results from the integration along the LOS of the Stokes parameters. 
We take this into account by separating the filament and background emission using the spatial information of the \planck\ maps.
We  describe the dust emission observed towards the filaments  as a simplified model with two components.
 One component corresponds to the filament, for $|r|<R_{\rm out}$, where   $r$ is  the radial position relative to the filament axis ($r=0$)  and $R_{\rm out}$ is the outer radius. 
The other component represents the background. 

We define as {\it background} the  emission that is observed in the vicinity of a filament. It comprises  the emission from the  
 molecular cloud where the filament is located and from the diffuse ISM  on larger scales. 
We argue that the former is the dominant contribution. 
Indeed, the Taurus B211 and L1506 filaments and the  lower column density gas surrounding them are detected  at similar velocities in $^{13}$CO and $^{12}$CO  \citep[between 2 and 9\,km\,s$^{-1}$,][]{Goldsmith2008}, indicating  that most of the background emission is  associated with the filaments.
CO emission is also detected around the Musca filament \citep{Mizuno2001}. 
\citet{planck2014-XXVIII} presents a map of the dark neutral medium in the Chameleon region derived from  the comparison of $\gamma$-ray emission measured by $Fermi$ with 
\hi\ and CO data. This map shows emission around the Musca filament
indicating that the background is not associated with the diffuse ISM traced by \hi\ emission. 
 
We separate the filament and background contributions to the \I, \Q, and \U\ maps  within the three fields defined by the white boxes in the Taurus and Musca images displayed 
in Fig.\,\ref{Pmaps}. The filaments have all a constant orientation on the POS within the selected fields.
For L1506 the field also excludes the star-forming parts of that filament (see Sect.\ref{L1506obs}).
Within each field, we separate pixels between filament and background areas using the \I\ map to delineate the position and width of the filament. 
We fit the pixels over the background  area  with a polynomial function in the direction perpendicular to the filament axis. 
The fits account for the variations of the background emission, most noticeable for \U\ in B211 (see Fig.\,\ref{Tau_map2}).
The spatial separation is illustrated on the  \I, \Q, and \U\ radial profiles  shown in 
Fig.\,\ref{IUQprof_bg}. 
These mean profiles are obtained by averaging  data within the selected fields in the direction parallel to the filament axes. 
They are related to the profiles presented in Sect.\,\ref{obs} 
as follows. The mean profile of Musca corresponds to the averaging of profiles  4 and 5 in Fig.\,\ref{Musca_profiles}, that of 
the B211 filament  to profiles 1 to 4 in Fig.\,\ref{Tau_profiles}, and 
 for L1506 to the profiles  2 to 4 in Fig.\,\ref{L1506_profiles}.
Figure\,\ref{IUQprof_bg} displays the fits of the \I, \Q\ and \U\ profiles for $|r|>R_{\rm out}$ with polynomial functions of degree three and interpolated for $|r|<R_{\rm out}$.
The fits reproduce well the variations of the background emission outside of the filaments (Fig.~\ref{IUQprof_bg}).   
 
\subsection{Derivation of the polarization properties}\label{pol_maps}

We use the spatial separation of the filament and background contributions to the Stokes maps to 
derive the polarization properties of the filaments and their backgrounds listed in Table\,\ref{table}.
We detail how the various entries in the Table have been computed.  

The fits provide  estimates of the background values interpolated at $r=0$. The entries 
 \I$_{\rm bg}$,  \Q$_{\rm bg}$, and  \U$_{\rm bg}$ in Table\,\ref{table} are 
mean values averaged along the filament crests. 
The polarization angles ($\psi_{\rm bg}$) and fractions ($p_{\rm bg}$) for the background are computed from \I$_{\rm bg}$,  \Q$_{\rm bg}$, and  \U$_{\rm bg}$. 
The error-bars are statistical uncertainties. There is also an uncertainty associated with our specific choice for the degree of the polynomial function, which we quantify giving values derived from a third degree polynomial (pol3) and a linear (pol1) fit of the background.    
To compute the dispersion of the polarization angle ($\sigma_{\psi_{\rm bg}}$),  we smooth the 
\Q\ and \U\ background subtracted maps with a $3\times3$ pixels boxcar average. 
The values of $\sigma_{\psi_{\rm bg}}$ in Table\,\ref{table} are noise corrected. 
They correspond to the square root of the difference between the variance of the polarization angles on the background and that of the noise. 
The noise variance is computed from the dispersion of \Q\ and \U\ in  reference, low brightness, areas within the Taurus and Musca maps, 
outside the molecular clouds. It comprises both the data noise and the fluctuations of the Galactic emission.
The uncertainty on the noise correction is not a significant source of error on $\sigma_{\psi_{\rm bg}}$ after smoothing the data with a $3\times3$ pixels boxcar. 
We have also checked that we obtain values for $\sigma_{\psi_{\rm bg}}$ within the quoted error-bars using a $5\times5$ pixels boxcar average. 

We compute the filament \I, \Q, and \U\ emission averaging 
pixels of the background-subtracted maps along cuts perpendicular to the filament. 
The data averaging, done to reduce the noise, yields  about 20 values of each Stokes parameters along the crest of each filament, 
spaced by $2\arcm$ for an angular resolution of $4\parcm8$. The mean Stokes parameters
(\I$_{\rm fil}$, \Q$_{\rm fil}$, and  \U$_{\rm fil}$),  the mean polarization angle ($\psi_{\rm fil}$)  and fraction ($p_{\rm fil}$),  
the dispersion $\sigma_{\psi_{\rm fil}}$, and their error bars are computed from the average and the dispersion of these values. 
In Table\,\ref{table}, we also list the polarization properties computed for the total filament emission without background subtraction (i.e. the observed emission towards  the filament). 
The values of $\sigma_{\psi_{\rm fil}}$ are systematically greater with than without background subtraction, because the dispersion of polarization angles 
depend on the intensity of the polarized emission, which is reduced by the subtraction of the background.

The data analysis is illustrated in two figures. 
The profiles  of the polarization angle, the total polarized intensity, and the polarization fraction derived from the data and the fits to the background are shown in Fig.\,\ref{polarprof}.   This figure also displays the intrinsic polarized intensity of the filament after subtraction of the background emission. 
Fig.~\ref{psifil_profiles}-left shows the profile of $\psi_{\rm fil}$ along the crest for each of the three filaments. 
We have also averaged the background-subtracted \I, \Q, and \U\  maps in the direction parallel to the filament axis to plot the profiles of the polarization angle  
across the filaments (Fig.~\ref{psifil_profiles}-right).

\begin{table*}
\begingroup
\newdimen\tblskip \tblskip=5pt
    \caption{Polarization properties of the Musca,  B211, and  L1506 filaments (Fil)  and their backgrounds (Bg). The total and filament values  are computed on maps without and with background subtraction, respectively. They are  average  values across the filament area as explained in Sect.\,\ref{pol_maps}. The background values are at $r=0$.
 We give two sets of values derived from the polynomial fits of degree  three (pol3) 
  and one (pol1) for comparison.      } 
         \label{table} 
\nointerlineskip
\vskip -3mm
\setbox\tablebox=\vbox{
   \newdimen\digitwidth 
   \setbox0=\hbox{\rm 0} 
   \digitwidth=\wd0 
   \catcode`*=\active 
   \def*{\kern\digitwidth}
   \newdimen\signwidth 
   \setbox0=\hbox{+} 
   \signwidth=\wd0 
   \catcode`!=\active 
   \def!{\kern\signwidth}
   \newdimen\pointwidth
   \setbox0=\hbox{{.}}
   \pointwidth=\wd0
   \catcode`?=\active
   \def?{\kern\pointwidth}
\halign{
\hbox to 1.0 in{#\leaderfil}\tabskip=1.8em&
\hfil #\hfil&
\hfil #\hfil&
\hfil #\hfil&
\hfil #\hfil&
\hfil #\hfil&
\hfil #\hfil\tabskip=0pt\cr
\noalign{\doubleline \vskip 2pt}
\omit&  \I &\Q&\U&$\psi$\,$^{(1)}$&$\sigma_\psi$\,$^{(2)}$&$p$\cr
\omit&   [MJy sr$^{-1}$]&[MJy sr$^{-1}$]&[MJy sr$^{-1}$] &[deg]&[deg]&[$\%$]\cr 
\noalign{\vskip 4pt\hrule\vskip 6pt}
\multispan1\hfil Musca \hfil \cr
Total&$3.67\pm0.04$&$0.29\pm0.01$&$-0.17\pm0.01$&$15.4\pm0.5$&$2.1\pm0.3$&$\,\,9.1\pm0.5$\cr 
Fil(pol3)&$2.12\pm0.04$&$0.11\pm0.01$&$-0.11\pm0.01$&$22.2\pm1.3$&$5.1\pm0.6$&\,\,$7.3\pm0.4$\cr  
Fil(pol1)&$2.39\pm0.05$&$0.15\pm0.01$&$-0.12\pm0.01$&$19.3\pm1.0$&$3.8\pm0.4$&\,\,$8.2\pm0.4$\cr  
Bg(pol3)&$1.55\pm0.01$&$0.18\pm0.01$&$-0.06\pm0.01$&$\,\,9.8\pm0.3$&$2.7\pm0.2$&$12.1\pm0.1$\cr
Bg(pol1)&$1.28\pm0.01$&$0.13\pm0.01$&$-0.05\pm0.01$&$10.0\pm0.4$&$3.8\pm0.2$&$11.2\pm0.1$\cr  
\noalign{\vskip 4pt\hrule\vskip 6pt}
\multispan1\hfil B211 \hfil \cr
Total&$7.31\pm0.19$&$-0.24\pm0.01$&$-0.11\pm0.01$&$77.9\pm0.8$&$3.5\pm0.4$&$3.5\pm0.2$\cr 
Fil(pol3)&$4.50\pm0.19$&$-0.12\pm0.01$&$-0.04\pm0.01$&$81.0\pm1.8$&$7.8\pm0.9$&$2.7\pm0.2$\cr 
Fil(pol1)&$4.86\pm0.19$&$-0.14\pm0.01$&$-0.03\pm0.01$&$84.7\pm1.6$&$6.7\pm0.8$&$2.9\pm0.2$\cr 
Bg(pol3)&$2.81\pm0.03$&$-0.12\pm0.01$&$-0.07\pm0.01$&$75.3\pm0.8$&$5.7\pm0.6$&$4.9\pm0.2$\cr
Bg(pol1)&$2.45\pm0.02$&$-0.10\pm0.01$&$-0.07\pm0.01$&$70.1\pm1.0$&$7.9\pm0.6$&$5.1\pm0.1$\cr
\noalign{\vskip 4pt\hrule\vskip 6pt}
\multispan1\hfil L1506 \hfil \cr
Total&$3.28\pm0.09$&$-0.01\pm0.01$&$0.07\pm0.01$ &$132.3\pm2.5$&\,\,$6.5\pm2.0$&$2.1\pm0.2$\cr  
Fil(pol3)&$1.69\pm0.08$&$\,\,\,0.04\pm0.01$&$0.03\pm0.01$&$161.7\pm3.4$&\,\,$9.9\pm2.4$&$3.3\pm0.3$\cr 
Fil(pol1)&$1.77\pm0.09$&$\,\,\,0.05\pm0.01$&$0.03\pm0.01$&$164.0\pm2.8$&\,\,$7.8\pm2.1$&$3.6\pm0.4$\cr 
Bg(pol3)&$1.59\pm0.03$&$-0.05\pm0.01$&$0.04\pm0.01$&$107.8\pm2.3$&$15.1\pm1.5$&$3.9\pm0.3$\cr 
Bg(pol1)&$1.50\pm0.02$&$-0.06\pm0.01$&$0.04\pm0.01$&$105.2\pm2.0$&$13.5\pm1.3$&$4.7\pm0.2$\cr 
\noalign{\vskip 5pt\hrule\vskip 3pt}}}
\endPlancktablewide
%\par
\endgroup
\tablefoot{\tablefoottext{1}{The errors on the polarization angles, $\psi$, correspond to statistical errors on the mean value of $\psi$.} 
\tablefoottext{2}{The  dispersion of the polarization angles, $\sigma_\psi$,  are derived as explained in Sect.\,\ref{pol_maps}.}}
\end{table*}

 \begin{figure*}
   %\begin{tabular}{c}
   \hspace{0.5cm}
\includegraphics[width=9.cm]{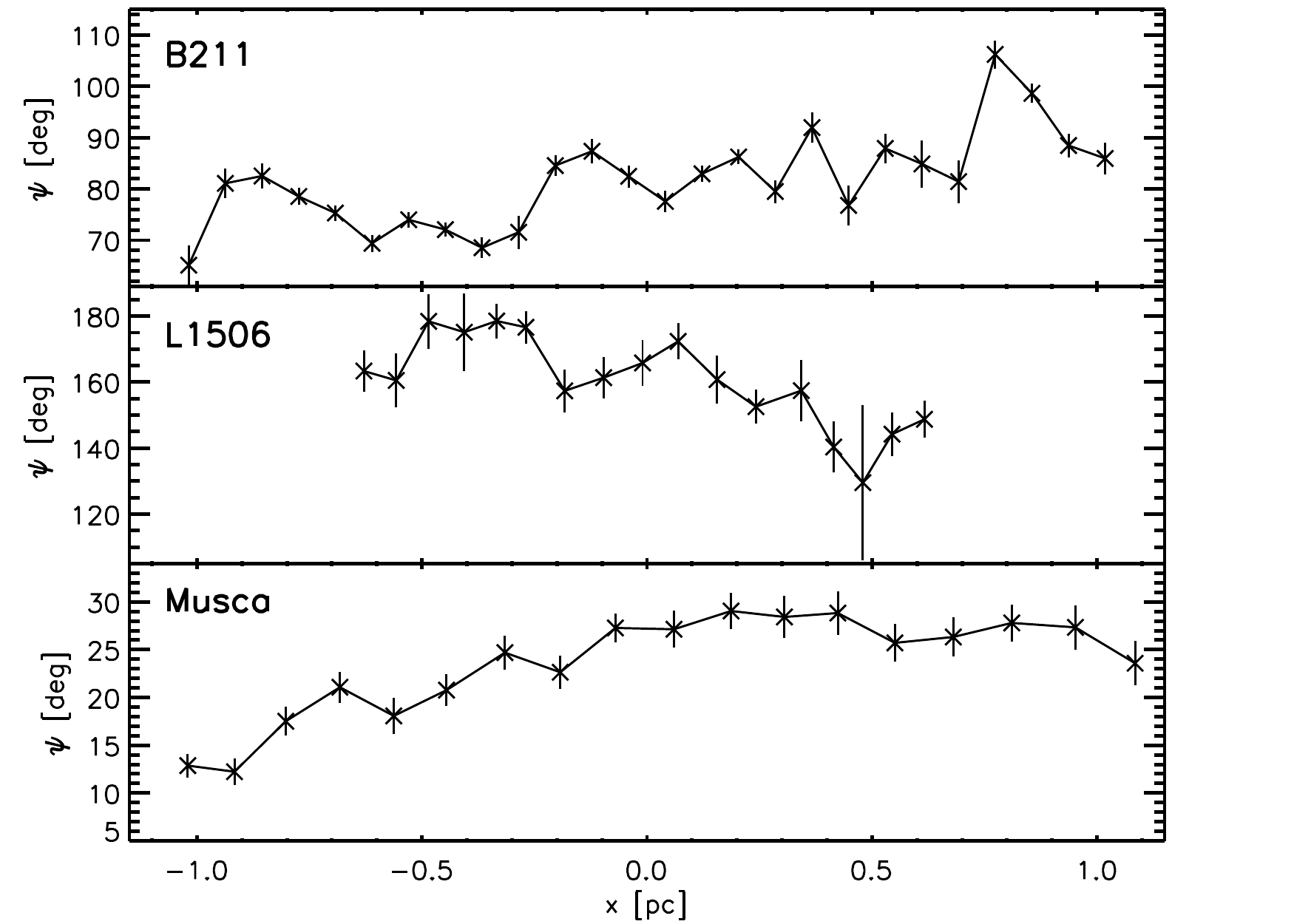}
\includegraphics[width=9.cm]{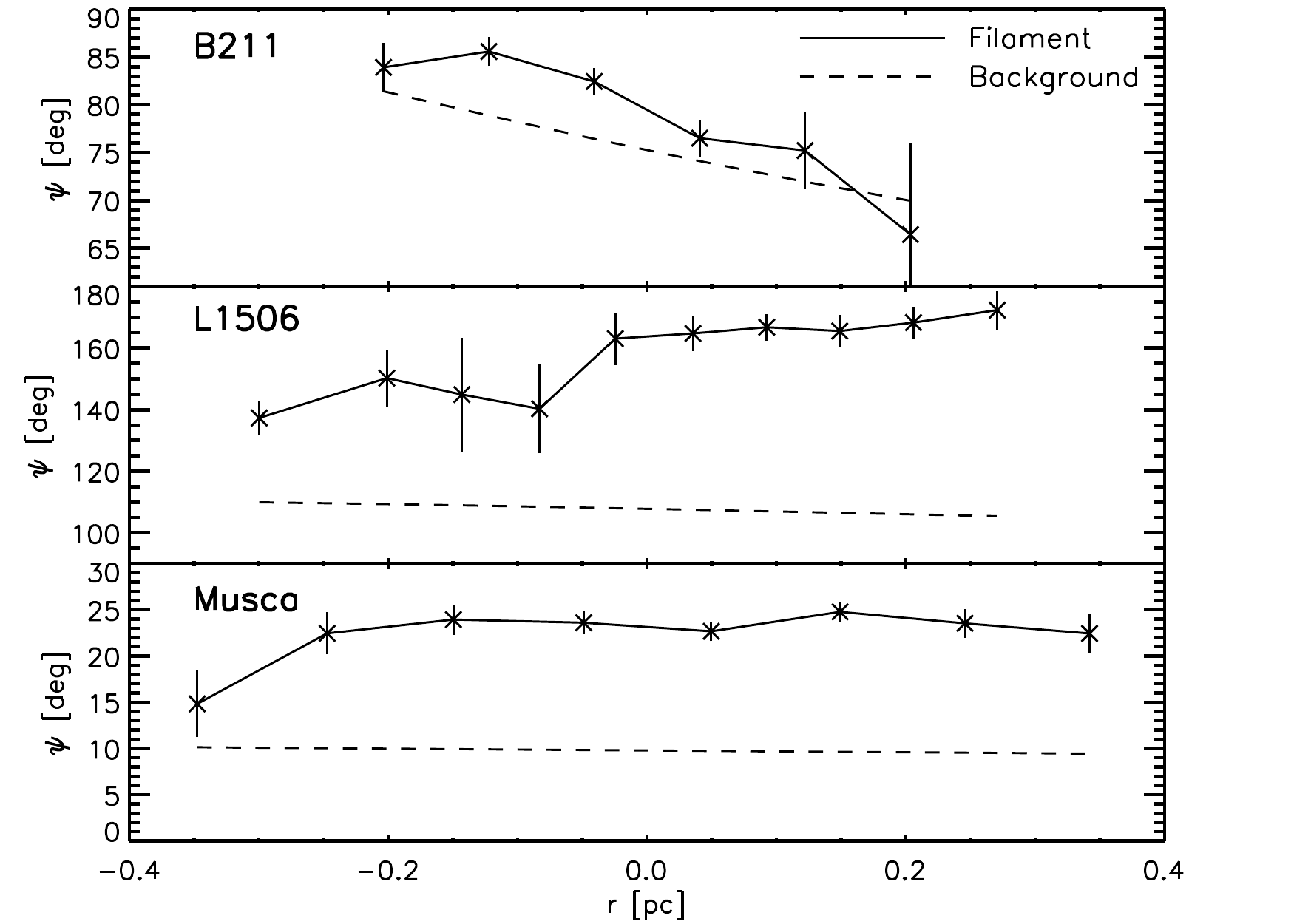}
 %\end{tabular}
   \caption{ Polarization angle along (left panel), and perpendicular (right panel) to, the  crests of the filaments.
   The crosses are 
   data points computed from \Q\ and \U\ values, after 
   averaging the background subtracted maps in the direction  perpendicular (left panel) and parallel (right) to the filament axes.
In the left panel, $x$ is a coordinate along the filament crest, while in the right panel, $r$ is the radial distance to the filament axes.
    The beam is $0.3\,$pc for Musca, and $0.2\,$pc for B211 and L1506. 
    }
              \label{psifil_profiles}
    \end{figure*}
      
\subsection{Comparing the filament properties with that of their backgrounds }\label{subsec:comparison}

We use the results of our data analysis to compare the polarization properties of the filaments to those of their backgrounds.

Figs.\,\ref{polarprof} and \ref{psifil_profiles} show that 
for the three filaments, $\psi_{\rm fil}$ differs from  $\psi_{\rm bg}$ by $12^\circ$, $6^\circ$ and 
 54$^\circ$ for the Musca, B211 and l1506 filaments, respectively (see Table\,\ref{table}). 
 In Appendix\,\ref{App2Layers}, we compute analytically the polarized emission resulting
from the superposition along the line of sight of two emission components with
different polarization angles.
This analytical model is used to compute the observed polarization angle
and fraction of the total emission 
as a function of the polarized intensity contrast and the difference in polarization angles. In the observations, the two components represent the filament and its background.  
Like in the model, the observed polarization angle  derived from the total emission at the position of the filament (without background subtraction)  differs from  $\psi_{\rm fil}$. 
The difference of $29^\circ$  for L1506 
is in good agreement with the value for the analytical model  in Fig.\,\ref{chi_df} for $\Delta \psi = 54^\circ$ and equal
contributions of the filament and background to the polarized emission.

The LOS integration of both components, for $\psi_{\rm fil}\ne\psi_{\rm bg}$, always depolarizes the total emission. 
This effect has been ignored in earlier studies
because it cannot be easily taken into account with stellar and sub-mm ground-based polarization observations. 
The L1506 filament illustrates the depolarization that results from the integration of the emission along the LOS:
the polarized emission peaks at the position of the filament only 
after subtraction of the background  (Fig.\,\ref{polarprof}).
For each of the filaments the effect of the LOS integration on $p$ is different.
The polarization fractions of the filaments  are smaller than the values derived from the total emission for Musca and B211, 
while  for L1506 it is greater. For the three filaments, 
the  polarization fraction is smaller than that of the background interpolated at $r=0$, as can be read in the right column 
of Table\,\ref{table} and seen in the
last row of Fig.\,\ref{polarprof}, but this decrease is small for L1506 ($p_{\rm fil}=3.3 \pm 0.3$ vs $p_{\rm bg}=3.9 \pm 0.3$).  
In Appendix\,\ref{depol}, we compute the depolarization factor $F$ as a function of the polarization angle difference and of the polarized 
intensity contrast. 
Figure\,\ref{DepolFig_App} shows that for $\Delta\psi=54^\circ$, $F \simeq  0.6$ for comparable contributions of the filament and background 
to the polarized emission as in  L1506 (see $P$ profile in 
Fig.\,\ref{polarprof}). This factor is in good agreement with the ratio between the two $p$ values,  without and with background subtraction, 
for  L1506 in Table\,\ref{table}. 

The differences between the filament position angles (PA) and 
$\chi$ ($\psi+90^\circ$) are listed in Table\,\ref{table_po_gamma}. We find that $\vec{B}_{\rm POS}$  in the backgrounds of Musca and B211 are close, within $20^\circ$,
to being orthogonal to the filament axis, while for L1506, the background  $\vec{B}_{\rm POS}$  is at 37$^\circ$.   In the filaments, $\vec{B}_{\rm POS}$ is 
perpendicular within $10^\circ$ to the axis of Musca and B211, while it is close to parallel in L1506. We point out, though,  
that  two orientations that are nearly perpendicular in 3D may be close to parallel on the POS \citep{planck2014-XXXII,planck2015-XXXV}.

 \section{Interpretation of the polarization fraction}\label{Interp}

We discuss possible interpretations of the polarization fraction and its variation from the backgrounds to the filaments.
The polarization fraction depends on dust grain properties and on the magnetic field structure expressed as the sum of mean and
turbulent components. The observed polarization fraction is empirically parametrized as 
 \begin{align}
p= p_{\rm dust}\,R\,F\,\cos^2\gamma,\label{p_eq}
\end{align}
\noindent
 to distinguish four different effects due to both the local properties of dust and magnetic fields, and the LOS integration.
The polarization properties of dust are taken into account with $p_{\rm dust}$ that depends on the composition, size, and shape of dust grains \citep{Lee1985,Hildebrand1988}.
The Rayleigh reduction factor, $R\le1$, 
characterizes the efficiency of grain alignment with the local magnetic field orientation.
The factor $F$ expresses the impact on the polarization fraction of  
the variation of the magnetic field orientation along the LOS  and within the beam.  
The role of the orientation of the mean magnetic field with respect to the POS is expressed by the  $\cos^2\gamma$ factor.
The polarization fraction is maximal when the magnetic field is uniform and in the POS ($\gamma =0$), 
while there is no linear polarization if the magnetic field is along  the LOS ($\gamma = 90^\circ$).    
A main difficulty in the interpretaion of polarization observations is
that these four quantities cannot be determined independently. 
In particular, the product $p_{\rm dust} \, R \, F $ is degenerated with the
orientation of the mean magnetic field.   

The interpretation of the polarization fraction presented  in Sect.\,\ref{Brole} focuses  on the structure of the magnetic field.  
The factors $p_{\rm dust}$ and $ R$ 
are  discussed in Sect.\,\ref{sec:dust_factors}.

 \subsection{The structure of the magnetic field}\label{Brole}

 \subsubsection{Mean magnetic field orientation}\label{Bmean}

 For each of the three filaments,  we find that the polarization angles vary from the background to the filament. 
 These variations reflect changes in the 3D structure of the  magnetic field, which impact $p$  in two ways. First,  changes of $\psi$ along the LOS  depolarize the emission lowering the observed $p$ (see Appendix\,\ref{App2Layers}). 
Second, $p$ depends on the  angle of \vec{B} with respect to the POS ($\gamma$), which statistically must vary as much as $\psi$.  
We quantify these two aspects.

(1) L1506 illustrates the depolarization due to the superposition of emitting layers with different polarization angles (Sect.\,\ref{subsec:comparison}). 
For this filament, the decrease of $p$ versus $N_{\rm H}$ in  Fig.\,\ref{pVsnh_3fil}  can be almost entirely explained by the change of the $\vec{B}_{\rm POS}$  orientation 
between the filament and its background. For Musca and B211, the $\psi $ differences are too small to account for the observed decrease of $p$ in the filaments (see 
Fig.\,\ref{DepolFig_App}).

(2) The smooth variations of $\psi_{\rm bg}$ in the background of B211 and Musca, by about $60^\circ$ and $20^\circ$ respectively, 
are associated with variations of $p$  by $3$-$5\,\%$ (Fig.\,\ref{polarprof}). The variations of $\psi $ are likely to be associated with variations 
of $\gamma $ of comparable amplitude that could contribute to the variations of $p$. We build on this idea to quantify
the variations  of $\gamma $ that would be needed to account for the difference between  $p$ values
for the filament and the background (at $r=0$),  $p_{\rm fil}$ and $p_{\rm bg}$,  listed in Table\,\ref{table}.
The angles $\gamma_{\rm fil}$ and $\gamma_{\rm bg}$ of \vec{B} with respect to the POS for the filament and the background 
are calculated, within the range $0^\circ$ to $90^\circ$, from  Eq.\,(\ref{p_eq}) written as 
\begin{equation}
p =p_0 \cos\gamma^2,\label{po}
\end{equation}
where $p_0 = p_{\rm dust} \, R \, F $ may differ between  the filament ($p_0^{\rm fil}$) and the background ($p_0^{\rm bg}$).
The difference $\Delta \gamma = \gamma_{\rm fil} - \gamma_{\rm bg}$ depends on both $p_0^{\rm bg}$ and $f_0 = p_0^{\rm fil}/p_0^{\rm bg}$.  
For illustration we discuss two cases.
First,  in Table\,\ref{table_po_gamma} for $f_0 = 1$, we give two values of 
 $\Delta \gamma $ computed  for extreme values of $ p_0^{\rm bg}$: the minimum value set by the observed polarization fraction at $r=0$ and a maximum value of $ 0.2$. 
Second, in Fig.\,\ref{Varpo}, we plot  $\gamma_{\rm bg} $ and $\gamma_{\rm fil} $  
versus  $f_0$  for $p_0^{\rm bg} = 0.15$. In this figure, the sign of $\Delta \gamma $ changes from negative to positive at different values
of $f_0$ for each of the filaments. 
 The orientation of \vec{B} contributes  to  the decrease of the polarization fraction in the filament
when $\Delta \gamma > 0$, 
i.e., $\vec{B}$ is  closer to the  LOS in the filament than in the background.
By no way, this could be  the rule to explain the low values of the
polarization fraction that have been observed in all other filaments in
dark clouds \citep[e.g.][]{Goodman1995,Sugitani2011,Cashman2014}.   We conclude that other factors than the mean magnetic field orientation contribute to the decrease of $p$ in the filaments.

     \begin{table}
\begingroup
\newdimen\tblskip \tblskip=5pt
    \caption{
Columns 2 and 3 give the values of   $p_0^{\rm bg}$  and the corresponding $\Delta\gamma=\gamma_{\rm fil}-\gamma_{\rm bg}$ computed for $p_0^{\rm bg}=p_0^{\rm fil}$ (see Sect.\,\ref{Bmean}). 
 Column 4 gives the values of $\Delta\chi=|\chi_{\rm fil}-\chi_{\rm bg}|$ corresponding to the $\psi_{\rm fil}$ and $\psi_{\rm bg}$ values of Table\,\ref{table} for the polynomial fits (pol3). Columns 5 and 6 give the relative angle between the PA of the filament and the $\vec{B}_{\rm POS}$ angle in the filament and in the background, respectively. The PA  of the filaments are given in Table\,\ref{table_param}.   }
         \label{table_po_gamma} 
\nointerlineskip
\vskip -3mm
\setbox\tablebox=\vbox{
   \newdimen\digitwidth 
   \setbox0=\hbox{\rm 0} 
   \digitwidth=\wd0 
   \catcode`*=\active 
   \def*{\kern\digitwidth}
   \newdimen\signwidth 
   \setbox0=\hbox{+} 
   \signwidth=\wd0 
   \catcode`!=\active 
   \def!{\kern\signwidth}
   \newdimen\pointwidth
   \setbox0=\hbox{{.}}
   \pointwidth=\wd0
   \catcode`?=\active
   \def?{\kern\pointwidth}
\halign{
\hbox to 0.6 in{#\leaderfil}\tabskip=1.5em&
\hfil #\hfil&
\hfil #\hfil&
\hfil #\hfil&
\hfil #\hfil&
\hfil #\hfil\tabskip=0pt\cr
\noalign{\doubleline \vskip 2pt}
\omit &\,\,\,\,$p_0^{\rm bg}$\,\,$^{(1)}$&$\Delta\gamma$&$\Delta\chi$&$|\chi_{\rm fil}-{\rm PA}|$&$|\chi_{\rm bg}-{\rm PA}|$\cr 
\omit & &[deg]&[deg]&[deg]&[deg]\cr 
\noalign{\vskip 4pt\hrule\vskip 6pt}
Musca & 0.12--0.2 &\,\,\,39--14 &12&82&70\cr
\noalign{\vskip 4pt\hrule\vskip 6pt}
B211 & 0.049--0.2 &42--8 &6&81&75\cr
\noalign{\vskip 4pt\hrule\vskip 6pt}
L1506 & 0.039--0.2 &25--3 &54&17&37\cr
\noalign{\vskip 5pt\hrule\vskip 3pt}}}
\endPlancktablewide
%\par
\endgroup
\tablefoot{\tablefoottext{1}{The smallest  value of $p_0^{\rm bg}$ is  the observed $p_{\rm bg}$. The maximum value of  0.2  is close to the  maximum dust polarization fraction observed by \planck\ at 353\,GHz \citep[][]{planck2014-XIX}.}}
\end{table}

  \begin{figure}
   \centerline{
 \resizebox{0.9\hsize}{!}{\includegraphics{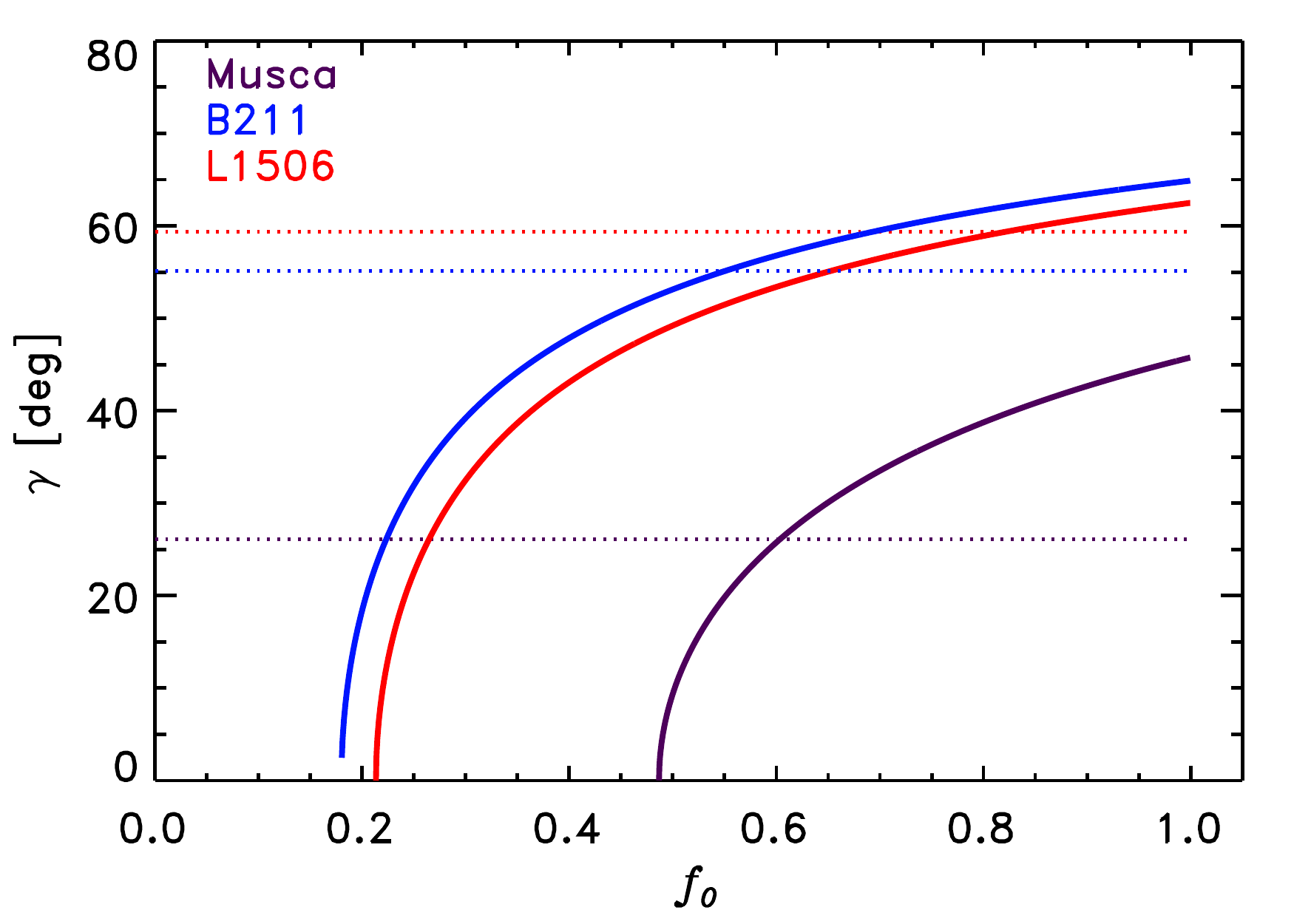}}}
   \caption{ Plot  of the angles of   \vec{B} with respect to the POS  $\gamma_{\rm bg} $ (dotted line) and $\gamma_{\rm fil}$ (solid line)  computed with Eq.\, \ref{po}, as a function of  $f_0 = p_0^{\rm fil}/p_0^{\rm bg}$   for $p_0^{\rm bg} = 0.15$.  }
              \label{Varpo}
    \end{figure}

\subsubsection{Fluctuation of the magnetic field orientation}

Depolarization in filaments could result from 
the integration along the LOS of a large number of emission layers with different  orientations of the magnetic field. 
Assuming the number of layers is proportional to 
the total column density $N_{\rm H}$ \citep{Myers1991},  
we expect $p$ to decrease  for increasing  $N_{\rm H}$ due to the averaging of the random component of the magnetic field \citep[][]{Jones2015}. 
These models depend on the ratio between the strength of the random and 
of the mean components of the magnetic field \citep[][]{Jones1992}. The steepest dependence, $p$ scaling as $N_{\rm H}^{-0.5}$, is obtained when the random component is 
dominant, in which case the dispersion of the polarization angle reaches its maximum value $\sigma_\psi^{\rm max}$ of $52^\circ$ \citep{planck2014-XIX}.
In such a model, we expect the dispersion of $\psi$ to be close to $\sigma_\psi^{\rm max}$ in the background and much smaller  in the filament \citep[see Fig.\,9 in][]{Jones1992}. This 
prediction is inconsistent with our data because  (1) $\sigma_{\psi_{\rm bg}}$ is much smaller than  $\sigma_\psi^{\rm max}$, and  (2) it is comparable to and even smaller than 
$\sigma_{\psi_{\rm fil}}$ (Table\,\ref{table}).
 
Moreover Fig.\,\ref{psifil_profiles} shows that the variations of the magnetic field orientation
 are not random. Systematic variations  of the polarization angle along and across the filaments
must also exist on scales unresolved by \planck\ observations. Hence, we expect some 
depolarization from coherent changes of the field orientation in the  beam  and along the LOS.
Spectroscopic observations of B211 show density and velocity structures on scales five times smaller than 
the \planck\ beam, and coherent over lengths of $\sim0.4\,$pc, i.e., two \planck\ beams \citep{Hacar2013}.
Similar structures are anticipated to exist for the magnetic field.

Theoretical modelling is warranted 
to quantify the depolarization within the beam and along the LOS, and to test whether  the structure of the magnetic field may account for the observed decrease of 
$p$ in the filaments. 
The decrease of $p$ due to  the structure of the magnetic field  has already been quantified for helical fields 
 \citep[e.g.,][]{Carlqvist1997,Fiege2000pol}. Other models could be investigated such as the
 magneto-hydrostatic configuration presented by \citet{Tomisaka2014}. \citet{Tomisaka2015} describes 
 the change in the polarization angle and the decrease of the polarization fraction produced 
by  the pinching of the \vec{B}-field lines by gravity within this model. Since the angular resolution of \planck\  does not 
resolve the inner structure of the filaments, observations with a higher angular resolution would be needed to fully test  such models.
This is the research path to understanding 
the role played by the magnetic field in the formation of star forming filaments.

\subsection{Grain alignment efficiency and dust growth}
\label{sec:dust_factors} 

Different mechanisms have been proposed to explain how the spin axes of grains can  become aligned  with  the magnetic field.  
Alignment could result from 
magnetic relaxation  \citep{Davis1951,Jones1967,Purcell1979,Spitzer1979}. 
However, the most recent theory stresses  the role of radiative  torques   \citep[RAT,][]{Dolginov1976,Draine1996,Draine1997,Lazarian1997,Lazarian2007,LazarianHoang2007,Hoang2014}.
A number of studies interpret polarization observations in the framework of  this theory \citep{Gerakines1995,Whittet2001,Whittet2008, Andersson2010,Andersson2011,Andersson2013,Cashman2014}.
 The observed drop of $p$ with column density has been interpreted as evidence of the progressive loss of grain alignment with increasing column 
density \citep{Andersson2015}.  
However, this interpretation cannot be validated without also considering the impact of gas density on the grain size and shape. 

Dust observations, both in extinction and in emission, provide a wealth of evidence for grain growth in dense gas within molecular clouds \citep[e.g.,][as recent references]{Ysard2013,Roy2013,Lefevre2014}. This increase of the typical size of large grains may contribute to the observed dependence of $\lambda_{\rm max}$, the peak of the polarization curve in extinction,  on the visual extinction $A_{\rm V}$  \citep{Wurm2002,Voshchinnikov2013,Voshchinnikov2014}.
 Dust growth through coagulation and accretion also modifies the shape of grains, therefore their polarization cross-sections, and $p_{\rm dust}$.  
So the study of the variation of polarization with $A_{\rm V}$ in dense shielded regions requires the modelling of both grain growth and alignment efficiency.
Grain growth may allow for a sustained alignment up to high column densities \citep{Andersson2015}.  
Thus, the product $p_{\rm dust}R$ may not be changing much from the backgrounds to the filaments.

\section{Conclusions}\label{conclusion}

We  have presented and analysed the \planck\ dust polarization maps towards three nearby star forming filaments:  
the Musca, B211, and L1506 filaments.These filaments 
can be recognized in the maps of intensity and Stokes \Q\ and \U\ parameters.
We use these maps to separate the filament 
emission from its background, and infer the structure of the magnetic field from the polarization properties.
This focussed study complements statistical analysis of \planck\ polarization observations of molecular clouds \citep{planck2014-XX,planck2014-XXXII,planck2015-XXXV}.
  
 \planck\ images allow us to describe the observed Stokes parameters with a two-component model, the filaments and their backgrounds.  
 We show that it is important to remove the background emission in all  three Stokes parameters, \I, \Q, and \U\ to properly measure the polarization properties ($p$ and $\psi$) intrinsic to  the filaments.   Both the polarization angle and fraction measured at the intensity peak of the filaments differ from their intrinsic values. 
 
In all three cases,  we measure variations in the polarization angle of the filaments ($\psi_{\rm fil}$) with respect to that of their 
backgrounds ($\psi_{\rm bg}$) and these variations are found to be coherent along the pc-scale length of the filaments.  
The  differences between $\psi_{\rm fil}$ and  $\psi_{\rm bg}$ for two of the three filaments are larger 
than the dispersion of the polarization angles.  Hence, these differences are not random fluctuations and they indicate a change in the orientation 
of the POS component of the magnetic field between the filaments and their backgrounds.
We also observe coherent variations of $\psi$ across the background and within the filaments. 
These observational results are all evidence for changes  of the 3D magnetic field structure. 

Like in earlier studies, we find a systematic decrease of the polarization fraction for increasing gas column density.
 For L1506 the change  of $\psi$ in the filament with respect to that of the background accounts for most of the observed drop of $p$ in the filament. 
We argue that the magnetic field structure contributes to the observed decrease of the polarization fraction in the filaments.  
We show that the depolarization in the filaments cannot be due to random fluctuations of $\psi$ 
because (1) the dispersion of $\psi$ is small ($10^\circ$) and much smaller than the value of $52^\circ$ for a random distribution, and (2) it is 
comparable in the filaments and their corresponding backgrounds.

Variations of the angle of \vec{B} with respect to the POS cannot explain the systematic decrease of $p$ with $N_{\rm H}$ either, but  unresolved  
structure of the magnetic field within the filaments may contribute to that decrease. 
Indeed, we find that the dispersion of $\psi$ in the filaments is comparable to, and even larger than, that in the background. 
These fluctuations of $\psi$ are not random but due to coherent variations along and across the filaments that trace the structure of the magnetic field within the filaments.
The drop in $p$ expected from the magnetic field structure does not preclude some contribution from variations of grain alignment with column density. 
Theoretical modelling is needed 
to test whether  the inner structure of the magnetic field may account for the observed decrease of 
$p$ in the filaments. 
Modelling is  also crucial to quantify the role that the magnetic field plays in the formation and evolution of star forming filaments.

The  ordered magnetic fields implied by the  small dispersion of the polarization angle measured inside and around the three filaments suggest that the magnetic field is  dynamically significant at the scale of the clouds. This is consistent with 
 recent studies showing that the relative orientation between \vec{B}$_{\rm POS}$ and the column density structures changes systematically with column density
\citep{planck2014-XXXII,planck2015-XXXV}. 
These results are also in agreement with the ordered morphology of magnetic fields observed from 100 pc to sub-pc scales \citep[see e.g.,][]{LiHb2014}.
  
Further analyses of a larger sample of filaments observed by \planck, but also 
higher angular resolution observations, are required 
to investigate the magnetic field structure in filaments.  
More extensive molecular line mapping of a larger sample of filaments is very desirable, in order to set stronger observational constraints on the dynamics of these structures, as well as to investigate the link between the velocity and the magnetic fields in molecular clouds.
Comparison with  dedicated numerical simulations will  also be  valuable in our understanding and interpretation of the observational results.

\begin{acknowledgements}
The Planck Collaboration acknowledges the support of: ESA; CNES, and
CNRS/INSU-IN2P3-INP (France); ASI, CNR, and INAF (Italy); NASA and DoE
(USA); STFC and UKSA (UK); CSIC, MINECO, JA and RES (Spain); Tekes, AoF,
and CSC (Finland); DLR and MPG (Germany); CSA (Canada); DTU Space
(Denmark); SER/SSO (Switzerland); RCN (Norway); SFI (Ireland);
FCT/MCTES (Portugal); ERC and PRACE (EU). A description of the Planck
Collaboration and a list of its members, indicating which technical
or scientific activities they have been involved in, can be found at
\url{http://www.cosmos.esa.int/web/planck/planck-collaboration}.
The research leading to these results has received funding from the European Research Council under the European Union's Seventh Framework Programme (FP7/2007-2013) / ERC grant agreement No. 267934.
\end{acknowledgements}

%
%\bibliography{AA,Planck_bib_4polar_17mar,Planck_bib}
\bibliographystyle{aa}
\bibliography{aa,Planck_bib}%,FB_bib}Planck_bib_4polar_17mar,
%\input{PlanckXXXIII_arXiv_July2015.bbl}

%-------------------------------------------------------------------

\begin{appendix}
\section{A two-component model along the LOS}\label{App2Layers}

The observed polarized emission results from the integration along the LOS of Stokes \Q\ and \U\ parameters of linearly polarized emission. Thus the observed polarization angle and fraction  correspond to mean values of the emission along the LOS, which  can be represented by the superposition of various layers/components with independent properties. 
A simplified  two-component model  applies as a first approximation to interstellar filaments and  their backgrounds.

 \subsection{Mean polarization angle along the LOS}\label{MeanPolAngle}

   Here, we estimate 
   the difference between the polarization angle $\psi$ observed towards the filament,  resulting from the integration of the emission along the LOS, and the intrinsic polarization angle of the filament $\psi_{\rm fil}$. This difference depends on   the polarized intensity contrast of the filament with respect to that of the background.  
 
The observed \Q\ and \U\ emission integrated over the filament and the background are
\begin{eqnarray}
Q =  P_{\rm bg} \, (f_{\rm d}\cos2\psi_{\rm fil} +\cos2\psi_{\rm bg}),\\
U =  P_{\rm bg} \, (f_{\rm d}\sin2\psi_{\rm fil}+\sin2\psi_{\rm bg}),\,\,\,
 \end{eqnarray}
\noindent
where $P_{\rm bg}$ is the polarized emission of the background,
and $f_{\rm d}=P_{\rm fil}/P_{\rm bg}$ the polarized intensity contrast between the two components.
If the two components have the same polarized intensities ($f_{\rm d}=1$) the observed polarization angle $\psi$ is equal to the average value, as can be seen in Fig.\,\ref{chi_df}. 
When the filament polarized intensity contrast is larger than one ($f_{\rm d}>1$) and $|\psi_{\rm fil}-\psi_{\rm bg}|>0$,  the difference between $\psi_{\rm fil}$ and $\psi$ 
decreases when  $f_{\rm d}$ increases.  For $f_{\rm d}>1$, this difference first increases with  $|\psi_{\rm fil}-\psi_{\rm bg}|$ and then decreases when  $|\psi_{\rm fil}-\psi_{\rm bg}|$ approaches $90^\circ$, while 
for $f_{\rm d}<1$, it increases with  $|\psi_{\rm fil}-\psi_{\rm bg}|$.
It is thus important to separate the components along the LOS,  to access  the underlying magnetic field orientation of the filament.

\begin{figure}
   \centerline{
   \hspace*{-0.3 cm}
 \resizebox{0.9\hsize}{!}{\includegraphics{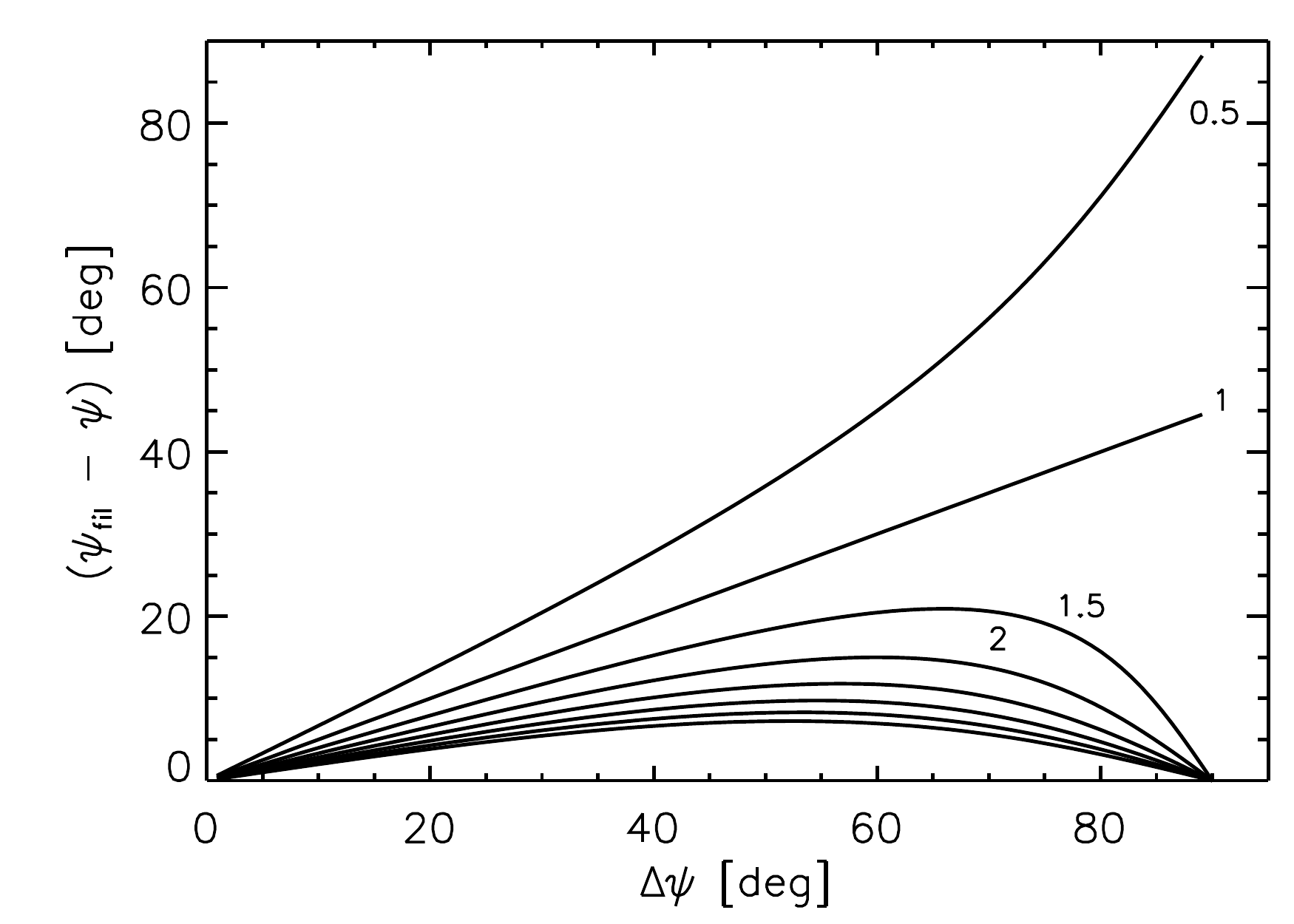}}}
   \caption{
   Difference between  the intrinsic polarization angle of the filament, $\psi_{\rm fil}$, and the observed polarization angle, $\psi$, as a function of $\Delta\psi=|\psi_{\rm fil}-\psi_{\rm bg}|$. 
    The polarized intensity contrast of the filament is indicated close to the corresponding curves.      Note that for $\Delta\psi=90^\circ$ and $f_{\rm d}=1$, $\psi$ is not defined and the resulting observed polarization fraction is null.         }
              \label{chi_df}
    \end{figure}
    
\subsection{Depolarization from  rotation of the POS component of the magnetic field}\label{depol} 

Here, we estimate the decrease  of the polarized emission with respect to the total emission 
 when two components  with different field orientations overlap along the LOS.

For the simplified two-component model, 
the  depolarization factor, $F$, 
 can be   expressed as
\begin{equation}
F=\frac{P}{P_{\rm fil}+P_{\rm bg}},\label{Feq}
\end{equation}
\noindent
where $P$ is the observed polarized emission summed over the filament and the background. 
Equation\,(\ref{Feq}) is equivalent to 
\begin{equation}
F^2=\frac{P^2}{P_{\rm fil}^2+P_{\rm bg}^2+2P_{\rm fil}P_{\rm bg}},
\end{equation}
\noindent
where $P^2=U^2+Q^2$ with $U=U_{\rm fil}+U_{\rm bg}$ and $Q=Q_{\rm fil}+Q_{\rm bg}$. Here \Q\ and \U\ are given by Eqs.\,(\ref{eq1}) and (\ref{eq2}) with the corresponding subscripts for the filament and the background.
Thus
\begin{equation}
P^2=P_{\rm fil}^2+P_{\rm bg}^2+2P_{\rm fil}P_{\rm bg}\cos2\Delta\psi\,,
\end{equation}
\noindent
where $\Delta\psi=|\psi_{\rm fil}-\psi_{\rm bg}|$.
These relations lead to
\begin{equation}
F^2=1-\frac{2P_{\rm fil}P_{\rm bg}(1-\cos2\Delta\psi)}{P_{\rm fil}^2+P_{\rm bg}^2+2P_{\rm fil}P_{\rm bg}},
\end{equation}
\noindent
which is equivalent to
\begin{equation}
F^2=1-2\frac{1-\cos2\Delta\psi}{(1+\beta)^2}\beta\,,
\end{equation}
\noindent
for $\beta=P_{\rm bg}/P_{\rm fil}$.
The maximal depolarization, i.e., the smallest $F$ value  $|\cos\Delta\psi|$,  is obtained for $ P_{\rm bg}=P_{\rm fil}$.

The observed polarized emission of two components, due to the combination of the  Stokes \Q\ and \U\ parameters  along the LOS,  results in depolarization if the POS magnetic field rotates in one component with respect to the other. $P$  is zero  if  $\Delta\psi=90^\circ$ and  the two components have the same  polarized emission ($P_{\rm fil}=P_{\rm bg}$, see Fig.\,\ref{DepolFig_App}). 
The depolarization factor towards the L1506 filament  is $F \simeq  0.6$ for $\Delta\psi=54^\circ$ and a contrast of polarized emission of about one.  For the Musca and B211 filament, $F$ is close to unity since the POS angles in the filaments are close ($\Delta\psi\sim10^\circ$) to that  of their backgrounds (cf. Table\,\ref{table}).

\begin{figure}
	          \centerline{ 
 \resizebox{0.9\hsize}{!}{\includegraphics{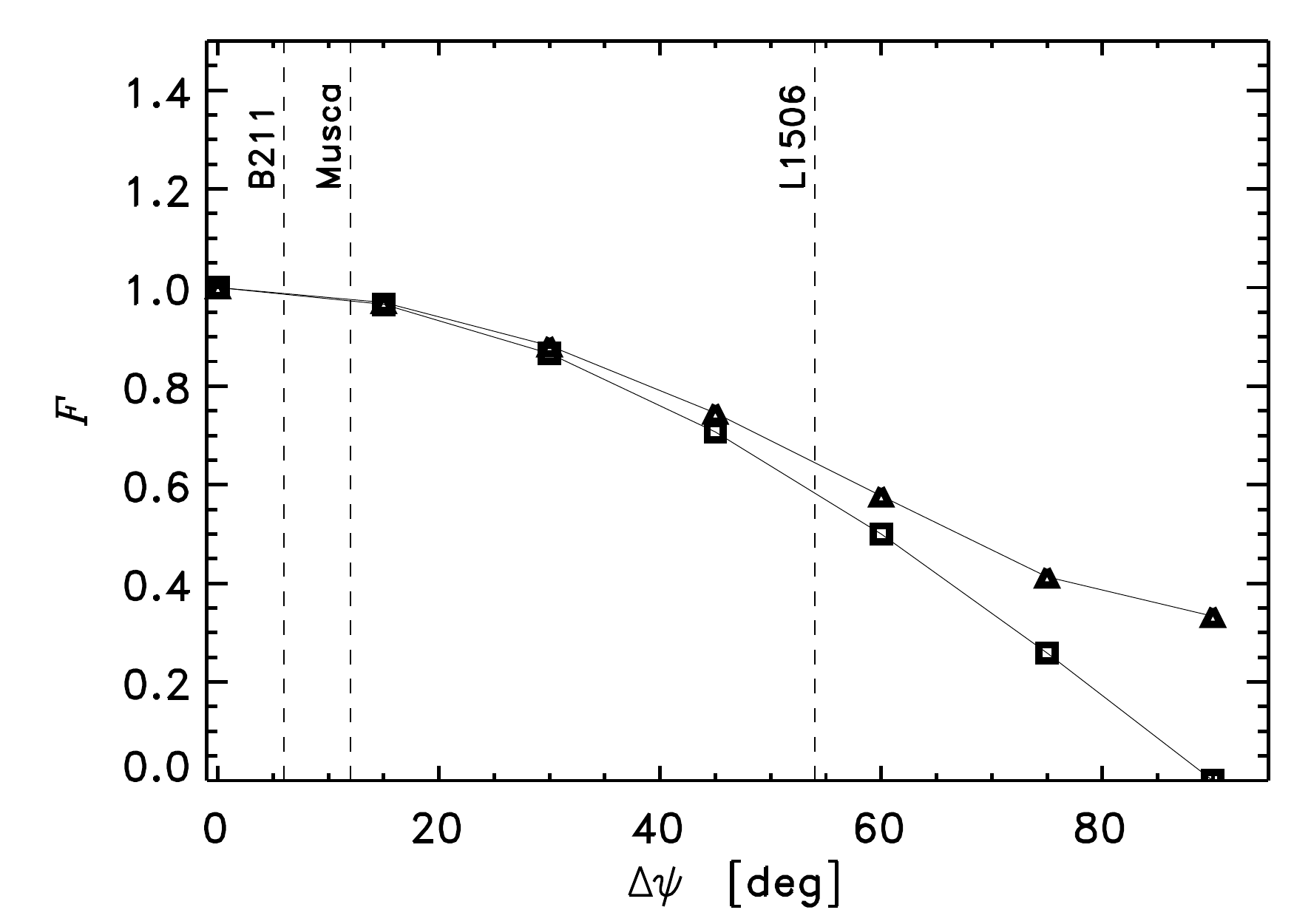}}
 }
% %\vspace*{-0.25 cm}
   \caption{ 
Depolarization factor due to the rotation of the field on the POS  for two components with the same polarized intensity (squares) and with a  contrast of two (triangles).
     }
              \label{DepolFig_App}
    \end{figure}

\end{appendix}

\end{document}

%% file: Planck.tex
\def\setsymbol#1#2{\expandafter\def\csname #1\endcsname{#2}}
\def\getsymbol#1{\csname #1\endcsname}

\def\Planck{\textit{Planck}}

\def\all2013resultspapers{\nocite{planck2013-p01, planck2013-p02, planck2013-p02a, planck2013-p02d, planck2013-p02b, planck2013-p03, planck2013-p03c, planck2013-p03f, planck2013-p03d, planck2013-p03e, planck2013-p01a, planck2013-p06, planck2013-p03a, planck2013-pip88, planck2013-p08, planck2013-p11, planck2013-p12, planck2013-p13, planck2013-p14, planck2013-p15, planck2013-p05b, planck2013-p17, planck2013-p09, planck2013-p09a, planck2013-p20, planck2013-p19, planck2013-pipaberration, planck2013-p05, planck2013-p05a, planck2013-pip56, planck2013-p06b}}

\newbox\tablebox    \newdimen\tablewidth
\def\leaderfil{\leaders\hbox to 5pt{\hss.\hss}\hfil}
\def\endPlancktablewide{\tablewidth=\textwidth 
    $$\hss\copy\tablebox\hss$$
    \vskip-\lastskip\vskip -2pt}
\def\tablenote#1 #2\par{\begingroup \parindent=0.8em
    \abovedisplayshortskip=0pt\belowdisplayshortskip=0pt
    \noindent
    $$\hss\vbox{\hsize\tablewidth \hangindent=\parindent \hangafter=1 \noindent
    \hbox to \parindent{$^#1$\hss}\strut#2\strut\par}\hss$$
    \endgroup}
\def\doubleline{\vskip 3pt\hrule \vskip 1.5pt \hrule \vskip 5pt}

\def\L2{\ifmmode L_2\else $L_2$\fi}

\def\DeltaT{\ifmmode \Delta T\else $\Delta T$\fi}
\def\deltat{\ifmmode \Delta t\else $\Delta t$\fi}
\def\fknee{\ifmmode f_{\rm knee}\else $f_{\rm knee}$\fi}
\def\Fmax{\ifmmode F_{\rm max}\else $F_{\rm max}$\fi}
\def\solar{\ifmmode{\rm M}_{\mathord\odot}\else${\rm M}_{\mathord\odot}$\fi}
\def\Msolar{\ifmmode{\rm M}_{\mathord\odot}\else${\rm M}_{\mathord\odot}$\fi}
\def\Lsolar{\ifmmode{\rm L}_{\mathord\odot}\else${\rm L}_{\mathord\odot}$\fi}

\def\inv{\ifmmode^{-1}\else$^{-1}$\fi}
\def\mo{\ifmmode^{-1}\else$^{-1}$\fi}
\def\sup#1{\ifmmode ^{\rm #1}\else $^{\rm #1}$\fi}
\def\expo#1{\ifmmode \times 10^{#1}\else $\times 10^{#1}$\fi}
\def\,{\thinspace}
\def\lsim{\mathrel{\raise .4ex\hbox{\rlap{$<$}\lower 1.2ex\hbox{$\sim$}}}}
\def\gsim{\mathrel{\raise .4ex\hbox{\rlap{$>$}\lower 1.2ex\hbox{$\sim$}}}}

\def\simprop{\mathrel{\raise .4ex\hbox{\rlap{$\propto$}\lower 1.2ex\hbox{$\sim$}}}}
\def\deg{\ifmmode^\circ\else$^\circ$\fi}
\def\pdeg{\ifmmode $\setbox0=\hbox{$^{\circ}$}\rlap{\hskip.11\wd0 .}$^{\circ}
          \else \setbox0=\hbox{$^{\circ}$}\rlap{\hskip.11\wd0 .}$^{\circ}$\fi}
\def\arcs{\ifmmode {^{\scriptstyle\prime\prime}}
          \else $^{\scriptstyle\prime\prime}$\fi}
\def\arcm{\ifmmode {^{\scriptstyle\prime}}
          \else $^{\scriptstyle\prime}$\fi}
\newdimen\sa  \newdimen\sb
\def\parcs{\sa=.07em \sb=.03em
     \ifmmode \hbox{\rlap{.}}^{\scriptstyle\prime\kern -\sb\prime}\hbox{\kern -\sa}
     \else \rlap{.}$^{\scriptstyle\prime\kern -\sb\prime}$\kern -\sa\fi}
\def\parcm{\sa=.08em \sb=.03em
     \ifmmode \hbox{\rlap{.}\kern\sa}^{\scriptstyle\prime}\hbox{\kern-\sb}
     \else \rlap{.}\kern\sa$^{\scriptstyle\prime}$\kern-\sb\fi}
\def\ra[#1 #2 #3.#4]{#1\sup{h}#2\sup{m}#3\sup{s}\llap.#4}
\def\dec[#1 #2 #3.#4]{#1\deg#2\arcm#3\arcs\llap.#4}
\def\deco[#1 #2 #3]{#1\deg#2\arcm#3\arcs}
\def\rra[#1 #2]{#1\sup{h}#2\sup{m}}

\def\dots{\relax\ifmmode \ldots\else $\ldots$\fi}
\def\WHzsr{\ifmmode $W\,Hz\mo\,sr\mo$\else W\,Hz\mo\,sr\mo\fi}
\def\mHz{\ifmmode $\,mHz$\else \,mHz\fi}
\def\GHz{\ifmmode $\,GHz$\else \,GHz\fi}
\def\mKs{\ifmmode $\,mK\,s$^{1/2}\else \,mK\,s$^{1/2}$\fi}
\def\muKs{\ifmmode \,\mu$K\,s$^{1/2}\else \,$\mu$K\,s$^{1/2}$\fi}
\def\muKRJs{\ifmmode \,\mu$K$_{\rm RJ}$\,s$^{1/2}\else \,$\mu$K$_{\rm RJ}$\,s$^{1/2}$\fi}
\def\muKHz{\ifmmode \,\mu$K\,Hz$^{-1/2}\else \,$\mu$K\,Hz$^{-1/2}$\fi}
\def\MJysr{\ifmmode \,$MJy\,sr\mo$\else \,MJy\,sr\mo\fi}
\def\MJysrmK{\ifmmode \,$MJy\,sr\mo$\,mK$_{\rm CMB}\mo\else \,MJy\,sr\mo\,mK$_{\rm CMB}\mo$\fi}
\def\microns{\ifmmode \,\mu$m$\else \,$\mu$m\fi}

\def\muK{\ifmmode \,\mu$K$\else \,$\mu$\hbox{K}\fi}
\def\microK{\ifmmode \,\mu$K$\else \,$\mu$\hbox{K}\fi}
\def\muW{\ifmmode \,\mu$W$\else \,$\mu$\hbox{W}\fi}
\def\kms{\ifmmode $\,km\,s$^{-1}\else \,km\,s$^{-1}$\fi}
\def\kmsMpc{\ifmmode $\,\kms\,Mpc\mo$\else \,\kms\,Mpc\mo\fi}
\setsymbol{LFI:center:frequency:70GHz:units}{70.3\,GHz}
\setsymbol{LFI:center:frequency:44GHz:units}{44.1\,GHz}
\setsymbol{LFI:center:frequency:30GHz:units}{28.5\,GHz}

\setsymbol{LFI:center:frequency:70GHz}{70.3}
\setsymbol{LFI:center:frequency:44GHz}{44.1}
\setsymbol{LFI:center:frequency:30GHz}{28.5}

\setsymbol{LFI:center:frequency:LFI18:Rad:M:units}{71.7\GHz}
\setsymbol{LFI:center:frequency:LFI19:Rad:M:units}{67.5\GHz}
\setsymbol{LFI:center:frequency:LFI20:Rad:M:units}{69.2\GHz}
\setsymbol{LFI:center:frequency:LFI21:Rad:M:units}{70.4\GHz}
\setsymbol{LFI:center:frequency:LFI22:Rad:M:units}{71.5\GHz}
\setsymbol{LFI:center:frequency:LFI23:Rad:M:units}{70.8\GHz}
\setsymbol{LFI:center:frequency:LFI24:Rad:M:units}{44.4\GHz}
\setsymbol{LFI:center:frequency:LFI25:Rad:M:units}{44.0\GHz}
\setsymbol{LFI:center:frequency:LFI26:Rad:M:units}{43.9\GHz}
\setsymbol{LFI:center:frequency:LFI27:Rad:M:units}{28.3\GHz}
\setsymbol{LFI:center:frequency:LFI28:Rad:M:units}{28.8\GHz}
\setsymbol{LFI:center:frequency:LFI18:Rad:S:units}{70.1\GHz}
\setsymbol{LFI:center:frequency:LFI19:Rad:S:units}{69.6\GHz}
\setsymbol{LFI:center:frequency:LFI20:Rad:S:units}{69.5\GHz}
\setsymbol{LFI:center:frequency:LFI21:Rad:S:units}{69.5\GHz}
\setsymbol{LFI:center:frequency:LFI22:Rad:S:units}{72.8\GHz}
\setsymbol{LFI:center:frequency:LFI23:Rad:S:units}{71.3\GHz}
\setsymbol{LFI:center:frequency:LFI24:Rad:S:units}{44.1\GHz}
\setsymbol{LFI:center:frequency:LFI25:Rad:S:units}{44.1\GHz}
\setsymbol{LFI:center:frequency:LFI26:Rad:S:units}{44.1\GHz}
\setsymbol{LFI:center:frequency:LFI27:Rad:S:units}{28.5\GHz}
\setsymbol{LFI:center:frequency:LFI28:Rad:S:units}{28.2\GHz}

\setsymbol{LFI:center:frequency:LFI18:Rad:M}{71.7}
\setsymbol{LFI:center:frequency:LFI19:Rad:M}{67.5}
\setsymbol{LFI:center:frequency:LFI20:Rad:M}{69.2}
\setsymbol{LFI:center:frequency:LFI21:Rad:M}{70.4}
\setsymbol{LFI:center:frequency:LFI22:Rad:M}{71.5}
\setsymbol{LFI:center:frequency:LFI23:Rad:M}{70.8}
\setsymbol{LFI:center:frequency:LFI24:Rad:M}{44.4}
\setsymbol{LFI:center:frequency:LFI25:Rad:M}{44.0}
\setsymbol{LFI:center:frequency:LFI26:Rad:M}{43.9}
\setsymbol{LFI:center:frequency:LFI27:Rad:M}{28.3}
\setsymbol{LFI:center:frequency:LFI28:Rad:M}{28.8}
\setsymbol{LFI:center:frequency:LFI18:Rad:S}{70.1}
\setsymbol{LFI:center:frequency:LFI19:Rad:S}{69.6}
\setsymbol{LFI:center:frequency:LFI20:Rad:S}{69.5}
\setsymbol{LFI:center:frequency:LFI21:Rad:S}{69.5}
\setsymbol{LFI:center:frequency:LFI22:Rad:S}{72.8}
\setsymbol{LFI:center:frequency:LFI23:Rad:S}{71.3}
\setsymbol{LFI:center:frequency:LFI24:Rad:S}{44.1}
\setsymbol{LFI:center:frequency:LFI25:Rad:S}{44.1}
\setsymbol{LFI:center:frequency:LFI26:Rad:S}{44.1}
\setsymbol{LFI:center:frequency:LFI27:Rad:S}{28.5}
\setsymbol{LFI:center:frequency:LFI28:Rad:S}{28.2}

\setsymbol{LFI:white:noise:sensitivity:70GHz:units}{134.7\muKs}
\setsymbol{LFI:white:noise:sensitivity:44GHz:units}{164.7\muKs}
\setsymbol{LFI:white:noise:sensitivity:30GHz:units}{143.4\muKs}

\setsymbol{LFI:white:noise:sensitivity:70GHz}{134.7}
\setsymbol{LFI:white:noise:sensitivity:44GHz}{164.7}
\setsymbol{LFI:white:noise:sensitivity:30GHz}{143.4}

\setsymbol{LFI:white:noise:sensitivity:LFI18:Rad:M:units}{512.0\muKs}
\setsymbol{LFI:white:noise:sensitivity:LFI19:Rad:M:units}{581.4\muKs}
\setsymbol{LFI:white:noise:sensitivity:LFI20:Rad:M:units}{590.8\muKs}
\setsymbol{LFI:white:noise:sensitivity:LFI21:Rad:M:units}{455.2\muKs}
\setsymbol{LFI:white:noise:sensitivity:LFI22:Rad:M:units}{492.0\muKs}
\setsymbol{LFI:white:noise:sensitivity:LFI23:Rad:M:units}{507.7\muKs}
\setsymbol{LFI:white:noise:sensitivity:LFI24:Rad:M:units}{462.2\muKs}
\setsymbol{LFI:white:noise:sensitivity:LFI25:Rad:M:units}{413.6\muKs}
\setsymbol{LFI:white:noise:sensitivity:LFI26:Rad:M:units}{478.6\muKs}
\setsymbol{LFI:white:noise:sensitivity:LFI27:Rad:M:units}{277.7\muKs}
\setsymbol{LFI:white:noise:sensitivity:LFI28:Rad:M:units}{312.3\muKs}
\setsymbol{LFI:white:noise:sensitivity:LFI18:Rad:S:units}{465.7\muKs}
\setsymbol{LFI:white:noise:sensitivity:LFI19:Rad:S:units}{555.6\muKs}
\setsymbol{LFI:white:noise:sensitivity:LFI20:Rad:S:units}{623.2\muKs}
\setsymbol{LFI:white:noise:sensitivity:LFI21:Rad:S:units}{564.1\muKs}
\setsymbol{LFI:white:noise:sensitivity:LFI22:Rad:S:units}{534.4\muKs}
\setsymbol{LFI:white:noise:sensitivity:LFI23:Rad:S:units}{542.4\muKs}
\setsymbol{LFI:white:noise:sensitivity:LFI24:Rad:S:units}{399.2\muKs}
\setsymbol{LFI:white:noise:sensitivity:LFI25:Rad:S:units}{392.6\muKs}
\setsymbol{LFI:white:noise:sensitivity:LFI26:Rad:S:units}{418.6\muKs}
\setsymbol{LFI:white:noise:sensitivity:LFI27:Rad:S:units}{302.9\muKs}
\setsymbol{LFI:white:noise:sensitivity:LFI28:Rad:S:units}{285.3\muKs}

\setsymbol{LFI:white:noise:sensitivity:LFI18:Rad:M}{512.0}
\setsymbol{LFI:white:noise:sensitivity:LFI19:Rad:M}{581.4}
\setsymbol{LFI:white:noise:sensitivity:LFI20:Rad:M}{590.8}
\setsymbol{LFI:white:noise:sensitivity:LFI21:Rad:M}{455.2}
\setsymbol{LFI:white:noise:sensitivity:LFI22:Rad:M}{492.0}
\setsymbol{LFI:white:noise:sensitivity:LFI23:Rad:M}{507.7}
\setsymbol{LFI:white:noise:sensitivity:LFI24:Rad:M}{462.2}
\setsymbol{LFI:white:noise:sensitivity:LFI25:Rad:M}{413.6}
\setsymbol{LFI:white:noise:sensitivity:LFI26:Rad:M}{478.6}
\setsymbol{LFI:white:noise:sensitivity:LFI27:Rad:M}{277.7}
\setsymbol{LFI:white:noise:sensitivity:LFI28:Rad:M}{312.3}
\setsymbol{LFI:white:noise:sensitivity:LFI18:Rad:S}{465.7}
\setsymbol{LFI:white:noise:sensitivity:LFI19:Rad:S}{555.6}
\setsymbol{LFI:white:noise:sensitivity:LFI20:Rad:S}{623.2}
\setsymbol{LFI:white:noise:sensitivity:LFI21:Rad:S}{564.1}
\setsymbol{LFI:white:noise:sensitivity:LFI22:Rad:S}{534.4}
\setsymbol{LFI:white:noise:sensitivity:LFI23:Rad:S}{542.4}
\setsymbol{LFI:white:noise:sensitivity:LFI24:Rad:S}{399.2}
\setsymbol{LFI:white:noise:sensitivity:LFI25:Rad:S}{392.6}
\setsymbol{LFI:white:noise:sensitivity:LFI26:Rad:S}{418.6}
\setsymbol{LFI:white:noise:sensitivity:LFI27:Rad:S}{302.9}
\setsymbol{LFI:white:noise:sensitivity:LFI28:Rad:S}{285.3}

\setsymbol{LFI:knee:frequency:70GHz:units}{29.5\mHz}
\setsymbol{LFI:knee:frequency:44GHz:units}{56.2\mHz}
\setsymbol{LFI:knee:frequency:30GHz:units}{113.7\mHz}

\setsymbol{LFI:knee:frequency:70GHz}{29.5}
\setsymbol{LFI:knee:frequency:44GHz}{56.2}
\setsymbol{LFI:knee:frequency:30GHz}{113.7}

\setsymbol{LFI:knee:frequency:LFI18:Rad:M:units}{16.3\mHz}
\setsymbol{LFI:knee:frequency:LFI19:Rad:M:units}{15.1\mHz}
\setsymbol{LFI:knee:frequency:LFI20:Rad:M:units}{18.7\mHz}
\setsymbol{LFI:knee:frequency:LFI21:Rad:M:units}{37.2\mHz}
\setsymbol{LFI:knee:frequency:LFI22:Rad:M:units}{12.7\mHz}
\setsymbol{LFI:knee:frequency:LFI23:Rad:M:units}{34.6\mHz}
\setsymbol{LFI:knee:frequency:LFI24:Rad:M:units}{46.2\mHz}
\setsymbol{LFI:knee:frequency:LFI25:Rad:M:units}{24.9\mHz}
\setsymbol{LFI:knee:frequency:LFI26:Rad:M:units}{67.6\mHz}
\setsymbol{LFI:knee:frequency:LFI27:Rad:M:units}{187.4\mHz}
\setsymbol{LFI:knee:frequency:LFI28:Rad:M:units}{122.2\mHz}
\setsymbol{LFI:knee:frequency:LFI18:Rad:S:units}{17.7\mHz}
\setsymbol{LFI:knee:frequency:LFI19:Rad:S:units}{22.0\mHz}
\setsymbol{LFI:knee:frequency:LFI20:Rad:S:units}{8.7\mHz}
\setsymbol{LFI:knee:frequency:LFI21:Rad:S:units}{25.9\mHz}
\setsymbol{LFI:knee:frequency:LFI22:Rad:S:units}{15.8\mHz}
\setsymbol{LFI:knee:frequency:LFI23:Rad:S:units}{129.8\mHz}
\setsymbol{LFI:knee:frequency:LFI24:Rad:S:units}{100.9\mHz}
\setsymbol{LFI:knee:frequency:LFI25:Rad:S:units}{38.9\mHz}
\setsymbol{LFI:knee:frequency:LFI26:Rad:S:units}{58.9\mHz}
\setsymbol{LFI:knee:frequency:LFI27:Rad:S:units}{104.4\mHz}
\setsymbol{LFI:knee:frequency:LFI28:Rad:S:units}{40.7\mHz}

\setsymbol{LFI:knee:frequency:LFI18:Rad:M}{16.3}
\setsymbol{LFI:knee:frequency:LFI19:Rad:M}{15.1}
\setsymbol{LFI:knee:frequency:LFI20:Rad:M}{18.7}
\setsymbol{LFI:knee:frequency:LFI21:Rad:M}{37.2}
\setsymbol{LFI:knee:frequency:LFI22:Rad:M}{12.7}
\setsymbol{LFI:knee:frequency:LFI23:Rad:M}{34.6}
\setsymbol{LFI:knee:frequency:LFI24:Rad:M}{46.2}
\setsymbol{LFI:knee:frequency:LFI25:Rad:M}{24.9}
\setsymbol{LFI:knee:frequency:LFI26:Rad:M}{67.6}
\setsymbol{LFI:knee:frequency:LFI27:Rad:M}{187.4}
\setsymbol{LFI:knee:frequency:LFI28:Rad:M}{122.2}
\setsymbol{LFI:knee:frequency:LFI18:Rad:S}{17.7}
\setsymbol{LFI:knee:frequency:LFI19:Rad:S}{22.0}
\setsymbol{LFI:knee:frequency:LFI20:Rad:S}{8.7}
\setsymbol{LFI:knee:frequency:LFI21:Rad:S}{25.9}
\setsymbol{LFI:knee:frequency:LFI22:Rad:S}{15.8}
\setsymbol{LFI:knee:frequency:LFI23:Rad:S}{129.8}
\setsymbol{LFI:knee:frequency:LFI24:Rad:S}{100.9}
\setsymbol{LFI:knee:frequency:LFI25:Rad:S}{38.9}
\setsymbol{LFI:knee:frequency:LFI26:Rad:S}{58.9}
\setsymbol{LFI:knee:frequency:LFI27:Rad:S}{104.4}
\setsymbol{LFI:knee:frequency:LFI28:Rad:S}{40.7}

\setsymbol{LFI:slope:70GHz:units}{$-1.03$\mHz}
\setsymbol{LFI:slope:44GHz:units}{$-0.89$\mHz}
\setsymbol{LFI:slope:30GHz:units}{$-0.87$\mHz}

\setsymbol{LFI:slope:70GHz}{$-1.03$}
\setsymbol{LFI:slope:44GHz}{$-0.89$}
\setsymbol{LFI:slope:30GHz}{$-0.87$}

\setsymbol{LFI:slope:LFI18:Rad:M:units}{$-1.04$\mHz}
\setsymbol{LFI:slope:LFI19:Rad:M:units}{$-1.09$\mHz}
\setsymbol{LFI:slope:LFI20:Rad:M:units}{$-0.69$\mHz}
\setsymbol{LFI:slope:LFI21:Rad:M:units}{$-1.56$\mHz}
\setsymbol{LFI:slope:LFI22:Rad:M:units}{$-1.01$\mHz}
\setsymbol{LFI:slope:LFI23:Rad:M:units}{$-0.96$\mHz}
\setsymbol{LFI:slope:LFI24:Rad:M:units}{$-0.83$\mHz}
\setsymbol{LFI:slope:LFI25:Rad:M:units}{$-0.91$\mHz}
\setsymbol{LFI:slope:LFI26:Rad:M:units}{$-0.95$\mHz}
\setsymbol{LFI:slope:LFI27:Rad:M:units}{$-0.87$\mHz}
\setsymbol{LFI:slope:LFI28:Rad:M:units}{$-0.88$\mHz}
\setsymbol{LFI:slope:LFI18:Rad:S:units}{$-1.15$\mHz}
\setsymbol{LFI:slope:LFI19:Rad:S:units}{$-1.00$\mHz}
\setsymbol{LFI:slope:LFI20:Rad:S:units}{$-0.95$\mHz}
\setsymbol{LFI:slope:LFI21:Rad:S:units}{$-0.92$\mHz}
\setsymbol{LFI:slope:LFI22:Rad:S:units}{$-1.01$\mHz}
\setsymbol{LFI:slope:LFI23:Rad:S:units}{$-0.95$\mHz}
\setsymbol{LFI:slope:LFI24:Rad:S:units}{$-0.73$\mHz}
\setsymbol{LFI:slope:LFI25:Rad:S:units}{$-1.16$\mHz}
\setsymbol{LFI:slope:LFI26:Rad:S:units}{$-0.79$\mHz}
\setsymbol{LFI:slope:LFI27:Rad:S:units}{$-0.82$\mHz}
\setsymbol{LFI:slope:LFI28:Rad:S:units}{$-0.91$\mHz}

\setsymbol{LFI:slope:LFI18:Rad:M}{$-1.04$}
\setsymbol{LFI:slope:LFI19:Rad:M}{$-1.09$}
\setsymbol{LFI:slope:LFI20:Rad:M}{$-0.69$}
\setsymbol{LFI:slope:LFI21:Rad:M}{$-1.56$}
\setsymbol{LFI:slope:LFI22:Rad:M}{$-1.01$}
\setsymbol{LFI:slope:LFI23:Rad:M}{$-0.96$}
\setsymbol{LFI:slope:LFI24:Rad:M}{$-0.83$}
\setsymbol{LFI:slope:LFI25:Rad:M}{$-0.91$}
\setsymbol{LFI:slope:LFI26:Rad:M}{$-0.95$}
\setsymbol{LFI:slope:LFI27:Rad:M}{$-0.87$}
\setsymbol{LFI:slope:LFI28:Rad:M}{$-0.88$}
\setsymbol{LFI:slope:LFI18:Rad:S}{$-1.15$}
\setsymbol{LFI:slope:LFI19:Rad:S}{$-1.00$}
\setsymbol{LFI:slope:LFI20:Rad:S}{$-0.95$}
\setsymbol{LFI:slope:LFI21:Rad:S}{$-0.92$}
\setsymbol{LFI:slope:LFI22:Rad:S}{$-1.01$}
\setsymbol{LFI:slope:LFI23:Rad:S}{$-0.95$}
\setsymbol{LFI:slope:LFI24:Rad:S}{$-0.73$}
\setsymbol{LFI:slope:LFI25:Rad:S}{$-1.16$}
\setsymbol{LFI:slope:LFI26:Rad:S}{$-0.79$}
\setsymbol{LFI:slope:LFI27:Rad:S}{$-0.82$}
\setsymbol{LFI:slope:LFI28:Rad:S}{$-0.91$}

\setsymbol{LFI:FWHM:70GHz:units}{13\parcm01}
\setsymbol{LFI:FWHM:44GHz:units}{27\parcm92}
\setsymbol{LFI:FWHM:30GHz:units}{32\parcm65}

\setsymbol{LFI:FWHM:70GHz}{13.01}
\setsymbol{LFI:FWHM:44GHz}{27.92}
\setsymbol{LFI:FWHM:30GHz}{32.65}

\setsymbol{LFI:FWHM:LFI18:units}{13\parcm39}
\setsymbol{LFI:FWHM:LFI19:units}{13\parcm01}
\setsymbol{LFI:FWHM:LFI20:units}{12\parcm75}
\setsymbol{LFI:FWHM:LFI21:units}{12\parcm74}
\setsymbol{LFI:FWHM:LFI22:units}{12\parcm87}
\setsymbol{LFI:FWHM:LFI23:units}{13\parcm27}
\setsymbol{LFI:FWHM:LFI24:units}{22\parcm98}
\setsymbol{LFI:FWHM:LFI25:units}{30\parcm46}
\setsymbol{LFI:FWHM:LFI26:units}{30\parcm31}
\setsymbol{LFI:FWHM:LFI27:units}{32\parcm65}
\setsymbol{LFI:FWHM:LFI28:units}{32\parcm66}

\setsymbol{LFI:FWHM:LFI18}{13.39}
\setsymbol{LFI:FWHM:LFI19}{13.01}
\setsymbol{LFI:FWHM:LFI20}{12.75}
\setsymbol{LFI:FWHM:LFI21}{12.74}
\setsymbol{LFI:FWHM:LFI22}{12.87}
\setsymbol{LFI:FWHM:LFI23}{13.27}
\setsymbol{LFI:FWHM:LFI24}{22.98}
\setsymbol{LFI:FWHM:LFI25}{30.46}
\setsymbol{LFI:FWHM:LFI26}{30.31}
\setsymbol{LFI:FWHM:LFI27}{32.65}
\setsymbol{LFI:FWHM:LFI28}{32.66}

\setsymbol{LFI:FWHM:uncertainty:LFI18:units}{0.170\arcm}
\setsymbol{LFI:FWHM:uncertainty:LFI19:units}{0.174\arcm}
\setsymbol{LFI:FWHM:uncertainty:LFI20:units}{0.170\arcm}
\setsymbol{LFI:FWHM:uncertainty:LFI21:units}{0.156\arcm}
\setsymbol{LFI:FWHM:uncertainty:LFI22:units}{0.164\arcm}
\setsymbol{LFI:FWHM:uncertainty:LFI23:units}{0.171\arcm}
\setsymbol{LFI:FWHM:uncertainty:LFI24:units}{0.652\arcm}
\setsymbol{LFI:FWHM:uncertainty:LFI25:units}{1.075\arcm}
\setsymbol{LFI:FWHM:uncertainty:LFI26:units}{1.131\arcm}
\setsymbol{LFI:FWHM:uncertainty:LFI27:units}{1.266\arcm}
\setsymbol{LFI:FWHM:uncertainty:LFI28:units}{1.287\arcm}

\setsymbol{LFI:FWHM:uncertainty:LFI18}{0.170}
\setsymbol{LFI:FWHM:uncertainty:LFI19}{0.174}
\setsymbol{LFI:FWHM:uncertainty:LFI20}{0.170}
\setsymbol{LFI:FWHM:uncertainty:LFI21}{0.156}
\setsymbol{LFI:FWHM:uncertainty:LFI22}{0.164}
\setsymbol{LFI:FWHM:uncertainty:LFI23}{0.171}
\setsymbol{LFI:FWHM:uncertainty:LFI24}{0.652}
\setsymbol{LFI:FWHM:uncertainty:LFI25}{1.075}
\setsymbol{LFI:FWHM:uncertainty:LFI26}{1.131}
\setsymbol{LFI:FWHM:uncertainty:LFI27}{1.266}
\setsymbol{LFI:FWHM:uncertainty:LFI28}{1.287}

\setsymbol{HFI:center:frequency:100GHz:units}{100\,GHz}
\setsymbol{HFI:center:frequency:143GHz:units}{143\,GHz}
\setsymbol{HFI:center:frequency:217GHz:units}{217\,GHz}
\setsymbol{HFI:center:frequency:353GHz:units}{353\,GHz}
\setsymbol{HFI:center:frequency:545GHz:units}{545\,GHz}
\setsymbol{HFI:center:frequency:857GHz:units}{857\,GHz}

\setsymbol{HFI:center:frequency:100GHz}{100}
\setsymbol{HFI:center:frequency:143GHz}{143}
\setsymbol{HFI:center:frequency:217GHz}{217}
\setsymbol{HFI:center:frequency:353GHz}{353}
\setsymbol{HFI:center:frequency:545GHz}{545}
\setsymbol{HFI:center:frequency:857GHz}{857}

\setsymbol{HFI:Ndetectors:100GHz}{8}
\setsymbol{HFI:Ndetectors:143GHz}{11}
\setsymbol{HFI:Ndetectors:217GHz}{12}
\setsymbol{HFI:Ndetectors:353GHz}{12}
\setsymbol{HFI:Ndetectors:545GHz}{3}
\setsymbol{HFI:Ndetectors:857GHz}{4}

\setsymbol{HFI:FWHM:Maps:100GHz:units}{9\parcm88}
\setsymbol{HFI:FWHM:Maps:143GHz:units}{7\parcm18}
\setsymbol{HFI:FWHM:Maps:217GHz:units}{4\parcm87}
\setsymbol{HFI:FWHM:Maps:353GHz:units}{4\parcm65}
\setsymbol{HFI:FWHM:Maps:545GHz:units}{4\parcm72}
\setsymbol{HFI:FWHM:Maps:857GHz:units}{4\parcm39}
\setsymbol{HFI:FWHM:Maps:100GHz}{9.88}
\setsymbol{HFI:FWHM:Maps:143GHz}{7.18}
\setsymbol{HFI:FWHM:Maps:217GHz}{4.87}
\setsymbol{HFI:FWHM:Maps:353GHz}{4.65}
\setsymbol{HFI:FWHM:Maps:545GHz}{4.72}
\setsymbol{HFI:FWHM:Maps:857GHz}{4.39}

\setsymbol{HFI:beam:ellipticity:Maps:100GHz}{1.15}
\setsymbol{HFI:beam:ellipticity:Maps:143GHz}{1.01}
\setsymbol{HFI:beam:ellipticity:Maps:217GHz}{1.06}
\setsymbol{HFI:beam:ellipticity:Maps:353GHz}{1.05}
\setsymbol{HFI:beam:ellipticity:Maps:545GHz}{1.14}
\setsymbol{HFI:beam:ellipticity:Maps:857GHz}{1.19}

\setsymbol{HFI:FWHM:Mars:100GHz:units}{9\parcm37}
\setsymbol{HFI:FWHM:Mars:143GHz:units}{7\parcm04}
\setsymbol{HFI:FWHM:Mars:217GHz:units}{4\parcm68}
\setsymbol{HFI:FWHM:Mars:353GHz:units}{4\parcm43}
\setsymbol{HFI:FWHM:Mars:545GHz:units}{3\parcm80}
\setsymbol{HFI:FWHM:Mars:857GHz:units}{3\parcm67}

\setsymbol{HFI:FWHM:Mars:100GHz}{9.37}
\setsymbol{HFI:FWHM:Mars:143GHz}{7.04}
\setsymbol{HFI:FWHM:Mars:217GHz}{4.68}
\setsymbol{HFI:FWHM:Mars:353GHz}{4.43}
\setsymbol{HFI:FWHM:Mars:545GHz}{3.80}
\setsymbol{HFI:FWHM:Mars:857GHz}{3.67}

\setsymbol{HFI:beam:ellipticity:Mars:100GHz}{1.18}
\setsymbol{HFI:beam:ellipticity:Mars:143GHz}{1.03}
\setsymbol{HFI:beam:ellipticity:Mars:217GHz}{1.14}
\setsymbol{HFI:beam:ellipticity:Mars:353GHz}{1.09}
\setsymbol{HFI:beam:ellipticity:Mars:545GHz}{1.25}
\setsymbol{HFI:beam:ellipticity:Mars:857GHz}{1.03}

\setsymbol{HFI:CMB:relative:calibration:100GHz}{$\lsim 1\%$}
\setsymbol{HFI:CMB:relative:calibration:143GHz}{$\lsim 1\%$}
\setsymbol{HFI:CMB:relative:calibration:217GHz}{$\lsim 1\%$}
\setsymbol{HFI:CMB:relative:calibration:353GHz}{$\lsim 1\%$}
\setsymbol{HFI:CMB:relative:calibration:545GHz}{}
\setsymbol{HFI:CMB:relative:calibration:857GHz}{}

\setsymbol{HFI:CMB:absolute:calibration:100GHz}{$\lsim 2\%$}
\setsymbol{HFI:CMB:absolute:calibration:143GHz}{$\lsim 2\%$}
\setsymbol{HFI:CMB:absolute:calibration:217GHz}{$\lsim 2\%$}
\setsymbol{HFI:CMB:absolute:calibration:353GHz}{$\lsim 2\%$}
\setsymbol{HFI:CMB:absolute:calibration:545GHz}{}
\setsymbol{HFI:CMB:absolute:calibration:857GHz}{}

\setsymbol{HFI:FIRAS:gain:calibration:accuracy:statistical:100GHz}{}
\setsymbol{HFI:FIRAS:gain:calibration:accuracy:statistical:143GHz}{}
\setsymbol{HFI:FIRAS:gain:calibration:accuracy:statistical:217GHz}{}
\setsymbol{HFI:FIRAS:gain:calibration:accuracy:statistical:353GHz}{2.5\%}
\setsymbol{HFI:FIRAS:gain:calibration:accuracy:statistical:545GHz}{1\%}
\setsymbol{HFI:FIRAS:gain:calibration:accuracy:statistical:857GHz}{0.5\%}

\setsymbol{HFI:FIRAS:gain:calibration:accuracy:systematic:100GHz}{}
\setsymbol{HFI:FIRAS:gain:calibration:accuracy:systematic:143GHz}{}
\setsymbol{HFI:FIRAS:gain:calibration:accuracy:systematic:217GHz}{}
\setsymbol{HFI:FIRAS:gain:calibration:accuracy:systematic:353GHz}{}
\setsymbol{HFI:FIRAS:gain:calibration:accuracy:systematic:545GHz}{7\%}
\setsymbol{HFI:FIRAS:gain:calibration:accuracy:systematic:857GHz}{7\%}

\setsymbol{HFI:FIRAS:zero:point:accuracy:100GHz:units}{0.8\MJysr}
\setsymbol{HFI:FIRAS:zero:point:accuracy:143GHz:units}{}
\setsymbol{HFI:FIRAS:zero:point:accuracy:217GHz:units}{}
\setsymbol{HFI:FIRAS:zero:point:accuracy:353GHz:units}{1.4\MJysr}
\setsymbol{HFI:FIRAS:zero:point:accuracy:545GHz:units}{2.2\MJysr}
\setsymbol{HFI:FIRAS:zero:point:accuracy:857GHz:units}{1.7\MJysr}

\setsymbol{HFI:FIRAS:zero:point:accuracy:100GHz}{0.8}
\setsymbol{HFI:FIRAS:zero:point:accuracy:143GHz}{}
\setsymbol{HFI:FIRAS:zero:point:accuracy:217GHz}{}
\setsymbol{HFI:FIRAS:zero:point:accuracy:353GHz}{1.4}
\setsymbol{HFI:FIRAS:zero:point:accuracy:545GHz}{2.2}
\setsymbol{HFI:FIRAS:zero:point:accuracy:857GHz}{1.7}

\setsymbol{HFI:unit:conversion:100GHz:units}{0.2415\MJysrmK}
\setsymbol{HFI:unit:conversion:143GHz:units}{0.3694\MJysrmK}
\setsymbol{HFI:unit:conversion:217GHz:units}{0.4811\MJysrmK}
\setsymbol{HFI:unit:conversion:353GHz:units}{0.2883\MJysrmK}
\setsymbol{HFI:unit:conversion:545GHz:units}{0.05826\MJysrmK}
\setsymbol{HFI:unit:conversion:857GHz:units}{0.002238\MJysrmK}

\setsymbol{HFI:unit:conversion:100GHz}{0.2415}
\setsymbol{HFI:unit:conversion:143GHz}{0.3694}
\setsymbol{HFI:unit:conversion:217GHz}{0.4811}
\setsymbol{HFI:unit:conversion:353GHz}{0.2883}
\setsymbol{HFI:unit:conversion:545GHz}{0.05826}
\setsymbol{HFI:unit:conversion:857GHz}{0.002238}

\setsymbol{HFI:colour:correction:alpha=-2:V1.01:100GHz}{0.9893}
\setsymbol{HFI:colour:correction:alpha=-2:V1.01:143GHz}{0.9759}
\setsymbol{HFI:colour:correction:alpha=-2:V1.01:217GHz}{1.0007}
\setsymbol{HFI:colour:correction:alpha=-2:V1.01:353GHz}{1.0028}
\setsymbol{HFI:colour:correction:alpha=-2:V1.01:545GHz}{1.0019}
\setsymbol{HFI:colour:correction:alpha=-2:V1.01:857GHz}{0.9889}

\setsymbol{HFI:colour:correction:alpha=0:V1.01:100GHz}{1.0008}
\setsymbol{HFI:colour:correction:alpha=0:V1.01:143GHz}{1.0148}
\setsymbol{HFI:colour:correction:alpha=0:V1.01:217GHz}{0.9909}
\setsymbol{HFI:colour:correction:alpha=0:V1.01:353GHz}{0.9888}
\setsymbol{HFI:colour:correction:alpha=0:V1.01:545GHz}{0.9878}
\setsymbol{HFI:colour:correction:alpha=0:V1.01:857GHz}{1.0014}

\providecommand{\sorthelp}[1]{}

%% file: PIP_108_Arzoumanian_authors_and_institutes.tex
%This author list corresponds to \title{Author list for SVN PIP\_108\_Arzoumanian: Signature of the magnetic field geometry of interstellar filaments in the maps of dust polarization}
%Prepared by R. Leonardi (rleonardi@sciops.esa.int), ESAC/ESA
%This version is from Thu Nov 06 16:37:22 2014 CET
%\subtitle{There are 189 co-authors in this list}
\author{\small
Planck Collaboration:
P.~A.~R.~Ade\inst{78}
\and
N.~Aghanim\inst{54}
\and
M.~I.~R.~Alves\inst{54}
\and
M.~Arnaud\inst{66}
\and
D.~Arzoumanian\inst{54}\thanks{Corresponding\,\,author:\,\,D.\,\,Arzoumanian,\,\,\url{doris.arzoumanian@ias.u-psud.fr}}
\and
J.~Aumont\inst{54}
\and
C.~Baccigalupi\inst{77}
\and
A.~J.~Banday\inst{83, 9}
\and
R.~B.~Barreiro\inst{59}
% ...
}
\institute{\small
APC, AstroParticule et Cosmologie, Universit\'{e} Paris Diderot, CNRS/IN2P3, CEA/lrfu, Observatoire de Paris, Sorbonne Paris Cit\'{e}, 10, rue Alice Domon et L\'{e}onie Duquet, 75205 Paris Cedex 13, France\goodbreak
\and
African Institute for Mathematical Sciences, 6-8 Melrose Road, Muizenberg, Cape Town, South Africa\goodbreak
\and
Agenzia Spaziale Italiana Science Data Center, Via del Politecnico snc, 00133, Roma, Italy\goodbreak
\and
Agenzia Spaziale Italiana, Viale Liegi 26, Roma, Italy\goodbreak
\and
Aix Marseille Universit\'{e}, CNRS, LAM (Laboratoire d'Astrophysique de Marseille) UMR 7326, 13388, Marseille, France\goodbreak
\and
Astrophysics Group, Cavendish Laboratory, University of Cambridge, J J Thomson Avenue, Cambridge CB3 0HE, U.K.\goodbreak
\and
Astrophysics \& Cosmology Research Unit, School of Mathematics, Statistics \& Computer Science, University of KwaZulu-Natal, Westville Campus, Private Bag X54001, Durban 4000, South Africa\goodbreak
\and
CITA, University of Toronto, 60 St. George St., Toronto, ON M5S 3H8, Canada\goodbreak
\and
CNRS, IRAP, 9 Av. colonel Roche, BP 44346, F-31028 Toulouse cedex 4, France\goodbreak
\and
California Institute of Technology, Pasadena, California, U.S.A.\goodbreak
\and
Centro de Estudios de F\'{i}sica del Cosmos de Arag\'{o}n (CEFCA), Plaza San Juan, 1, planta 2, E-44001, Teruel, Spain\goodbreak
\and
Computational Cosmology Center, Lawrence Berkeley National Laboratory, Berkeley, California, U.S.A.\goodbreak
\and
Consejo Superior de Investigaciones Cient\'{\i}ficas (CSIC), Madrid, Spain\goodbreak
\and
DSM/Irfu/SPP, CEA-Saclay, F-91191 Gif-sur-Yvette Cedex, France\goodbreak
\and
DTU Space, National Space Institute, Technical University of Denmark, Elektrovej 327, DK-2800 Kgs. Lyngby, Denmark\goodbreak
\and
D\'{e}partement de Physique Th\'{e}orique, Universit\'{e} de Gen\`{e}ve, 24, Quai E. Ansermet,1211 Gen\`{e}ve 4, Switzerland\goodbreak
\and
Departamento de F\'{\i}sica, Universidad de Oviedo, Avda. Calvo Sotelo s/n, Oviedo, Spain\goodbreak
\and
Department of Astrophysics/IMAPP, Radboud University Nijmegen, P.O. Box 9010, 6500 GL Nijmegen, The Netherlands\goodbreak
\and
Department of Physics \& Astronomy, University of British Columbia, 6224 Agricultural Road, Vancouver, British Columbia, Canada\goodbreak
\and
Department of Physics and Astronomy, Dana and David Dornsife College of Letter, Arts and Sciences, University of Southern California, Los Angeles, CA 90089, U.S.A.\goodbreak
\and
Department of Physics and Astronomy, University College London, London WC1E 6BT, U.K.\goodbreak
\and
Department of Physics, Florida State University, Keen Physics Building, 77 Chieftan Way, Tallahassee, Florida, U.S.A.\goodbreak
\and
Department of Physics, Gustaf H\"{a}llstr\"{o}min katu 2a, University of Helsinki, Helsinki, Finland\goodbreak
\and
Department of Physics, Princeton University, Princeton, New Jersey, U.S.A.\goodbreak
\and
Department of Physics, University of California, Santa Barbara, California, U.S.A.\goodbreak
\and
Department of Physics, University of Illinois at Urbana-Champaign, 1110 West Green Street, Urbana, Illinois, U.S.A.\goodbreak
\and
Dipartimento di Fisica e Astronomia G. Galilei, Universit\`{a} degli Studi di Padova, via Marzolo 8, 35131 Padova, Italy\goodbreak
\and
Dipartimento di Fisica e Scienze della Terra, Universit\`{a} di Ferrara, Via Saragat 1, 44122 Ferrara, Italy\goodbreak
\and
Dipartimento di Fisica, Universit\`{a} La Sapienza, P. le A. Moro 2, Roma, Italy\goodbreak
\and
Dipartimento di Fisica, Universit\`{a} degli Studi di Milano, Via Celoria, 16, Milano, Italy\goodbreak
\and
Dipartimento di Fisica, Universit\`{a} degli Studi di Trieste, via A. Valerio 2, Trieste, Italy\goodbreak
\and
Dipartimento di Fisica, Universit\`{a} di Roma Tor Vergata, Via della Ricerca Scientifica, 1, Roma, Italy\goodbreak
\and
Dipartimento di Matematica, Universit\`{a} di Roma Tor Vergata, Via della Ricerca Scientifica, 1, Roma, Italy\goodbreak
\and
Discovery Center, Niels Bohr Institute, Blegdamsvej 17, Copenhagen, Denmark\goodbreak
\and
Dpto. Astrof\'{i}sica, Universidad de La Laguna (ULL), E-38206 La Laguna, Tenerife, Spain\goodbreak
\and
European Space Agency, ESAC, Planck Science Office, Camino bajo del Castillo, s/n, Urbanizaci\'{o}n Villafranca del Castillo, Villanueva de la Ca\~{n}ada, Madrid, Spain\goodbreak
\and
European Space Agency, ESTEC, Keplerlaan 1, 2201 AZ Noordwijk, The Netherlands\goodbreak
\and
Facolt\`{a} di Ingegneria, Universit\`{a} degli Studi e-Campus, Via Isimbardi 10, Novedrate (CO), 22060, Italy\goodbreak
\and
Gran Sasso Science Institute, INFN, viale F. Crispi 7, 67100 L'Aquila, Italy\goodbreak
\and
HGSFP and University of Heidelberg, Theoretical Physics Department, Philosophenweg 16, 69120, Heidelberg, Germany\goodbreak
\and
Helsinki Institute of Physics, Gustaf H\"{a}llstr\"{o}min katu 2, University of Helsinki, Helsinki, Finland\goodbreak
\and
INAF - Osservatorio Astronomico di Padova, Vicolo dell'Osservatorio 5, Padova, Italy\goodbreak
\and
INAF - Osservatorio Astronomico di Roma, via di Frascati 33, Monte Porzio Catone, Italy\goodbreak
\and
INAF - Osservatorio Astronomico di Trieste, Via G.B. Tiepolo 11, Trieste, Italy\goodbreak
\and
INAF/IASF Bologna, Via Gobetti 101, Bologna, Italy\goodbreak
\and
INAF/IASF Milano, Via E. Bassini 15, Milano, Italy\goodbreak
\and
INFN, Sezione di Bologna, Via Irnerio 46, I-40126, Bologna, Italy\goodbreak
\and
INFN, Sezione di Roma 1, Universit\`{a} di Roma Sapienza, Piazzale Aldo Moro 2, 00185, Roma, Italy\goodbreak
\and
INFN, Sezione di Roma 2, Universit\`{a} di Roma Tor Vergata, Via della Ricerca Scientifica, 1, Roma, Italy\goodbreak
\and
INFN/National Institute for Nuclear Physics, Via Valerio 2, I-34127 Trieste, Italy\goodbreak
\and
IUCAA, Post Bag 4, Ganeshkhind, Pune University Campus, Pune 411 007, India\goodbreak
\and
Imperial College London, Astrophysics group, Blackett Laboratory, Prince Consort Road, London, SW7 2AZ, U.K.\goodbreak
\and
Infrared Processing and Analysis Center, California Institute of Technology, Pasadena, CA 91125, U.S.A.\goodbreak
\and
Institut d'Astrophysique Spatiale, CNRS (UMR8617) Universit\'{e} Paris-Sud 11, B\^{a}timent 121, Orsay, France\goodbreak
\and
Institut d'Astrophysique de Paris, CNRS (UMR7095), 98 bis Boulevard Arago, F-75014, Paris, France\goodbreak
\and
Institute of Astronomy, University of Cambridge, Madingley Road, Cambridge CB3 0HA, U.K.\goodbreak
\and
Institute of Theoretical Astrophysics, University of Oslo, Blindern, Oslo, Norway\goodbreak
\and
Instituto de Astrof\'{\i}sica de Canarias, C/V\'{\i}a L\'{a}ctea s/n, La Laguna, Tenerife, Spain\goodbreak
\and
Instituto de F\'{\i}sica de Cantabria (CSIC-Universidad de Cantabria), Avda. de los Castros s/n, Santander, Spain\goodbreak
\and
Istituto Nazionale di Fisica Nucleare, Sezione di Padova, via Marzolo 8, I-35131 Padova, Italy\goodbreak
\and
Jet Propulsion Laboratory, California Institute of Technology, 4800 Oak Grove Drive, Pasadena, California, U.S.A.\goodbreak
\and
Jodrell Bank Centre for Astrophysics, Alan Turing Building, School of Physics and Astronomy, The University of Manchester, Oxford Road, Manchester, M13 9PL, U.K.\goodbreak
\and
Kavli Institute for Cosmology Cambridge, Madingley Road, Cambridge, CB3 0HA, U.K.\goodbreak
\and
LAL, Universit\'{e} Paris-Sud, CNRS/IN2P3, Orsay, France\goodbreak
\and
LERMA, CNRS, Observatoire de Paris, 61 Avenue de l'Observatoire, Paris, France\goodbreak
\and
Laboratoire AIM, IRFU/Service d'Astrophysique - CEA/DSM - CNRS - Universit\'{e} Paris Diderot, B\^{a}t. 709, CEA-Saclay, F-91191 Gif-sur-Yvette Cedex, France\goodbreak
\and
Laboratoire Traitement et Communication de l'Information, CNRS (UMR 5141) and T\'{e}l\'{e}com ParisTech, 46 rue Barrault F-75634 Paris Cedex 13, France\goodbreak
\and
Laboratoire de Physique Subatomique et de Cosmologie, Universit\'{e} Joseph Fourier Grenoble I, CNRS/IN2P3, Institut National Polytechnique de Grenoble, 53 rue des Martyrs, 38026 Grenoble cedex, France\goodbreak
\and
Laboratoire de Physique Th\'{e}orique, Universit\'{e} Paris-Sud 11 \& CNRS, B\^{a}timent 210, 91405 Orsay, France\goodbreak
\and
Lawrence Berkeley National Laboratory, Berkeley, California, U.S.A.\goodbreak
\and
Max-Planck-Institut f\"{u}r Astrophysik, Karl-Schwarzschild-Str. 1, 85741 Garching, Germany\goodbreak
\and
McGill Physics, Ernest Rutherford Physics Building, McGill University, 3600 rue University, Montr\'{e}al, QC, H3A 2T8, Canada\goodbreak
\and
National University of Ireland, Department of Experimental Physics, Maynooth, Co. Kildare, Ireland\goodbreak
\and
Niels Bohr Institute, Blegdamsvej 17, Copenhagen, Denmark\goodbreak
\and
Observational Cosmology, Mail Stop 367-17, California Institute of Technology, Pasadena, CA, 91125, U.S.A.\goodbreak
\and
Optical Science Laboratory, University College London, Gower Street, London, U.K.\goodbreak
\and
SISSA, Astrophysics Sector, via Bonomea 265, 34136, Trieste, Italy\goodbreak
\and
School of Physics and Astronomy, Cardiff University, Queens Buildings, The Parade, Cardiff, CF24 3AA, U.K.\goodbreak
\and
Space Sciences Laboratory, University of California, Berkeley, California, U.S.A.\goodbreak
\and
Special Astrophysical Observatory, Russian Academy of Sciences, Nizhnij Arkhyz, Zelenchukskiy region, Karachai-Cherkessian Republic, 369167, Russia\goodbreak
\and
Sub-Department of Astrophysics, University of Oxford, Keble Road, Oxford OX1 3RH, U.K.\goodbreak
\and
UPMC Univ Paris 06, UMR7095, 98 bis Boulevard Arago, F-75014, Paris, France\goodbreak
\and
Universit\'{e} de Toulouse, UPS-OMP, IRAP, F-31028 Toulouse cedex 4, France\goodbreak
% ...
}

%% file: PlanckXXXIII_arXiv_July2015.bbl
\begin{thebibliography}{101}
\expandafter\ifx\csname natexlab\endcsname\relax\def\natexlab#1{#1}\fi

\bibitem[{{Abergel} {et~al.}(1994){Abergel}, {Boulanger}, {Mizuno}, \&
  {Fukui}}]{Abergel1994}
{Abergel}, A., {Boulanger}, F., {Mizuno}, A., \& {Fukui}, Y. 1994, ApJ, 423,
  L59

\bibitem[{{Andersson}(2015)}]{Andersson2015}
{Andersson}, B.-G. 2015, in Astrophysics and Space Science Library, Vol. 407,
  Astrophysics and Space Science Library, ed. A.~{Lazarian}, E.~M. {de Gouveia
  Dal Pino}, \& C.~{Melioli}, 59

\bibitem[{{Andersson} {et~al.}(2013){Andersson}, {Piirola}, {De Buizer},
  {Clemens}, {Uomoto}, {Charcos-Llorens}, {Geballe}, {Lazarian}, {Hoang}, \&
  {Vornanen}}]{Andersson2013}
{Andersson}, B.-G., {Piirola}, V., {De Buizer}, J., {et~al.} 2013, \apj, 775,
  84

\bibitem[{{Andersson} {et~al.}(2011){Andersson}, {Pintado}, {Potter}, {Strai{\v
  z}ys}, \& {Charcos-Llorens}}]{Andersson2011}
{Andersson}, B.-G., {Pintado}, O., {Potter}, S.~B., {Strai{\v z}ys}, V., \&
  {Charcos-Llorens}, M. 2011, \aap, 534, A19

\bibitem[{{Andersson} \& {Potter}(2010)}]{Andersson2010}
{Andersson}, B.-G. \& {Potter}, S.~B. 2010, \apj, 720, 1045

\bibitem[{{Andr{\'e}} {et~al.}(2014){Andr{\'e}}, {Di Francesco},
  {Ward-Thompson}, {Inutsuka}, {Pudritz}, \& {Pineda}}]{Andre2014}
{Andr{\'e}}, P., {Di Francesco}, J., {Ward-Thompson}, D., {et~al.} 2014,
  Protostars and Planets VI, 27

\bibitem[{{Andr{\'e}} {et~al.}(2010){Andr{\'e}}, {Men'shchikov}, {Bontemps},
  {K{\"o}nyves}, {Motte}, {Schneider}, {Didelon}, {Minier}, {Saraceno},
  {Ward-Thompson}, {di Francesco}, {White}, {Molinari}, {Testi}, {Abergel},
  {Griffin}, {Henning}, {Royer}, {Mer{\'{\i}}n}, {Vavrek}, {Attard},
  {Arzoumanian}, {Wilson}, {Ade}, {Aussel}, {Baluteau}, {Benedettini},
  {Bernard}, {Blommaert}, {Cambr{\'e}sy}, {Cox}, {di Giorgio}, {Hargrave},
  {Hennemann}, {Huang}, {Kirk}, {Krause}, {Launhardt}, {Leeks}, {Le Pennec},
  {Li}, {Martin}, {Maury}, {Olofsson}, {Omont}, {Peretto}, {Pezzuto}, {Prusti},
  {Roussel}, {Russeil}, {Sauvage}, {Sibthorpe}, {Sicilia-Aguilar}, {Spinoglio},
  {Waelkens}, {Woodcraft}, \& {Zavagno}}]{Andre2010}
{Andr{\'e}}, P., {Men'shchikov}, A., {Bontemps}, S., {et~al.} 2010, A{\&}A,
  518, L102

\bibitem[{{Arzoumanian} {et~al.}(2011){Arzoumanian}, {Andr{\'e}}, {Didelon},
  {K{\"o}nyves}, {Schneider}, {Men'shchikov}, {Sousbie}, {Zavagno}, {Bontemps},
  {di Francesco}, {Griffin}, {Hennemann}, {Hill}, {Kirk}, {Martin}, {Minier},
  {Molinari}, {Motte}, {Peretto}, {Pezzuto}, {Spinoglio}, {Ward-Thompson},
  {White}, \& {Wilson}}]{Arzoumanian2011}
{Arzoumanian}, D., {Andr{\'e}}, P., {Didelon}, P., {et~al.} 2011, A{\&}A, 529,
  L6

\bibitem[{{Attard} {et~al.}(2009){Attard}, {Houde}, {Novak}, {Li},
  {Vaillancourt}, {Dowell}, {Davidson}, \& {Shinnaga}}]{Attard2009}
{Attard}, M., {Houde}, M., {Novak}, G., {et~al.} 2009, \apj, 702, 1584

\bibitem[{{Bally} {et~al.}(1987){Bally}, {Lanber}, {Stark}, \&
  {Wilson}}]{Bally1987}
{Bally}, J., {Lanber}, W.~D., {Stark}, A.~A., \& {Wilson}, R.~W. 1987, ApJ,
  312, L45

\bibitem[{{Cambr{\'e}sy}(1999)}]{Cambresy1999}
{Cambr{\'e}sy}, L. 1999, A{\&}A, 345, 965

\bibitem[{{Carlqvist} \& {Kristen}(1997)}]{Carlqvist1997}
{Carlqvist}, P. \& {Kristen}, H. 1997, \aap, 324, 1115

\bibitem[{{Cashman} \& {Clemens}(2014)}]{Cashman2014}
{Cashman}, L.~R. \& {Clemens}, D.~P. 2014, \apj, 793, 126

\bibitem[{{Chapman} {et~al.}(2011){Chapman}, {Goldsmith}, {Pineda}, {Clemens},
  {Li}, \& {Kr{\v c}o}}]{Chapman2011}
{Chapman}, N.~L., {Goldsmith}, P.~F., {Pineda}, J.~L., {et~al.} 2011, \apj,
  741, 21

\bibitem[{{Crutcher} {et~al.}(2004){Crutcher}, {Nutter}, {Ward-Thompson}, \&
  {Kirk}}]{Crutcher2004}
{Crutcher}, R.~M., {Nutter}, D.~J., {Ward-Thompson}, D., \& {Kirk}, J.~M. 2004,
  \apj, 600, 279

\bibitem[{{Davis} \& {Greenstein}(1951)}]{Davis1951}
{Davis}, Jr., L. \& {Greenstein}, J.~L. 1951, \apj, 114, 206

\bibitem[{{Dolginov} \& {Mitrofanov}(1976)}]{Dolginov1976}
{Dolginov}, A.~Z. \& {Mitrofanov}, I.~G. 1976, \apss, 43, 291

\bibitem[{{Draine} \& {Weingartner}(1996)}]{Draine1996}
{Draine}, B.~T. \& {Weingartner}, J.~C. 1996, \apj, 470, 551

\bibitem[{{Draine} \& {Weingartner}(1997)}]{Draine1997}
{Draine}, B.~T. \& {Weingartner}, J.~C. 1997, \apj, 480, 633

\bibitem[{{Elias}(1978)}]{Elias1978}
{Elias}, J.~H. 1978, \apj, 224, 857

\bibitem[{{Falceta-Gon{\c c}alves} {et~al.}(2008){Falceta-Gon{\c c}alves},
  {Lazarian}, \& {Kowal}}]{Falceta-Goncalves2008}
{Falceta-Gon{\c c}alves}, D., {Lazarian}, A., \& {Kowal}, G. 2008, \apj, 679,
  537

\bibitem[{{Falceta-Gon{\c c}alves} {et~al.}(2009){Falceta-Gon{\c c}alves},
  {Lazarian}, \& {Kowal}}]{Falceta-Goncalves2009}
{Falceta-Gon{\c c}alves}, D., {Lazarian}, A., \& {Kowal}, G. 2009, in Revista
  Mexicana de Astronomia y Astrofisica Conference Series, Vol.~36, 37--44

\bibitem[{{Falgarone} {et~al.}(2001){Falgarone}, {Pety}, \&
  {Phillips}}]{Falgarone2001}
{Falgarone}, E., {Pety}, J., \& {Phillips}, T.~G. 2001, ApJ, 555, 178

\bibitem[{{Fiege} \& {Pudritz}(2000)}]{Fiege2000pol}
{Fiege}, J.~D. \& {Pudritz}, R.~E. 2000, \apj, 544, 830

\bibitem[{{Franco}(1991)}]{Franco1991}
{Franco}, G.~A.~P. 1991, \aap, 251, 581

\bibitem[{{Gerakines} {et~al.}(1995){Gerakines}, {Whittet}, \&
  {Lazarian}}]{Gerakines1995}
{Gerakines}, P.~A., {Whittet}, D.~C.~B., \& {Lazarian}, A. 1995, \apjl, 455,
  L171

\bibitem[{{Goldsmith} {et~al.}(2008){Goldsmith}, {Heyer}, {Narayanan}, {Snell},
  {Li}, \& {Brunt}}]{Goldsmith2008}
{Goldsmith}, P.~F., {Heyer}, M., {Narayanan}, G., {et~al.} 2008, ApJ, 680, 428

\bibitem[{{Goodman} {et~al.}(1990){Goodman}, {Bastien}, {Menard}, \&
  {Myers}}]{Goodman1990}
{Goodman}, A.~A., {Bastien}, P., {Menard}, F., \& {Myers}, P.~C. 1990, \apj,
  359, 363

\bibitem[{{Goodman} {et~al.}(1995){Goodman}, {Jones}, {Lada}, \&
  {Myers}}]{Goodman1995}
{Goodman}, A.~A., {Jones}, T.~J., {Lada}, E.~A., \& {Myers}, P.~C. 1995, \apj,
  448, 748

\bibitem[{{G{\'o}rski} {et~al.}(2005){G{\'o}rski}, {Hivon}, {Banday},
  {Wandelt}, {Hansen}, {Reinecke}, \& {Bartelmann}}]{Gorski2005}
{G{\'o}rski}, K.~M., {Hivon}, E., {Banday}, A.~J., {et~al.} 2005, \apj, 622,
  759

\bibitem[{{Gregorio Hetem} {et~al.}(1988){Gregorio Hetem}, {Sanzovo}, \&
  {Lepine}}]{Gregorio-hetem1988}
{Gregorio Hetem}, J.~C., {Sanzovo}, G.~C., \& {Lepine}, J.~R.~D. 1988, \aaps,
  76, 347

\bibitem[{{Hacar} {et~al.}(2013){Hacar}, {Tafalla}, {Kauffmann}, \&
  {Kov{\'a}cs}}]{Hacar2013}
{Hacar}, A., {Tafalla}, M., {Kauffmann}, J., \& {Kov{\'a}cs}, A. 2013, \aap,
  554, A55

\bibitem[{{Heyer} {et~al.}(2008){Heyer}, {Gong}, {Ostriker}, \&
  {Brunt}}]{Heyer2008}
{Heyer}, M., {Gong}, H., {Ostriker}, E., \& {Brunt}, C. 2008, ApJ, 680, 420

\bibitem[{{Hildebrand}(1983)}]{Hildebrand1983}
{Hildebrand}, R.~H. 1983, QJRAS, 24, 267

\bibitem[{{Hildebrand}(1988)}]{Hildebrand1988}
{Hildebrand}, R.~H. 1988, \qjras, 29, 327

\bibitem[{{Hily-Blant} \& {Falgarone}(2009)}]{HilyBlant2009}
{Hily-Blant}, P. \& {Falgarone}, E. 2009, A{\&}A, 500, L29

\bibitem[{{Hoang} \& {Lazarian}(2014)}]{Hoang2014}
{Hoang}, T. \& {Lazarian}, A. 2014, \mnras, 438, 680

\bibitem[{{Joncas} {et~al.}(1992){Joncas}, {Boulanger}, \&
  {Dewdney}}]{Joncas1992}
{Joncas}, G., {Boulanger}, F., \& {Dewdney}, P.~E. 1992, \apj, 397, 165

\bibitem[{{Jones} \& {Spitzer}(1967)}]{Jones1967}
{Jones}, R.~V. \& {Spitzer}, Jr., L. 1967, \apj, 147, 943

\bibitem[{{Jones} {et~al.}(2015){Jones}, {Bagley}, {Krejny}, {Andersson}, \&
  {Bastien}}]{Jones2015}
{Jones}, T.~J., {Bagley}, M., {Krejny}, M., {Andersson}, B.-G., \& {Bastien},
  P. 2015, \aj, 149, 31

\bibitem[{{Jones} {et~al.}(1992){Jones}, {Klebe}, \& {Dickey}}]{Jones1992}
{Jones}, T.~J., {Klebe}, D., \& {Dickey}, J.~M. 1992, \apj, 389, 602

\bibitem[{{Kenyon} {et~al.}(1994){Kenyon}, {Dobrzycka}, \&
  {Hartmann}}]{Kenyon1994}
{Kenyon}, S.~J., {Dobrzycka}, D., \& {Hartmann}, L. 1994, \aj, 108, 1872

\bibitem[{{Lazarian}(2007)}]{Lazarian2007}
{Lazarian}, A. 2007, \jqsrt, 106, 225

\bibitem[{{Lazarian} {et~al.}(1997){Lazarian}, {Goodman}, \&
  {Myers}}]{Lazarian1997}
{Lazarian}, A., {Goodman}, A.~A., \& {Myers}, P.~C. 1997, \apj, 490, 273

\bibitem[{{Lazarian} \& {Hoang}(2007)}]{LazarianHoang2007}
{Lazarian}, A. \& {Hoang}, T. 2007, \mnras, 378, 910

\bibitem[{{Lee} \& {Draine}(1985)}]{Lee1985}
{Lee}, H.~M. \& {Draine}, B.~T. 1985, \apj, 290, 211

\bibitem[{{Lef{\`e}vre} {et~al.}(2014){Lef{\`e}vre}, {Pagani}, {Juvela},
  {Paladini}, {Lallement}, {Marshall}, {Andersen}, {Bacmann}, {McGehee},
  {Montier}, {Noriega-Crespo}, {Pelkonen}, {Ristorcelli}, \&
  {Steinacker}}]{Lefevre2014}
{Lef{\`e}vre}, C., {Pagani}, L., {Juvela}, M., {et~al.} 2014, \aap, 572, A20

\bibitem[{{Li} \& {Goldsmith}(2012)}]{Li2012}
{Li}, D. \& {Goldsmith}, P.~F. 2012, \apj, 756, 12

\bibitem[{{Li} {et~al.}(2014){Li}, {Goodman}, {Sridharan}, {Houde}, {Li},
  {Novak}, \& {Tang}}]{LiHb2014}
{Li}, H.-B., {Goodman}, A., {Sridharan}, T.~K., {et~al.} 2014, Protostars and
  Planets VI, 101

\bibitem[{{Matthews} {et~al.}(2009){Matthews}, {McPhee}, {Fissel}, \&
  {Curran}}]{Matthews2009}
{Matthews}, B.~C., {McPhee}, C.~A., {Fissel}, L.~M., \& {Curran}, R.~L. 2009,
  \apjs, 182, 143

\bibitem[{{Matthews} {et~al.}(2001){Matthews}, {Wilson}, \&
  {Fiege}}]{Matthews2001}
{Matthews}, B.~C., {Wilson}, C.~D., \& {Fiege}, J.~D. 2001, \apj, 562, 400

\bibitem[{{Matthews} {et~al.}(2014){Matthews}, {Ade}, {Angil{\`e}}, {Benton},
  {Chapin}, {Chapman}, {Devlin}, {Fissel}, {Fukui}, {Gandilo}, {Gundersen},
  {Hargrave}, {Klein}, {Korotkov}, {Moncelsi}, {Mroczkowski}, {Netterfield},
  {Novak}, {Nutter}, {Olmi}, {Pascale}, {Poidevin}, {Savini}, {Scott},
  {Shariff}, {Soler}, {Tachihara}, {Thomas}, {Truch}, {Tucker}, {Tucker}, \&
  {Ward-Thompson}}]{Matthews2014}
{Matthews}, T.~G., {Ade}, P.~A.~R., {Angil{\`e}}, F.~E., {et~al.} 2014, \apj,
  784, 116

\bibitem[{{McClure-Griffiths} {et~al.}(2006){McClure-Griffiths}, {Dickey},
  {Gaensler}, {Green}, \& {Haverkorn}}]{McClure-Griffiths2006}
{McClure-Griffiths}, N.~M., {Dickey}, J.~M., {Gaensler}, B.~M., {Green}, A.~J.,
  \& {Haverkorn}, M. 2006, \apj, 652, 1339

\bibitem[{{Mizuno} {et~al.}(2001){Mizuno}, {Yamaguchi}, {Tachihara}, {Toyoda},
  {Aoyama}, {Yamamoto}, {Onishi}, \& {Fukui}}]{Mizuno2001}
{Mizuno}, A., {Yamaguchi}, R., {Tachihara}, K., {et~al.} 2001, \pasj, 53, 1071

\bibitem[{{Molinari} {et~al.}(2010){Molinari}, {Swinyard}, {Bally}, {Barlow},
  {Bernard}, {Martin}, {Moore}, {Noriega-Crespo}, {Plume}, {Testi}, {Zavagno},
  {Abergel}, {Ali}, {Anderson}, {Andr{\'e}}, {Baluteau}, {Battersby},
  {Beltr{\'a}n}, {Benedettini}, {Billot}, {Blommaert}, {Bontemps}, {Boulanger},
  {Brand}, {Brunt}, {Burton}, {Calzoletti}, {Carey}, {Caselli}, {Cesaroni},
  {Cernicharo}, {Chakrabarti}, {Chrysostomou}, {Cohen}, {Compiegne}, {de
  Bernardis}, {de Gasperis}, {di Giorgio}, {Elia}, {Faustini}, {Flagey},
  {Fukui}, {Fuller}, {Ganga}, {Garcia-Lario}, {Glenn}, {Goldsmith}, {Griffin},
  {Hoare}, {Huang}, {Ikhenaode}, {Joblin}, {Joncas}, {Juvela}, {Kirk},
  {Lagache}, {Li}, {Lim}, {Lord}, {Marengo}, {Marshall}, {Masi}, {Massi},
  {Matsuura}, {Minier}, {Miville-Desch{\^e}nes}, {Montier}, {Morgan}, {Motte},
  {Mottram}, {M{\"u}ller}, {Natoli}, {Neves}, {Olmi}, {Paladini}, {Paradis},
  {Parsons}, {Peretto}, {Pestalozzi}, {Pezzuto}, {Piacentini}, {Piazzo},
  {Polychroni}, {Pomar{\`e}s}, {Popescu}, {Reach}, {Ristorcelli}, {Robitaille},
  {Robitaille}, {Rod{\'o}n}, {Roy}, {Royer}, {Russeil}, {Saraceno}, {Sauvage},
  {Schilke}, {Schisano}, {Schneider}, {Schuller}, {Schulz}, {Sibthorpe},
  {Smith}, {Smith}, {Spinoglio}, {Stamatellos}, {Strafella}, {Stringfellow},
  {Sturm}, {Taylor}, {Thompson}, {Traficante}, {Tuffs}, {Umana}, {Valenziano},
  {Vavrek}, {Veneziani}, {Viti}, {Waelkens}, {Ward-Thompson}, {White},
  {Wilcock}, {Wyrowski}, {Yorke}, \& {Zhang}}]{Molinari2010}
{Molinari}, S., {Swinyard}, B., {Bally}, J., {et~al.} 2010, A{\&}A, 518, L100

\bibitem[{{Montier} {et~al.}(2015){Montier}, {Plaszczynski}, {Levrier},
  {Tristram}, {Alina}, {Ristorcelli}, {Bernard}, \& {Guillet}}]{Montier2014}
{Montier}, L., {Plaszczynski}, S., {Levrier}, F., {et~al.} 2015, \aap, 574,
  A136

\bibitem[{{Motte} {et~al.}(2010){Motte}, {Zavagno}, {Bontemps}, {Schneider},
  {Hennemann}, {di Francesco}, {Andr{\'e}}, {Saraceno}, {Griffin}, {Marston},
  {Ward-Thompson}, {White}, {Minier}, {Men'shchikov}, {Hill}, {Abergel},
  {Anderson}, {Aussel}, {Balog}, {Baluteau}, {Bernard}, {Cox}, {Csengeri},
  {Deharveng}, {Didelon}, {di Giorgio}, {Hargrave}, {Huang}, {Kirk}, {Leeks},
  {Li}, {Martin}, {Molinari}, {Nguyen-Luong}, {Olofsson}, {Persi}, {Peretto},
  {Pezzuto}, {Roussel}, {Russeil}, {Sadavoy}, {Sauvage}, {Sibthorpe},
  {Spinoglio}, {Testi}, {Teyssier}, {Vavrek}, {Wilson}, \&
  {Woodcraft}}]{Motte2010}
{Motte}, F., {Zavagno}, A., {Bontemps}, S., {et~al.} 2010, A{\&}A, 518, L77

\bibitem[{{Myers}(2009)}]{Myers2009}
{Myers}, P.~C. 2009, ApJ, 700, 1609

\bibitem[{{Myers} \& {Goodman}(1988)}]{Myers1988}
{Myers}, P.~C. \& {Goodman}, A.~A. 1988, ApJ, 326, L27

\bibitem[{{Myers} \& {Goodman}(1991)}]{Myers1991}
{Myers}, P.~C. \& {Goodman}, A.~A. 1991, \apj, 373, 509

\bibitem[{{Nakamura} \& {Li}(2008)}]{Nakamura2008}
{Nakamura}, F. \& {Li}, Z. 2008, ApJ, 687, 354

\bibitem[{{Ostriker} {et~al.}(2001){Ostriker}, {Stone}, \&
  {Gammie}}]{Ostriker2001}
{Ostriker}, E.~C., {Stone}, J.~M., \& {Gammie}, C.~F. 2001, \apj, 546, 980

\bibitem[{{Pagani} {et~al.}(2010){Pagani}, {Ristorcelli}, {Boudet}, {Giard},
  {Abergel}, \& {Bernard}}]{Pagani2010}
{Pagani}, L., {Ristorcelli}, I., {Boudet}, N., {et~al.} 2010, \aap, 512, A3

\bibitem[{{Palmeirim} {et~al.}(2013){Palmeirim}, {Andr{\'e}}, {Kirk},
  {Ward-Thompson}, {Arzoumanian}, {K{\"o}nyves}, {Didelon}, {Schneider},
  {Benedettini}, {Bontemps}, {Di Francesco}, {Elia}, {Griffin}, {Hennemann},
  {Hill}, {Martin}, {Men'shchikov}, {Molinari}, {Motte}, {Nguyen Luong},
  {Nutter}, {Peretto}, {Pezzuto}, {Roy}, {Rygl}, {Spinoglio}, \&
  {White}}]{Palmeirim2013}
{Palmeirim}, P., {Andr{\'e}}, P., {Kirk}, J., {et~al.} 2013, \aap, 550, A38

\bibitem[{{Pascale} {et~al.}(2012){Pascale}, {Ade}, {Angil{\`e}}, {Benton},
  {Devlin}, {Dober}, {Fissel}, {Fukui}, {Gandilo}, {Gundersen}, {Hargrave},
  {Klein}, {Korotkov}, {Matthews}, {Moncelsi}, {Mroczkowski}, {Netterfield},
  {Novak}, {Nutter}, {Olmi}, {Poidevin}, {Savini}, {Scott}, {Shariff}, {Soler},
  {Thomas}, {Truch}, {Tucker}, {Tucker}, \& {Ward-Thompson}}]{Pascale2012}
{Pascale}, E., {Ade}, P.~A.~R., {Angil{\`e}}, F.~E., {et~al.} 2012, in Society
  of Photo-Optical Instrumentation Engineers (SPIE) Conference Series, Vol.
  8444

\bibitem[{{Pereyra} \& {Magalh{\~a}es}(2004)}]{Pereyra2004}
{Pereyra}, A. \& {Magalh{\~a}es}, A.~M. 2004, ApJ, 603, 584

\bibitem[{{Planck HFI Core Team}(2011)}]{planck2011-1.5}
{Planck HFI Core Team}. 2011, \aap, 536, A4

\bibitem[{{\sorthelp{Planck Collaboration 2014A}}{Planck Collaboration
  I}(2014)}]{planck2013-p01}
{\sorthelp{Planck Collaboration 2014A}}{Planck Collaboration I}. 2014, \aap,
  571, A1

\bibitem[{{\sorthelp{Planck Collaboration 2014B}}{Planck Collaboration
  II}(2014)}]{planck2013-p02}
{\sorthelp{Planck Collaboration 2014B}}{Planck Collaboration II}. 2014, \aap,
  571, A2

\bibitem[{{\sorthelp{Planck Collaboration 2014K}}{Planck Collaboration
  XI}(2014)}]{planck2013-p06b}
{\sorthelp{Planck Collaboration 2014K}}{Planck Collaboration XI}. 2014, \aap,
  571, A11

\bibitem[{{\sorthelp{Planck Collaboration 2015A}}{Planck Collaboration
  I}(2015)}]{planck2014-a01}
{\sorthelp{Planck Collaboration 2015A}}{Planck Collaboration I}. 2015, \aap,
  submitted

\bibitem[{{\sorthelp{Planck Collaboration 2015J}}{Planck Collaboration
  X}(2015)}]{planck2014-a12}
{\sorthelp{Planck Collaboration 2015J}}{Planck Collaboration X}. 2015, \aap,
  submitted

\bibitem[{{\sorthelp{Planck Collaboration IntS}}{Planck Collaboration Int.
  XIX}(2015)}]{planck2014-XIX}
{\sorthelp{Planck Collaboration IntS}}{Planck Collaboration Int. XIX}. 2015,
  \aap, 576, A104

\bibitem[{{\sorthelp{Planck Collaboration IntT}}{Planck Collaboration Int.
  XX}(2015)}]{planck2014-XX}
{\sorthelp{Planck Collaboration IntT}}{Planck Collaboration Int. XX}. 2015,
  \aap, 576, A105

\bibitem[{{\sorthelp{Planck Collaboration IntU}}{Planck Collaboration Int.
  XXI}(2015)}]{planck2014-XXI}
{\sorthelp{Planck Collaboration IntU}}{Planck Collaboration Int. XXI}. 2015,
  \aap, 576, A106

\bibitem[{{\sorthelp{Planck Collaboration IntV}}{Planck Collaboration Int.
  XXII}(2015)}]{planck2014-XXII}
{\sorthelp{Planck Collaboration IntV}}{Planck Collaboration Int. XXII}. 2015,
  \aap, submitted, 576, A107

\bibitem[{{\sorthelp{Planck Collaboration IntZC}}{Planck Collaboration Int.
  XXVIII}(2014)}]{planck2014-XXVIII}
{\sorthelp{Planck Collaboration IntZC}}{Planck Collaboration Int. XXVIII}.
  2014, \aap, submitted

\bibitem[{{\sorthelp{Planck Collaboration IntZD}}{Planck Collaboration Int.
  XXIX}(2014)}]{planck2014-XXIX}
{\sorthelp{Planck Collaboration IntZD}}{Planck Collaboration Int. XXIX}. 2014,
  \aap, submitted

\bibitem[{{\sorthelp{Planck Collaboration IntZE}}{Planck Collaboration Int.
  XXX}(2014)}]{planck2014-XXX}
{\sorthelp{Planck Collaboration IntZE}}{Planck Collaboration Int. XXX}. 2014,
  \aap, in press

\bibitem[{{\sorthelp{Planck Collaboration IntZG}}{Planck Collaboration Int.
  XXXII}(2014)}]{planck2014-XXXII}
{\sorthelp{Planck Collaboration IntZG}}{Planck Collaboration Int. XXXII}. 2014,
  \aap, submitted

\bibitem[{{\sorthelp{Planck Collaboration IntZJ}}{Planck Collaboration Int.
  XXXV}(2015)}]{planck2015-XXXV}
{\sorthelp{Planck Collaboration IntZJ}}{Planck Collaboration Int. XXXV}. 2015,
  \aap, submitted

\bibitem[{{Plaszczynski} {et~al.}(2014){Plaszczynski}, {Montier}, {Levrier}, \&
  {Tristram}}]{Plaszczynski2014}
{Plaszczynski}, S., {Montier}, L., {Levrier}, F., \& {Tristram}, M. 2014,
  \mnras, 439, 4048

\bibitem[{{Purcell}(1979)}]{Purcell1979}
{Purcell}, E.~M. 1979, \apj, 231, 404

\bibitem[{{Roy} {et~al.}(2013){Roy}, {Martin}, {Polychroni}, {Bontemps},
  {Abergel}, {Andr{\'e}}, {Arzoumanian}, {Di Francesco}, {Hill}, {Konyves},
  {Nguyen-Luong}, {Pezzuto}, {Schneider}, {Testi}, \& {White}}]{Roy2013}
{Roy}, A., {Martin}, P.~G., {Polychroni}, D., {et~al.} 2013, \apj, 763, 55

\bibitem[{{Schlafly} {et~al.}(2014){Schlafly}, {Green}, {Finkbeiner}, {Rix},
  {Bell}, {Burgett}, {Chambers}, {Draper}, {Hodapp}, {Kaiser}, {Magnier},
  {Martin}, {Metcalfe}, {Price}, \& {Tonry}}]{Schlafly2014}
{Schlafly}, E.~F., {Green}, G., {Finkbeiner}, D.~P., {et~al.} 2014, \apj, 786,
  29

\bibitem[{{Schmalzl} {et~al.}(2010){Schmalzl}, {Kainulainen}, {Quanz}, {Alves},
  {Goodman}, {Henning}, {Launhardt}, {Pineda}, \&
  {Rom{\'a}n-Z{\'u}{\~n}iga}}]{Schmalzl2010}
{Schmalzl}, M., {Kainulainen}, J., {Quanz}, S.~P., {et~al.} 2010, ApJ, 725,
  1327

\bibitem[{{Schneider} \& {Elmegreen}(1979)}]{Schneider1979}
{Schneider}, S. \& {Elmegreen}, B.~G. 1979, ApJS, 41, 87

\bibitem[{{Shu} {et~al.}(1987){Shu}, {Adams}, \& {Lizano}}]{Shu1987}
{Shu}, F.~H., {Adams}, F.~C., \& {Lizano}, S. 1987, A{\&}A, 25, 23

\bibitem[{{Sousbie}(2011)}]{Sousbie2011}
{Sousbie}, T. 2011, MNRAS, 414, 350

\bibitem[{{Spitzer} \& {McGlynn}(1979)}]{Spitzer1979}
{Spitzer}, Jr., L. \& {McGlynn}, T.~A. 1979, \apj, 231, 417

\bibitem[{{Stepnik} {et~al.}(2003){Stepnik}, {Abergel}, {Bernard}, {Boulanger},
  {Cambr{\'e}sy}, {Giard}, {Jones}, {Lagache}, {Lamarre}, {Meny}, {Pajot}, {Le
  Peintre}, {Ristorcelli}, {Serra}, \& {Torre}}]{Stepnik2003}
{Stepnik}, B., {Abergel}, A., {Bernard}, J.-P., {et~al.} 2003, \aap, 398, 551

\bibitem[{{Sugitani} {et~al.}(2011){Sugitani}, {Nakamura}, {Watanabe},
  {Tamura}, {Nishiyama}, {Nagayama}, {Kandori}, {Nagata}, {Sato}, {Gutermuth},
  {Wilson}, \& {Kawabe}}]{Sugitani2011}
{Sugitani}, K., {Nakamura}, F., {Watanabe}, M., {et~al.} 2011, \apj, 734, 63

\bibitem[{{Tomisaka}(2014)}]{Tomisaka2014}
{Tomisaka}, K. 2014, \apj, 785, 24

\bibitem[{{Tomisaka}(2015)}]{Tomisaka2015}
{Tomisaka}, K. 2015, ArXiv e-prints

\bibitem[{{Voshchinnikov} {et~al.}(2013){Voshchinnikov}, {Das}, {Yakovlev}, \&
  {Il'in}}]{Voshchinnikov2013}
{Voshchinnikov}, N.~V., {Das}, H.~K., {Yakovlev}, I.~S., \& {Il'in}, V.~B.
  2013, Astronomy Letters, 39, 421

\bibitem[{{Voshchinnikov} \& {Hirashita}(2014)}]{Voshchinnikov2014}
{Voshchinnikov}, N.~V. \& {Hirashita}, H. 2014, \mnras, 445, 301

\bibitem[{{Ward-Thompson} {et~al.}(2000){Ward-Thompson}, {Kirk}, {Crutcher},
  {Greaves}, {Holland}, \& {Andr{\'e}}}]{Ward-Thompson2000}
{Ward-Thompson}, D., {Kirk}, J.~M., {Crutcher}, R.~M., {et~al.} 2000, \apjl,
  537, L135

\bibitem[{{Whittet} {et~al.}(2001){Whittet}, {Gerakines}, {Hough}, \&
  {Shenoy}}]{Whittet2001}
{Whittet}, D.~C.~B., {Gerakines}, P.~A., {Hough}, J.~H., \& {Shenoy}, S.~S.
  2001, \apj, 547, 872

\bibitem[{{Whittet} {et~al.}(2008){Whittet}, {Hough}, {Lazarian}, \&
  {Hoang}}]{Whittet2008}
{Whittet}, D.~C.~B., {Hough}, J.~H., {Lazarian}, A., \& {Hoang}, T. 2008, \apj,
  674, 304

\bibitem[{{Wurm} \& {Schnaiter}(2002)}]{Wurm2002}
{Wurm}, G. \& {Schnaiter}, M. 2002, \apj, 567, 370

\bibitem[{{Ysard} {et~al.}(2013){Ysard}, {Abergel}, {Ristorcelli}, {Juvela},
  {Pagani}, {K{\"o}nyves}, {Spencer}, {White}, \& {Zavagno}}]{Ysard2013}
{Ysard}, N., {Abergel}, A., {Ristorcelli}, I., {et~al.} 2013, \aap, 559, A133

\end{thebibliography}
